\documentclass[preprintnumbers,amsmath,amssymb,10pt]{revtex4}

\usepackage{bm,amsfonts,amssymb,amscd,amsmath}
\usepackage{graphicx}

\long\def\comment#1{ }
\newcommand{\order}[1]{\mathcal{O}{(#1)}}
\newcommand{\rmI}{{\rm I}}
\newcommand{\rmJ}{{\rm J}}

 \newcommand\beq{\begin{equation}}
 \newcommand\eeq{\end{equation}}
 \newcommand\beqn{\begin{eqnarray}}
 \newcommand\eeqn{\end{eqnarray}}

\newcommand{\nn}{\nonumber\\}

\newcommand{\rmd}{{\rm d}}
\newcommand{\rme}{{\rm e}}
\newcommand{\rmi}{{\rm i}}

\def\eq#1{{Eq.~(\ref{#1})}}
\def\fig#1{{Fig.~\ref{#1}}}

\def\k{{\bm k}}

\def\p{{\bm p}}
\def\r{{\bm r}}
\def\D{{\bm \Delta}}
\def\b{{\bm b}}
\def\B{{\bm B}}

\def\q{{\bm q}}
\def\x{{\bm x}}
\def\y{{\bm y}}
\def\z{{\bm z}}

\begin{document}

\title{Elliptic flow from color-dipole orientation in $pp$ and $pA$ collisions}
\author{Edmond Iancu$^{1}$ and Amir H. Rezaeian$^{2,3}$}

\affiliation{
$^1$ Institut de physique th\'eorique, Universit\'e Paris Saclay, CNRS, CEA, F-91191 Gif-sur-Yvette, France\\
$^2$ Departamento de F\'\i sica, Universidad T\'ecnica
Federico Santa Mar\'\i a, Avda. Espa\~na 1680,
Casilla 110-V, Valparaiso, Chile\\
$^3$  Centro Cient\'\i fico Tecnol\'ogico de Valpara\'\i so (CCTVal), Universidad T\'ecnica
Federico Santa Mar\'\i a, Casilla 110-V, Valpara\'\i so, Chile
}

\begin{abstract} 
For ultrarelativistic proton-proton and proton-nucleus collisions, we perform an
exploratory study of the contribution to the elliptic flow $v_2$
coming from the orientation of the momentum of the produced particles with respect to the reaction plane.
Via the CGC factorization valid at high energies, this contribution is related to the orientation of a color dipole with
respect to its impact parameter, which in turn probes the transverse inhomogeneity in the target.
Using the McLerran-Venugopalan model (with impact-parameter dependence) 
as an effective description for the soft gluon distribution in the (proton or nuclear) target, 
we present a semi-analytic calculation of the dipole scattering amplitude, including its angular dependence. 
We find that the angular dependence is controlled by soft gluon exchanges and hence is 
genuinely non-perturbative. The effects of multiple scattering turn out to be essential (in particular,
they change the sign of $v_2$). We  find that sizable values for $v_2$, comparable
to those observed in the LHC data and having a similar dependence upon the transverse momenta
of the produced particles, can be easily generated via peripheral collisions. 
In particular, $v_2$ develops a peak at a transverse momentum
which scales with the saturation momentum in the target.

\end{abstract}

\maketitle
\date{\today}

\section{Introduction}
 
The unexpectedly large azimuthal asymmetries in hadron production observed in high-multiplicity 
events in proton-proton ($pp$) and proton(deuteron)-nucleus ($pA$) collisions at the LHC 
and RHIC \cite{Khachatryan:2010gv,CMS:2012qk,Chatrchyan:2013nka,Abelev:2012ola,ABELEV:2013wsa,Abelev:2014mda,Aad:2012gla,Aad:2013fja,Aad:2014lta,Adare:2013piz,Adare:2014keg,Adamczyk:2014fcx,Adamczyk:2015xjc} have triggered intense debates 
concerning the physical origin of such phenomena. It is indeed an outstanding problem to understand how  
a small system like that produced in $pp$ or $pA$ collisions, an order of magnitude smaller than in
nucleus-nucleus ($AA$) collisions, can develop a collective behavior which is quite similar to that observed 
in $AA$ collisions, both in terms of its magnitude and in terms of its dependences upon the transverse 
momenta, the rapidities, and the masses of the produced hadrons 
\cite{Abelev:2012ola,ABELEV:2013wsa,Abelev:2014mda,Aad:2012gla,Aad:2013fja,Aad:2014lta}. 
Roughly speaking, the associated scientific debate opposes two paradigms. The first of them, which
is closer to the generally accepted interpretation of the corresponding phenomena in $AA$ collisions,
relates the azimuthal correlations observed in $pp$ and $pA$ collisions to `hydrodynamic flow', 
i.e. collective effects caused by strong interactions in the final state. Whereas such scenarios may
indeed lead to reasonable descriptions of the data (at least for sufficiently small transverse momenta
and with suitable choices for the initial conditions) \cite{dEnterria:2010xip,Bozek:2010pb,Bozek:2012gr,Bozek:2013uha,Bozek:2013ska,Qin:2013bha,Werner:2013ipa,Bzdak:2014dia,Weller:2017tsr}, 
it seems nevertheless difficult to conceive that  hydrodynamic flow may develop in such
small systems. This motivated the second paradigm, which rather builds upon the `initial state' physics, 
i.e. the collective phenomena associated with high parton densities in the wavefunctions of 
(one or both of) the incoming hadrons, prior to their collision  
\cite{Kovchegov:1999ep,Teaney:2002kn,Kovchegov:2002nf,Dumitru:2008wn,Gavin:2008ev,Dumitru:2010iy,Avsar:2010rf,Kovner:2010xk,Kovner:2011pe,Levin:2011fb,Iancu:2011ns,Schenke:2012wb,Schenke:2012fw,Dusling:2012iga,Dusling:2013qoz,Dusling:2015rja,
Kovchegov:2013ewa,Kovner:2012jm,Kovner:2014qea,Kovner:2015rna,Schenke:2015aqa,Lappi:2015vha,Lappi:2015vta,Rezaeian:2016szi,Schenke:2016ksl,Schenke:2016lrs}.

In practice, the azimuthal asymmetries are most conveniently measured via multi-particle angular 
correlation. But at a conceptual level, it is often preferable to think
in terms of the single-inclusive particle distribution event-by-event and its dependence upon 
the azimuthal angle $\phi$, as measured  w.r.t. the `reaction plane'. 
More precisely,  $\phi$ is the angle between the direction 
of motion of a produced hadron in the transverse plane and its impact factor.
Then the azimuthal asymmetries are encoded in the  `flow coefficients' $v_n(p_T)$ --- 
the $\cos(n\phi)$ Fourier moments of the single-inclusive distribution in $\phi$
(see e.g.  \cite{Snellings:2011sz}). From this perspective, 
the azimuthal asymmetries reflect a spontaneous breaking of rotational symmetry 
in the transverse plane, which may have various origins. The best known example is that of
non-central $AA$ collisions, where the rotational symmetry is broken by the elliptic shape
of the `interaction region' (the overlapping region between the two nuclei) in the transverse plane
\cite{Snellings:2011sz}. More generally (and including for central $AA$ collisions), azimuthal anisotropies can
be generated by fluctuations in the distribution of particles (nucleons, or even
gluons inside the participating nucleons) within the incoming nuclei \cite{Alver:2010gr}.
Clearly, the typical transverse sizes will be different for nucleon number, respectively, gluon
number fluctuations, potentially leading to different laws for the $p_T$-dependence of the
coefficients $v_n(p_T)$.

In the context of $AA$ collisions, the theoretical ideas concerning the particle (nucleon or
parton) number fluctuations are naturally embedded in the initial conditions for the hydrodynamical
equations. But such fluctuations can generate momentum-space azimuthal asymmetries 
 {\em already by themselves},
that is, even in the absence of interactions in the final state leading to (hydrodynamic) flow. This is
particularly interesting for the $pp$ and $pA$ collisions, where the importance of the final-state
interactions is far from being established. In this context, most of the calculations associated with the
`initial-state' paradigm alluded to above relied on the effective theory for
the color glass condensate (CGC) \cite{Iancu:2003xm,Gelis:2010nm}, which predicts the
event-by-event formation of `saturation domains'  inside a dense hadronic target.  These are regions 
with a typical transverse size $1/Q_s$ ($Q_s$ is the target saturation momentum) 
where gluons have large occupation numbers and are coherent with each other,
so they can also be described as condensates of strong chromo-electric fields.
These domains introduce a preferred direction in the transverse plane --- the orientation
of the color fields --- thus breaking the rotational symmetry.  In turn, this leads to azimuthal 
correlations in the particle production
in the collision between a dilute proton (the `projectile') and the
dense target: if 2 partons from the projectile have similar impact parameters (so that they scatter
off a same saturation domain) and they are in the same color state, then
they will receive similar kicks and hence emerge along nearby angles. 
The orientations of the saturation domains are of course random, so their effects will be washed out 
(by the averaging over the events) in the calculation of single-inclusive particle production.
But non-trivial correlations survive in the production of two (or more) particles, 
with very interesting features: the respective spectra are naturally `semi-hard' 
(the flow coefficients $v_n(p_T)$ are peaked around the saturation scale $Q_s$), their strength decreases
with increasing $Q_s$ (hence, in particular, with increasing energy), and they are
suppressed in the limit where the number of colors $N_c$ is large (they scale like $1/N_c^2$).
The `color field domain' model proposed in \cite{Kovner:2011pe,Dumitru:2014dra,Dumitru:2014vka,Dumitru:2014yza}
can be viewed too a (rather extreme) variant of this scenario: as shown in \cite{Lappi:2015vta}, 
the effects of these `color field domains' can be reproduced by fluctuating color 
fields in CGC provided non-Gaussian correlations are assumed to be important.

As it should be clear from the above discussion, the CGC-based approaches assume the $pp$ and $pA$ collisions
to be of the `dilute-dense' type. This is quite natural when the target is a large nucleus, such as
Pb with $A=208$, and it is also justified for $pp$ collisions so long as one considers particle production
at very forward rapidities --- meaning that the gluon distribution from the target proton has been subjected to
the high-energy (or small-$x$) evolution and hence is much denser than that of the projectile proton. 

In this paper, we shall remain within the general CGC
framework of `dilute-dense scattering', but we shall explore a more elementary mechanism for generating
azimuthal correlations: the case where the rotational symmetry in the transverse plane is broken simply
by the impact parameter (a 2-dimensional vector $\b$) of an impinging parton from the projectile, i.e. by 
the very fact that a parton hits the target disk at some point $\b$ which is away from the center
($b\ne 0$). Via its scattering, the parton will acquire some transverse momentum $\p$ and the respective
cross-section will generally dependent upon the angle $\phi$ made by the vectors $\p$ and $\b$.
 (From now on, we suppress the  subscript $T$ on transverse momenta or coordinates, to simplify writing.) 
For this to be the case, the cross-section must depend upon $\b$ in the first place, that is, 
the target should have some inhomogeneity in the transverse plane. Accordingly, this mechanism
naturally generates azimuthal asymmetries which probe the {\em variation} of the
transverse distribution of matter in the target.  These asymmetries do not require non-planar 
gluon exchanges, hence they admit a non-zero limit when $N_c\to\infty$.

The basic idea is not new --- it has been originally proposed in 
Refs.~\cite{Kopeliovich:2007fv,Kopeliovich:2007sd,Kopeliovich:2008nx} and more recently revisited
 in Ref.~\cite{Levin:2011fb,Zhou:2016rnt,Hagiwara:2017ofm}. As in these previous works,
we shall use the high-energy factorization for particle production in `dilute-dense' collisions in which
the cross-section for the production of a parton with transverse momentum $\p$ at $\b$
is related to the Fourier transform ($\r\to\p$) of the $S$-matrix $S(\r,\b)$ for the elastic scattering
a color dipole with transverse size $\r$ and  impact parameter $\b$ (say, 
a quark-antiquark dipole for the case of quark production, that we shall focus here on, for definiteness).
In this framework, the azimuthal asymmetry results from the dependence of the function $S(\r,\b)$ upon the 
dipole orientation (the angle $\theta$ made by the vectors $\r$ and $\b$). 

As compared to the previous
literature, we shall use a different theoretical description for the gluon distribution in the target,
namely, the McLerran-Venugopalan (MV) model \cite{McLerran:1993ka}, that we here extend
to include inhomogeneity in the transverse plane. Our respective extension will be inspired by `saturation
models' like IP-Sat \cite{Kowalski:2003hm,Kowalski:2008sa,Rezaeian:2012ji} 
and bCGC  \cite{Iancu:2003ge,Kowalski:2006hc,Rezaeian:2013tka}, 
which include a non-trivial $b$-dependence that has been tested
and calibrated via fits to the HERA data for diffractive vector meson production.
(See also Refs.~\cite{Munier:2001nr,Berger:2011ew,Berger:2012wx,Armesto:2014sma} for other
studies of the HERA phenomenology which explore the impact-parameter dependence
of the saturation physics.)
Note that one cannot directly use those `saturation models' for the present purposes, since
they are formulated as parametrizations for the dipole $S$-matrix which do not include
any angular dependence. By contrast, the MV model offers a description for the distribution
of the `valence' color sources in the target and allows for an explicit calculation of the dipole
scattering off the color fields produced by those sources. In the previous literature, this calculation
has been performed for a homogeneous target, but here we shall extend it to the case where
the `valence' color sources have a Gaussian distribution in impact parameter (including
lumpiness effect for a large nucleus). Due to the
formal simplicity of the model, we will be able to obtain quasi-analytic results for the dipole
 $S$-matrix $S(\r,\b)$, including its angular dependence, for both a smooth target (`a dense proton')
 and a lumpy one (`a large nucleus').  Such a quasi-analytic treatment turns out to be very useful 
 for the physical interpretation of our results.

One of the main conclusions of our analysis is that the azimuthal asymmetry
is controlled by soft exchanged momenta, of the same order as the transverse momentum scale
introduced by the inhomogeneity of the target. So, strictly speaking, these effects can only be
marginally addressed within our present, semi-classical, formalism, which is inspired by perturbative QCD.
But this also shows the importance of having a realistic model for the $b$-dependence, that was
already tested against the phenomenology. In fact, due to the dominance of soft exchanges,
we will also be led to consider the influence of a `gluon mass', namely a mass parameter
which controls the exponential decay of the gluon fields at large values of $b$ and mimics confinement.
Not surprisingly, this influence appears to be important, on the same footing as that of 
the other parameters of the model --- the transverse scale which characterizes the Gaussian distribution 
of the color sources and the target saturation scale at $b=0$.  These non-perturbative
aspects --- the transverse inhomogeneity of the target and the gluon
confinement --- have been differently modeled in the previous related studies 
\cite{Kopeliovich:2007fv,Kopeliovich:2007sd,Kopeliovich:2008nx,Levin:2011fb,Zhou:2016rnt,Hagiwara:2017ofm}. This
may explain the significant differences that can be observed between our new results and the
previous ones in the literature.

A priori, our model produces non-zero values for all the even Fourier coefficients
$v_{2n}(p,b)$ with $n\ge 1$, but with a strong hierarchy among them: $|v_2|
\gg |v_4|\gg |v_6|$ etc.  In order to render the model tractable, we shall 
perform additional approximations, which will preserve only the
information about the largest such a coefficient, the elliptic flow $v_2(p,b)$.
Using our (quasi)analytic results for the dipole $S$-matrix, it will be quite easy to compute
and study this coefficient. We shall thus find that $v_2(p,b)$ can be
quite large, $v_2\gtrsim 0.1$, in peripheral collisions, but it rapidly
decreases when moving towards more central collisions.
This corresponds to the fact that, in our model, the transverse inhomogeneity 
is peaked at the edge of the target.

We shall furthermore find that the effects of multiple scattering are truly essential:  
they can even change the sign of $v_2(p,b)$. 
Namely, $v_2(p)$ is found to be negative (but also tiny) for very large values of $p$,
where the single-scattering approximation applies,
but it turns positive, due to multiple scattering, at lower values 
--- including the most interesting physical regime where  $p$ is soft or semi-hard. 
A positive $v_2(p,b)$ means 
that the preferred direction of motion for a produced particle is along its impact parameter $\b$. 
In terms of dipole scattering, it means that the scattering is stronger (for a given dipole size $r$)
when the dipole is aligned along $\b$ rather than perpendicular on it. 

Finally, as a function of $p$, $v_2$ shows a maximum at a value proportional to the target saturation 
momentum $Q_s$ (say, as measured at $b=0$). Interestingly, this maximum becomes less pronounced 
(broader and smaller) when increasing $Q_s$, i.e. when the target becomes denser. 
This $p$-dependence looks acceptable from the viewpoint of the phenomenology and in fact
it can even be adjusted to reasonably describe the data in p+Pb collisions at the LHC 
with reasonable choices for the parameters.

So far, our discussion refers to a fixed value of the impact parameter $\b$: the quantity $v_2(p,b)$
characterizes the distribution of the produced particles w.r.t. the reaction plane in a particular event.
Since the direction of $\b$ (the `reaction plane angle') is not an observable, it is important to notice that
the azimuthal asymmetry under consideration can also be measured via multi-particle correlations.
Indeed, as previously mentioned, this asymmetry is sizable only for sufficiently peripheral collisions, 
in which the interaction region is relatively small. 
In $pp$ collisions, this region should be much smaller than any of the colliding protons.
In $pA$ collisions, it could be as large as the size of projectile proton, but this is still small compared
to the relevant impact parameters, of the order of the nuclear radius. The partons from the projectile 
which participate in such peripheral collisions have similar impact parameters, 
hence after the scattering they preferentially propagate along nearby directions
--- namely, along their average impact parameter.
This in turn implies the existence of azimuthal asymmetries in the multi-particle correlations;
e.g.,  $v_2\{2\}$ --- the elliptic azimuthal coefficient extracted from 2-particle 
correlations --- should be non-zero and positive. 

The above considerations have consequences not only for the multi-particle correlations, but also for the
single inclusive particle spectrum that we shall focus on in this paper. They imply that the strength of the
azimuthal asymmetries is also controlled by the geometry of the interaction region. For instance,
we shall find that the elliptic flow $v_2$ is 
(roughly) proportional to the eccentricity $\varepsilon_2$, which is
a measure of the projection of the impact parameters of the participants quarks along the direction of
 their {\em average} impact parameter.
Such geometrical aspects are clearly reminiscent of the classical
discussion of hydrodynamic flow in $AA$ collisions --- in both cases, a flow of particles in
the final state is generated via peripheral collisions --- but the underlying dynamics
is of course different: whereas in peripheral $AA$ collisions the flow is driven by the `pressure gradient'
(the final state interactions) associated with the spatial asymmetry of the interaction
region, in the new mechanism of interest for us here, the flow is rather a consequence of the angular
dependence of the amplitude for dipole scattering. 

Although in this paper we shall discuss only the {\em average} target geometry, it is quite clear that a
similar mechanism should also act when the target inhomogeneity is associated with {\em fluctuations}
--- say, in the gluon distribution produced by the high-energy evolution, or in the distribution of nucleons
inside a lumpy nuclear target. In the presence of fluctuations, azimuthal asymmetries can be also 
generated via more central collisions, but they can be probed only via multi-particle correlations, 
which are suppressed in the multicolor limit $N_c\to\infty$ \cite{Dumitru:2008wn}.
This discussion suggests that the mechanism to be discussed here is closely connected to that
from the `glasma' scenario, where the azimuthal asymmetries 
are associated with fluctuations leading to  `saturation domains' \cite{Schenke:2012wb,Schenke:2012fw,Dusling:2012iga,Dusling:2013qoz,Dusling:2015rja,
Kovchegov:2013ewa,Kovner:2012jm,Kovner:2014qea,Kovner:2015rna,Schenke:2015aqa,Lappi:2015vha,Lappi:2015vta,Rezaeian:2016szi,Schenke:2016ksl,Schenke:2016lrs}.  Perhaps a new aspect which is specific
to our discussion is the emphasis on peripheral collisions: we show that such collisions can generate sizable
azimuthal asymmetries already in the absence of fluctuations. Since related to the (average) target geometry,
these asymmetries are expected to factorize in the calculation of 
multi-particle correlations (e.g. $c_2\{2\}\simeq v_2^2$, where $c_2\{2\}$ is the second-order cumulant
\cite{Snellings:2011sz}) and also to survive in the large-$N_c$ limit.

This paper is organized as follows: In Sect. II, we concisely describe the factorization scheme that we use for quark
production in `dilute-dense' collisions and the associated calculation of the azimuthal asymmetry coefficients 
$v_n(p,b)$ in a given event. Sect. III contains our new analytic results. After introducing the (impact-parameter
dependent) MV model for the gluon distribution in the dense target in Sect. III.A, we present the calculation of the dipole $S$-matrix with angular dependence, first in the single scattering approximation (in Sect. III.B), 
next by including the effects of multiple scattering,
separately for a proton (in Sect. III.C) and for a large nucleus 
viewed as a lumpy superposition of independent nucleons (in Sect. III.D).
In Sect. IV, we present our numerical results for $v_2(p,b)$ and discuss their dependence upon
various parameters of the model as well as possible
implications for the phenomenology.  We summarize our results in Sect. V.

\section{Color-dipole orientation as the origin of the azimuthal asymmetry}

Consider particle production in a dilute-dense collision, say a proton-nucleus ($pA$) collision,
for definiteness, but the target could also be another proton provided 
the produced particle propagates at very forward rapidity.
We shall view this process at partonic level to leading order in perturbative QCD
at high gluon density (i.e. in the CGC effective theory). For simplicity we shall ignore
the fragmentation of the produced parton into hadrons. That is, we shall only compute the cross-section
for parton production, with the parton chosen to be a quark. (The discussion of gluon
production in this particular set-up would be entirely similar.) To the
accuracy of interest, the correct physical picture is as follows: a quark
collinear with the projectile proton undergoes multiple scattering off the dense gluon distribution of the
target and thus acquires some transverse momentum $\p$. The multiple scattering can be resummed 
to all orders within the eikonal approximation, which is most conveniently formulated in impact parameter
space (since the transverse coordinate of the quark is not modified by the interactions).
The cross-section is proportional to the modulus squared of the amplitude and
the quark impact parameters in the two amplitudes, direct and conjugate, are different.
As a result, one can express the rapidity and $\p$-distribution at fixed impact parameter
in terms of an effective $q\bar q$ dipole $S$-matrix,
\beq
\frac{\rmd \sigma^{q}(qA\to qX) }{\rmd\eta\,
\rmd^{2}\p\, \rmd^{2}\b}=
x_p q(x_p) \int  \frac{\rmd^{2}\r }{(2\pi)^2}\,\rme^{-\rmi \p\cdot\r}
S(\b,\r, x_g)= \frac{1}{(2\pi)^2}\,
x_p q(x_p) \tilde{S}(\b,\p,x_g).
\label{main}
\eeq
Here, $\p$ and $\b$ are the transverse momentum and the impact parameter of the produced
quark and $\eta$ is its rapidity in the center-of-frame (COM) frame. Furthermore, $x_p$ and $x_g$ are the longitudinal
momentum fractions of the partons participating in the scattering: the `collinear' quark from the 
proton and a gluon from the wavefunction of the nucleus. Energy-momentum conservation 
implies
\beq
\label{COM}
x_p=\frac{p}{\sqrt{s}}\,\rme^\eta\,,\qquad x_g=\frac{p}{\sqrt{s}}\,\rme^{-\eta}\,,
\eeq
where $p\equiv |\p|$ and $s$ is the COM energy squared for the scattering between the proton
and one nucleon from the nucleus. The quantity $S(\b,\r, x_g)\equiv S(\x,\y, x_g)$, with
$\b\equiv (\x+\y)/2$ and $\r\equiv \x-\y$, is the 
forward $S$-matrix for the scattering between a quark-antiquark dipole (with the quark leg at
$\x$ and the antiquark one at $\y$) and the nucleus, for a rapidity separation $Y=\ln(1/x_g)$.
Its Fourier transform $ \tilde{S}(\b,\p,x_g)$ plays the role of a generalized unintegrated
gluon distribution (a.k.a. a gluon  `transverse momentum distribution', or TMD) in the target. 
Since the dipole has a finite size and an orientation,
its scattering will generally depend upon the angle $\theta$ between $\b$ and $\r$. Via the Fourier transform, this will introduce an anisotropy in the cross-section for quark
production, i.e. a dependence upon the angle $\phi$  between $\b$ and $\p$. This anisotropy
can be characterized by the ensemble of Fourier components $v_n$ (a.k.a. `flow coefficients'), defined as
\begin{equation} 
v_{n}(p,b)\equiv\frac{\int_{0}^{2\pi} \rmd\phi \cos(n\phi)\,  \frac{\rmd \sigma^{q}(qA\to qX) }{\rmd \eta
\rmd^{2}\p \rmd^{2}\b}}
{\int_{0}^{2\pi} \rmd\phi \,
\frac{\rmd \sigma^{q}(qA\to qX) }{\rmd\eta\rmd^{2}\p \rmd^{2}\b}  }\,. \label{v2-1} 
\end{equation}
The Fourier moments involving $\sin(n\phi)$ vanish because the cross-section is symmetric under the
parity transformation $\phi\to -\phi$ (the reflection w.r.t. the reaction plane).
The above expression can be also evaluated with the dipole amplitude in the
coordinate representation (this will be useful e.g. when including the effects of multiple scattering in the 
eikonal approximation). Rewriting Eq.~(\ref{main}) as
\begin{eqnarray}
\frac{\rmd \sigma^{q}(qA\to qX) }{\rmd\eta\,\rmd^{2}\p \,\rmd^{2}\b}&=&
\frac{x_p q(x_p)}{(2\pi)^2}  
\int_0^\infty \rmd r \,r \int_{0}^{2\pi} \rmd\theta\,\rme^{-\rmi p r \cos(\phi-\theta)} \,
S(b,r,\theta),
\label{mainv}
\end{eqnarray}
one can perform the integral over $\phi$ in \eq{v2-1} with the help
of the the following identity,
\begin{equation} 
\,\rme^{\rmi A \cos \phi}=\sum_{n=-\infty}^{+\infty} \rmi^n\rmJ_{n}(A) \,\rme^{\rmi n \phi}, \label{id}
\end{equation}
where $\rmJ_{n}(x)$ denotes the Bessel function of the first kind. One thus obtains e.g.
\begin{eqnarray}
v_{2}(p,b)&=&-\frac{  \int r\rmd r \rmd\theta\,\cos(2\theta)\,\rmJ_{2}(p r)\,S(b,r,\theta)}{ 
\int r\rmd r \rmd\theta\,\rmJ_{0}(p r) \,S(b,r,\theta)}, \label{sv2} \\*[.2cm]
v_{3}(p,b)&=&-\rmi\,
\frac{\int r\rmd r \rmd\theta\,\cos(3\theta)\,\rmJ_{3}(p r) \,S(b,r,\theta)}{\int r\rmd r\rmd\theta\,
\rmJ_{0}(p r) \,S(b,r,\theta)}.\
\end{eqnarray}
Notice that the quark distribution function of the proton has canceled in the ratio. But the
information about the gluon distribution in the nucleus is still preserved in $v_n$, via the
dipole $S$-matrix. If one neglects the angular dependence of the latter
($S(\b,\r)\to S(b,r)$), then $v_{n}=0$ for any $n\ge 1$ regardless of the precise shape of
the target profile in $\b$. Notice also that a real contribution to $v_3$ requires the existence of
an imaginary part in the dipole $S$-matrix; hence,  a non-zero $v_3$ can be related to the odderon contribution to 
dipole scattering \cite{Kovchegov:2003dm,Hatta:2005as}.

So far, we have implicitly treated the projectile proton as a pointlike objet (indeed, we assumed that
all its valence quarks have the same impact parameter $\b$). As we shall see, this is indeed
a good approximation when the target is a large nucleus and for relatively central collisions, where the
nuclear matter distributions is quasi-homogeneous. But this is less justified for the case of a proton
target, or for peripheral collisions off a nucleus, which are the cases of main interest for what follows.
Fortunately, this can be easily remedied (at least at a formal level)
by replacing the standard quark distribution $x_p q(x_p)$
with its generalized version (a ``generalized parton distribution'', or GPD), 
which includes impact parameter dependence inside the
projectile: $x_p q(x_p)\to x_p q(x_p, \b)$, where $\b$ now refers to the position of a quark relative
to the center of its parent proton. Then Eqs.~\eqref{main} and \eqref{v2-1}  should be then replaced by
\beq
\frac{\rmd \sigma^{q}(qA\to qX) }{\rmd\eta\,
\rmd^{2}\p\, \rmd^{2}\B}= \frac{1}{(2\pi)^2} \int {\rmd^2\b} \
x_p q(x_p,\b-\B)\, \tilde{S}(\b,\p,x_g)\,,
\label{main1}
\eeq
and respectively
\begin{equation} 
v_{n}(p,B)\equiv\frac{\int_{0}^{2\pi} \rmd\Phi \,\cos(n\Phi)\,  \frac{\rmd \sigma^{q}(qA\to qX) }{\rmd \eta
\rmd^{2}\p \rmd^{2}\B}}
{\int_{0}^{2\pi} \rmd\Phi \,
\frac{\rmd \sigma^{q}(qA\to qX) }{\rmd\eta\rmd^{2}\p \rmd^{2}\B}  }\,, \label{vnpB} 
\end{equation}
where $\B$ denotes the impact parameter of the proton w.r.t. the center of the target and $\Phi$ is
the angle made by the vectors $\p$ and $\B$. A common prescription in the literature, that we 
here adopt as well, is to assume the factorization of the $\b$-dependence inside the projectile:
$x_p q(x_p, \b)\simeq x_p q(x_p) f(\b)$, with $\int {\rmd^2\b} \,f(\b)=1$. Under this assumption,
the generalization of \eq{sv2} to an extended projectile is easily found as
  \begin{eqnarray}
v_{2}(p,B)&=&-\frac{ \int b\rmd b\rmd\alpha\,\cos (2\alpha)\,f(|\b-\B|)
\int  r\rmd r \rmd\theta\, \cos(2\theta)\,\rmJ_{2}(p r)\,S(b,r,\theta)}{\int b\rmd b\rmd\alpha\,f(|\b-\B|) 
 \int r\rmd r \rmd\theta \,\rmJ_{0}(p r) \,S(b,r,\theta)}, \label{v2f}
\end{eqnarray}
where $\alpha$ is the angle between $\b$ and $\B$ and we have also assumed that the proton
distribution is isotropic: $f(\b)=f(|\b|)$.

\begin{figure}[t]
 \includegraphics[width=.55\textwidth]{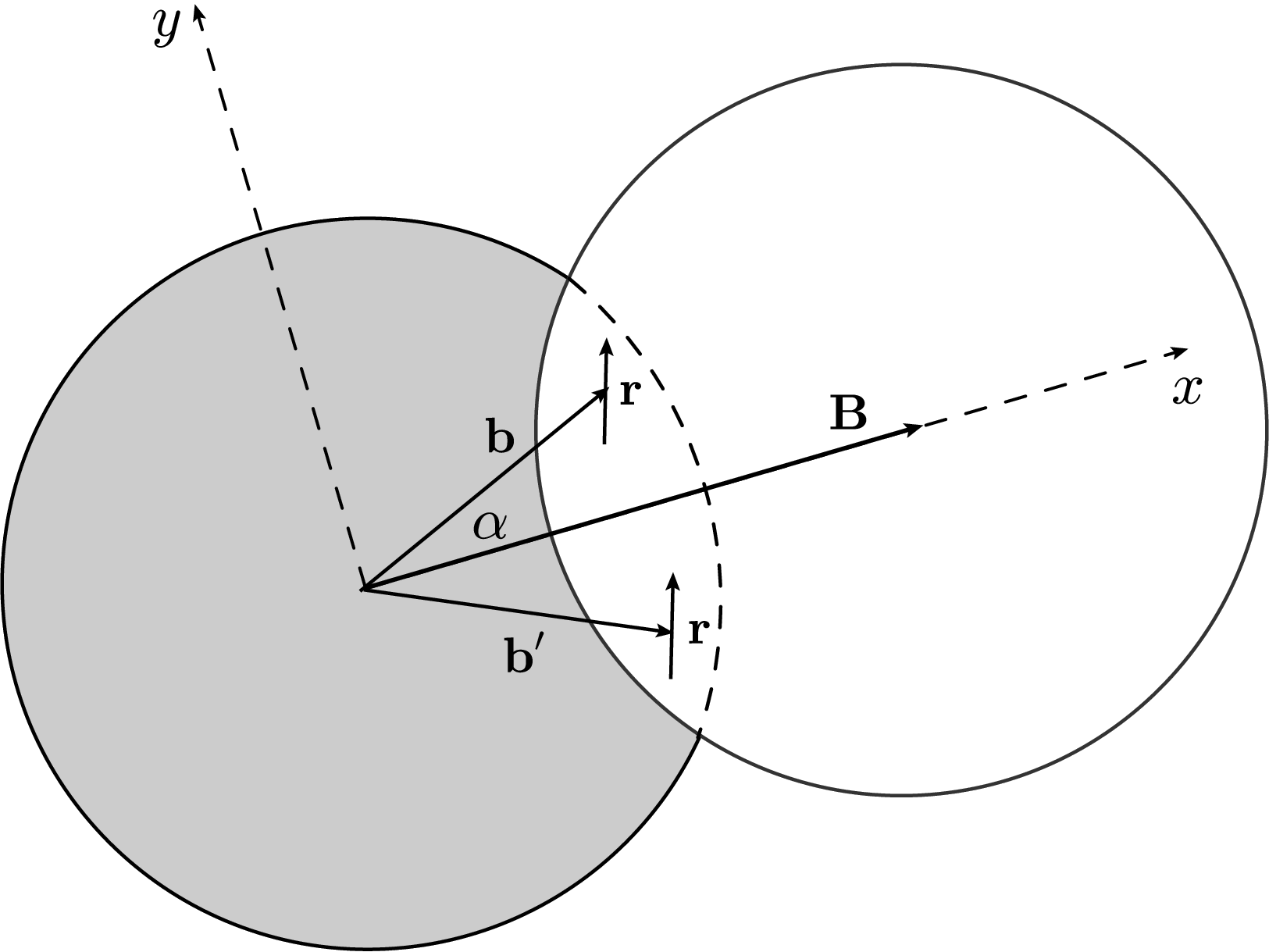}
  \caption{A peripheral $pp$ collision. The disk representing the target is filled with grey color, 
  to suggest the fact that this proton is denser.
   The dipole with size $\r$ can scatter off the target
  at any of the points within the almond-shaped interaction region.}
\label{fig:pp}
\end{figure}

We shall also need the eccentricity $\varepsilon_2$ of the interaction region. Writing the 2-dimensional
impact parameter of a participating quark as $\b=(x,y)$, where the $x$ and $z$ axes define the reaction
plane (that is, the $x$ axis is parallel to the direction of the vector $B$ --- the impact parameter of the
proton), then $\varepsilon_2$ can be estimated as
\beqn\label{epsilon}
\varepsilon_2(p,B)&\,=\,&\frac{\langle x^2-y^2\rangle}{\langle x^2+y^2\rangle}\,=\,
\frac{\langle b^2\cos(2\alpha)\rangle}{\langle b^2\rangle}\,,\nn
&\,=\,&\frac{\int b\rmd b\rmd\alpha\,b^2\cos(2\alpha)\,f(|\b-\B|)\, \int r\rmd r \rmd\theta \,\rmJ_{0}(p r) \,S(b,r,\theta)}
{\int b\rmd b\rmd\alpha\,b^2\,f(|\b-\B|)\, \int r\rmd r \rmd\theta \,\rmJ_{0}(p r) \,S(b,r,\theta)}\,,
\eeqn
where the brackets denote the averaging  over the quark impact parameters with weight-function
given by the local differential cross-section $\propto \tilde{S}(\b,\p,x_g)$ and also the angular average
 over $\Phi$ (i.e. over the direction of propagation of the produced particles). As before,
 $\alpha$ is the angle between $\b$ and $\B$, hence $x=b\cos\alpha$ and
$y=b\sin\alpha$ (see also Fig.~\ref{fig:pp}).

In fact, the quantity that is generally referred to as the `eccentricity' in the literature is the `momentum-integrated'
version of \eq{epsilon}, that is, the quantity $\varepsilon_2(B)$ which is obtained by separately integrating
the numerator and denominator over $p$, with measure $\int p\rmd p$. 
The result of this integration turns out to be very simple and in particular independent of the dipole scattering,
because of  `color transparency' : a color dipole with zero transverse size cannot scatter, 
meaning that $S(b,r,\theta)\to 1$ as $r\to 0$. This immediately implies the following
`sum-rule'  : 
\beq\label{SR}
\int\frac{\rmd^2\p}{(2\pi)^2}\,\tilde{S}(\b,\p)=1\ \ \Longrightarrow\ \ 
\int p\rmd p\int r\rmd r \rmd\theta \,\rmJ_{0}(p r) \,S(b,r,\theta)=2\pi\,.
\eeq
This `sum-rule' is in fact the expression of probability conservation: as manifest from \eq{main1},
$\tilde{S}(\b,\p)$ can be  interpreted as the probability density for a quark incident at $\b$ to acquire
a transverse momentum $\p$. Clearly, the total probability for the quark to emerge with {\em any}
momentum must be equal to one. Using Eqs.~\eqref{epsilon} and \eqref{SR}, one  finds
\beq\label{epsilonB}
\varepsilon_2(B)=\frac{\int b\rmd b\rmd\alpha\,b^2\cos(2\alpha)\,f(|\b-\B|)}
{\int b\rmd b\rmd\alpha\,b^2\,f(|\b-\B|)}\,.\eeq
As anticipated, this is a purely geometrical quantity, without any information about the scattering
of the dipole: it merely shows how the projectile is `seen' from the center of the target. It is quite obvious 
that $\varepsilon_2(B)$ vanishes for $B= 0$ and that it approaches to 1 when $B\to \infty$
(since in that limit the ratio $\langle y^2\rangle/\langle x^2\rangle$ approaches to zero, meaning that
$\alpha \to 0$ as well). This behavior will be confirmed by the explicit calculations to be presented later.

\section{Elliptic flow from dipole orientation in the MV model}

\subsection{Dipole scattering in the McLerran-Venugopalan model}
\label{sec:MV}

So long as the gluon energy fraction $x_g$ probed by the scattering is not too low,
one can ignore the high-energy evolution of the nuclear gluon distribution and describe the latter
within the McLerran-Venugopalan (MV) model \cite{McLerran:1993ka}. In this model, the nucleus is described
as a collection of independent color sources (`the valence quarks') with a Gaussian
color charge distribution in the transverse plane:
\begin{eqnarray}
\langle \rho^a(\x)\rho^b(\y)\rangle&=&\delta^{ab}\delta^{(2)}(\x-\y)\mu(\x)\,,
\end{eqnarray}
with $\mu(\x)$ the color charge squared per unit area in the transverse plane.
In the original formulation of the MV model, as applying to a (very) large nucleus, 
there was no explicit impact-parameter dependence: the color charge distribution was assumed
to be uniform, $\mu(\b)=\mu_0$, within a large disk with radius $R_A\propto A^{1/3}$, with
$A$ the atomic number. However, as already mentioned, the inhomogeneity of the target
in the transverse plane is essential for the physical effects that we are currently interested in.
So, in what follows we shall propose a generalization of the MV model
 which includes a physically motivated impact-parameter dependence (inspired by the
 fits to the HERA data).
We shall first present this dependence for the case where the target is a single nucleon
(say, any of the $A$ nucleons composing a large nucleus), and then extend the model to a nuclear 
target, in Sect.~\ref{sec:pA}. But the case of a proton target is also interesting by itself --- albeit the
applicability of the MV model to this case is of course questionable ---, in view of the phenomenology
of flow-like effects in high-multiplicity events in $pp$ collisions.

The profile $\mu(\b)$ of the proton color charge distribution in the transverse plane will be chosen to be
a Gaussian, in agreement with saturation fits 
\cite{Kowalski:2003hm,Kowalski:2008sa,Rezaeian:2012ji,Rezaeian:2013tka} 
to the HERA data on diffraction and vector meson production in DIS
(its Fourier transform $\tilde{\mu}(\Delta)$ will be later needed):
\begin{eqnarray}
\mu(b)&=&\mu_0 \,\rme^{-b^2/4R^2}\,,\nonumber\\
\tilde{\mu}(\Delta)&=& \int {\rmd^2\b} \,\rme^{-\rmi\b\cdot\D}\mu(b)= 4\pi R^2\mu_0 \,\rme^{-\Delta^2R^2}. \label{mub}\
\end{eqnarray}
The overall factor $\mu_0$, with dimensions of mass squared, is proportional to the total
color charge squared of the valence partons; e.g. for a proton target with
 $N_c$ valence quarks, one can write $4\pi R^2(N_c^2-1)\mu_0=g^2C^F N_c$, or $\mu_0=
\alpha_s/(2R^2)$. In practice, this quantity $\mu_0$ will be traded for the saturation momentum
in the center ($b=0$) of the target, a quantity which is constrained by the fits to the HERA data (see
\eq{qst} below). The scale $R$ which fixes the width of the $b$-distribution 
will be taken too from fits to the HERA data (more precisely from the fits using the IP-Sat model 
in Refs.~\cite{Kowalski:2003hm,Rezaeian:2012ji}). A typical value emerging from these fits is $R\simeq 0.3$~fm.
Notice that, with our present conventions, the `proton size' (in the sense of the region in impact
parameter space where the valence
color charges are distributed) is $2R\simeq 0.6$~fm, and {\em not} $R$. Indeed, the exponent of the Gaussian in 
\eq{mub} becomes equal to one when $b=2R$. Also, for $\Delta=0$, one has $\tilde{\mu}(0)=4\pi R^2\mu_0$,
showing that the natural `proton area' is $4\pi R^2$, and not $\pi R^2$.

The gauge potential 
created by the ultrarelativistic color charges is simply the 2-dimensional Coulomb field,
\beq\label{Coulomb}
A^-_a(\x)=\int \rmd^2 \z \,G(\x-\z)\rho^a(\z)\,,\qquad
G(\b)\equiv\int \frac{\rmd^2\q}{(2\pi)^2} \,\frac{\rme^{i\q\cdot\b}}{q^2}\,\simeq\,
\frac{1}{4\pi}\ln\frac{1}{b^2m^2}
\,,\eeq
where $m$ is an infrared cutoff, physically associated with confinement:
$m\sim \Lambda_{_{\rm QCD}}$. (So, the second estimate for $G(\b)$ given above applies
only for sufficiently small distances $b\lesssim 1/m$.)
This implies that the distribution of the color fields is Gaussian as well, with 2-point
correlation
\beq
\langle A^-_a(\x)A^-_b(\y)\rangle =\delta^{ab}\gamma(\x,\y)\,,
\eeq
where
\beq
\gamma(\x,\y)\equiv\int \rmd^2 \z \,G(\x-\z)G(\y-\z)\mu(\z)
= \int \frac{\rmd^2\q }{(2\pi)^2}\frac{\rmd^2\q^\prime }{(2\pi)^2} \,
\rme^{\rmi\q^\prime\cdot \x+\rmi\q\cdot\y}\,\frac{\tilde{\mu}(\q^\prime+\q)}{q^{\prime 2} q^2}\,.
\eeq

In the MV model and in the eikonal approximation, the projectile dipole independently scatters
off the color charges in the nucleus. Accordingly, its multiple scattering exponentiates,
as in the Glauber approximation: $S=\exp\{-N_{2g}\}$, where
$N_{2g}(\x,\y)$ is the amplitude for a single scattering via the exchange of two gluons:
\begin{eqnarray}
N_{2g}(\x,\y)&=&\frac{g^2C_F}{2} \Big[\gamma(\x,\x)+\gamma(\y,\y)-2\gamma(\x,\y)\Big]\,,\nn
&=&\frac{g^2C_F}{2}\int  \frac{\rmd^2\q^\prime }{(2\pi)^2}\frac{\rmd^2\q }{(2\pi)^2} \frac{\tilde{\mu}(\q^\prime+\q)}{ \q^{\prime 2} \q^2} \Big[\,\rme^{\rmi\q^\prime\cdot\x}-\,\rme^{\rmi\q^\prime\cdot\y}\Big]
\Big[\,\rme^{\rmi\q\cdot\x}-\,\rme^{\rmi\q\cdot\y}\Big].
\label{N2g}
\end{eqnarray}

Note that, unlike the valence color charges, which are effectively confined in the transverse plane 
within a disk with radius $R$, cf. \eq{mub}, the color fields created by these charges (the small-$x$
gluons) can be delocalized over much larger distances, due to the slow decay of the 2-dimensional
Coulomb propagator at large distances. In particular, it is easy to check that for very large
impact factors $x\sim y\gg R$ (with $x\equiv |\x|$ and $y\equiv |\y|$), the dipole amplitude predicted
by this model shows a power tail: $N_{2g}(\x,\y)\propto (\b\cdot\r)^2/b^4$, with $\r=\x-\y$ and $\b=(\x+\y)/2$.

At this stage, it is convenient to change variables, from $\x,\y$ to $\r,\b$, and from  $\q^\prime,\q$ to 
$\D, \k$, with $\k=({\q^\prime-\q})/{2}$ and $\D=\q^\prime+\q$. For the physical discussion to
follow, it is useful to keep in mind the physical meaning of the momenta $\k$ and $\D$ from
the viewpoint of our original problem, that of quark production: \texttt{(i)} $\k$ is the average transverse
momentum transmitted by the nucleus to the quark via a single collision;  \texttt{(ii)} $\D$ is the difference
between the transverse momenta acquired by the quark in the direct amplitude and the
complex conjugate amplitude, respectively; as such, it is a measure of the additional momentum transfer
associated with the inhomogeneity of the target. Clearly, $\D$ is a {\em soft} momentum, 
$\Delta\sim 1/R\sim\Lambda_{_{\rm QCD}}$, whereas is generally semi-hard, that is, it is either
comparable to the final momentum $p$ of the produced quark, or to the saturation momentum
$Q_s(b)$  of the target. Yet, soft values for $\k$ will be important too, when discussing the flow
coefficients in the presence of multiple scattering.

We thus obtain, 
\begin{equation}\label{T2g0}
N_{2g}(\b,\r)= \frac{g^2C_F}{2}\int \frac{\rmd^2\D }{(2\pi)^2}\frac{\rmd^2\k }{(2\pi)^2}  \frac{\tilde{\mu}(\D)}{(\k+\D/2)^2 (\k-\D/2)^2}
 \,\rme^{\rmi\D\cdot \b}\Big[\,\rme^{\rmi\D\cdot\r/2}+\,\rme^{-\rmi\D\cdot\r/2} -2\,\rme^{\rmi\k\cdot\r}\Big].
\end{equation}
The first two terms within the square brackets, which are independent of $\k$, represent `tadpole'
contributions where the two gluons exchanged with the target are attached to a same fermion
leg (the quark or the antiquark). 
The final term, which is negative, refers to `exchange' contributions,
where one gluon is attached to the quark leg and the other one, to the antiquark. 

Since $\tilde{\mu}(\D)$ is truly a function of $\Delta\equiv |\D|$, it is quite obvious that $N_{2g}(\b,\r)$ 
is an even function of $\b$ and also an even function of $\r$;
hence, it depends upon $\theta$ (the angle between $\b$ and $\r$) only via the squared dot product
$(\b\cdot\r)^2$. This in turn implies that all the odd `flow coefficients', like the `radial flow' $v_1$
and the `triangular' one $v_3$, must vanish. In what follows, we shall compute the elliptic flow $v_2$.
For pedagogy, we shall first present the respective calculation in the single scattering approximation.

\subsection{The single scattering approximation}
\label{sec:SSA}

The single scattering approximation $S(\b,\r)\!\simeq \!1 - N_{2g}(\b,\r)$ applies so long the dipole
is small enough for its transverse resolution $Q^2\equiv 1/r^2$ to be much larger than the (local)
saturation momentum $Q_s(\b)$ at its impact parameter. Equivalently (since $pr\sim 1$ by virtue
of the Fourier transform in \eq{main}), the produced quark is relatively hard, with a transverse momentum
$p\gg Q_s(\b)$. The saturation scale $Q_s(\b)$ in the MV model
will be more precisely defined in the next subsection, where we discuss multiple scattering. Here,
we anticipate that this is a {\em semi-hard} scale, comparable to, or larger than, the momentum
scale $1/R$ introduced by the impact-parameter distribution $\mu(b)$.

To compute $v_2$ in the single scattering approximation (SSA),
it is convenient to first perform the Fourier transform of the dipole amplitude, 
$N_{2g}(\b,\r)\to \tilde N_{2g}(\b,\p)$, and then use \eq{v2-1}.  It is quite clear that the `tadpole' pieces 
in \eq{T2g0} do not significantly contribute in the kinematics of interest: via the Fourier transform, 
the respective exponentials $\rme^{\pm\rmi\D\cdot \r/2}$ select $\Delta=2p$, but the function
 $\tilde\mu(2p)$ is exponentially suppressed for $2p\gg 1/R$.
As for the Fourier transform of the `exchange' piece in  \eq{T2g0}, this
is simply obtained by replacing $\k\to\p$. We deduce
\begin{eqnarray}
 \tilde{N}_{2g}(\b,\p)&=&\int \rmd^{2}\r \, \rme^{\rmi \p\cdot\r} \,N_{2g}(\b,\r)=-g^2C_F\int\frac{\rmd^2\Delta }{(2\pi)^2}  \frac{\tilde{\mu}(\Delta)}{(\p+\D/2)^2 (\p-\D/2)^2}
 \,\rme^{\rmi\D\cdot \b}   
\label{tildeT} \,.
\end{eqnarray}
Physically, the fact that $\k=\p$ means that the momentum carried by the final quark must be 
acquired via its only collision with the target.

\eq{tildeT} can be further simplified by using the fact that $pR\gg 1$, whereas the integral
is controlled by softer values $\Delta\lesssim 1/R$.  Accordingly, one can
expand the integrand in powers of $\Delta/p$ and keep only the leading order piece,
\begin{eqnarray}\label{Dexp}
\frac{1}{(\p+\D/2)^2 (\p-\D/2)^2} \,=\,  \frac{1}{\left(p^2 +  \frac{\Delta^2}{4}\right)^2 - (\p\cdot\D)^2}
\,\approx \,\frac{1}{p^4} \left(1- \frac{\Delta^2}{2p^2}+\frac{(\p\cdot\D)^2}{p^4}\right) +\dots \,,
\end{eqnarray}
where the dots stand for terms of order $({\Delta}/{p})^4$. After also using \eq{mub}, we are led to 
a Gaussian integral
\begin{eqnarray}
4\pi R^2
 \int\frac{\rmd^2\Delta }{(2\pi)^2}  \,\rme^{-\Delta^2R^2}
 \Delta^i \Delta^j
 \,\rme^{i\b\cdot\D} &=& -4\pi R^2\frac{\partial^2}{\partial b^i \partial b^j}
 \int\frac{\rmd^2\Delta }{(2\pi)^2}  \,\rme^{-\Delta^2R^2}
 \,\rme^{i\b\cdot\D}\,,\nonumber\\*[0.2cm]
 &=& -\frac{\partial^2}{\partial b^i \partial b^j} \,\rme^{-b^2/4R^2}\,=\,
 \frac{1}{2R^2}\left(\delta^{ij}-\frac{b^i b^j}{2R^2}\right) \,\rme^{-b^2/4R^2}\,.
\end{eqnarray}
Putting everything together and using the trigonometric identity $2\cos^2\phi-1=\cos(2\phi)$, we 
finally deduce 
\begin{eqnarray} \label{asy}
 -\tilde{N}_{2g}(\b,\p)\simeq \frac{g^2C_F}{p^4}\,\mu(b)\left[1-\frac{b^2}{8p^2R^4}\cos(2\phi)\right]\,.
 \end{eqnarray}
This holds up to terms suppressed by higher powers of $1/(pR)^2$. In this approximation,
the dipole amplitude is proportional to $\mu(b)$, hence it is as localized in $b$ as the `valence' color charges
from the target. This is so because the scattering involves the exchange of a hard gluon, with momentum
$p$, and this exchange is quasi-local.

The leading-order contribution at large $p$, proportional to $1/p^4$, is independent of $\phi$. 
This is recognized as the standard
result for the particle spectrum produced via a single, hard, scattering.
The angular dependence enters via the subleading term $\propto 1/p^6$, whose sign is quite remarkable:
this is such that the cross-section for quark production  (which in the present approximation
is proportional to $\tilde{S}(\b,\p)\simeq - \tilde{N}_{2g}(\b,\p)$) is largest when $\theta=\pi/2$.
Physically, this means that a quark produced via
a single scattering has more chances to propagate along a direction which is
perpendicular on its impact parameter ($\p\perp\b$), rather than parallel to it ($\p\parallel\b$).
In turn, this implies that the elliptic flow coefficient $v_2$ is {\em negative} in this regime. Namely,
by inserting \eq{asy} into \eq{v2-1}, one finds
 \beq \label{v2-s}
 v_2(p,b)\simeq \frac{\int_{-\pi}^{\pi} \rmd\phi \cos(2\phi)\,  
 \left[1-\frac{b^2}{8p^2R^4}\cos(2\phi)\right]}
 {\int_{-\pi}^{\pi} \rmd\phi \,\left[1-\frac{b^2}{8p^2R^4}\cos(2\phi)\right]}\,=\,
 -\frac{b^2}{16p^2R^4}
\,. 
\eeq
Except possibly for its sign, which is somewhat unexpected, the above 
result for $v_2$ shows the expected trends: it vanishes when $b\to 0$, since for such central
collisions the orientation of the incoming dipole plays no role, and it decreases quite fast
when increasing the momentum $p$ of the produced quarks, as this corresponds to exploring
dipoles with very small sizes $r\ll R$.  

The above calculation also illustrates another generic
feature of the $v_2$ (more generally, of the azimuthal anisotropy) generated by the
current mechanism: this is directly related to the target inhomogeneity in the transverse
plane, i.e. it is proportional to the derivatives of the $b$-distribution $\mu(\b)$.
It should be furthermore clear that the higher azimuthal harmonics
$\cos(2n\phi)$ with $n\ge 2$
would be generated via the higher-order terms in the large-$p$ expansion,
hence the corresponding Fourier coefficients are parametrically suppressed
--- by powers of $b^2/(p^2R^4)\sim 1/(pR)^2$ when $b\sim R$ ---
compared to the elliptic flow $v_2$.

Notice that, in this single-scattering approximation, 
the overall normalization $\mu_0$ of the charge-charge correlator, cf. \eq{mub}, 
and also the coupling constant $g^2C_F$, drop out from the calculation of $v_2$.
This last feature will be of course modified by the inclusion of multiple scattering,
which becomes compulsory for softer momenta $p\lesssim  Q_s(b)$ and will be
discussed in the  next section.

\subsection{Adding multiple scattering}
\label{sec:MS}

The multiple scattering between the quark projectile and the target becomes important
when the transverse momentum of the produced particle $p$ is comparable to, 
or smaller than, the nuclear saturation momentum $Q_s(b)$. This is actually the most interesting 
situation for the phenomenology of flow in $pp$ and $pA$ collisions at the LHC, as we shall see.
In that case, we must return to the general expression for the dipole $S$-matrix
(within the framework of the MV model, of course), namely 
\beq\label{st}
S(\b,\r)=\exp\{-N_{2g}(\b,\r)\}, 
\eeq
with $N_{2g}$ as given in Eq.~(\ref{T2g0}).
Due to the exponentiation, the Fourier transform ${S}(\b,\r)\to \tilde{S}(\b,\p)$ is more complicated.
Physically this reflects the fact that the momentum $\p$ of the produced quark
gets accumulated via several scatterings and hence needs not be identified with the momentum $\k$ transferred by a single collision. The typical situation, to be referred to as {\em soft multiple scattering}, 
is such that the number of quasi-independent scatterings is quite large, so that the typical value of
$k$ is much smaller than the final momentum $p$.

In order to isolate the angular dependence of the $S$-matrix, one may be tempted to perform 
the small-$\Delta$ expansion as in \eq{Dexp} already {\em before} performing the Fourier transform.
However, this manipulation, which corresponds to an expansion in powers of $\Delta^2/k^2$, would generate infrared divergences, leading to a result which is meaningless except for the leading order term, which has
no angular dependence. For instance, to first order in $\Delta^2/k^2$, one finds
\beqn\label{T2gexp}
N_{2g}(\b,\r)
 &\simeq&\frac{g^2C_F}{2}
 \int \frac{\rmd^2\Delta }{(2\pi)^2}\frac{\rmd^2\k }{(2\pi)^2}  \frac{\tilde{\mu}(\Delta)}{k^4}
  \left(1- \frac{\Delta^2}{2k^2}+\frac{(\k\cdot\D)^2}{k^4}\right)
    \Big[\,\rme^{\rmi\D\cdot\r/2}+\,\rme^{-\rmi\D\cdot\r/2} -2\,\rme^{\rmi\k\cdot\r}\Big]\,\rme^{\rmi\D\cdot \b}.
\eeqn
Here we have assumed that $k\gg \Delta$, yet if one attempts to compute the above integral over $\k$
(for fixed $\D$), one faces strong infrared ($k\to 0$) divergences, showing that this expansion 
is not really justified. To better see these divergences, notice that for sufficiently soft $k$ and $\Delta$,  
the $\r$-dependence within \eq{T2gexp} can be expanded out:
\beq\label{rexp}
\frac{1}{2}
\Big[\,\rme^{\rmi\D\cdot\r/2}+\,\rme^{-\rmi\D\cdot\r/2} -2\,\rme^{\rmi\k\cdot\r}\Big]
\,\simeq\,-\rmi\k\cdot\r +\frac{1}{2}(\k\cdot\r)^2 - \frac{1}{8}(\D\cdot\r)^2\,,\eeq
where the linear term in the r.h.s. vanishes (by parity) after the $\k$-integration.

Using the above, one sees 
that the dominant term $\propto 1/k^4$ in the large-$k$ expansion in \eq{T2gexp} gives rise
to a logarithmic integration for momenta $\k$ within the range $\Delta \ll k \ll 1/r$.
This is a well-known result  \cite{Iancu:2003xm}:
the (angle-averaged) scattering amplitude $N(b,r)$ for a small dipole in
the MV model is logarithmically sensitive to all transferred momenta within the interval
$m \ll k \ll 1/r$, where $m\sim \Lambda_{_{\rm QCD}}$ is the infrared cutoff introduced
in \eq{Coulomb}. In the present context, where the target is inhomogeneous, there is no
genuine infrared divergence in the calculation --- the associated momentum $\Delta$ 
effectively acts as an infrared cutoff on $k$ ---, but we recover the logarithmic enhancement
of the amplitude averaged over dipole orientations.

However, the second-order terms in the expansion in \eq{T2gexp}, which in particular carry an 
angular dependence, appear to develop a {\em quadratic} infrared divergence as $k\to 0$.
This shows that this particular effect --- the angular dependence of the dipole amplitude --- is in 
fact controlled by {\em soft} exchanged momenta, $k\sim\Delta$, whose contribution cannot
be computed via the expansion in powers of $\Delta^2/k^2$. Importantly, this also means that,
for semi-hard momenta $p\lesssim Q_s(b)$, one cannot perform a reliable calculation of $v_2$ 
from first principles --- not even within the limits of the MV model. Indeed, the soft momenta
$k\sim\Delta\lesssim 1/R$ lie within the realm of the non-perturbative, confinement, physics, 
so their description within the MV model is not really justified.
This being said, this model offers a convenient set-up for at least approaching the
physics of the dipole orientation, while at the same time being consistent with the
pQCD description of the angular-averaged amplitude. In that sense, we believe 
that the results of our subsequent analysis are still useful for
qualitative and even semi-quantitative studies of the phenomenology.

We thus conclude that, for the present purposes, one cannot expand
the double integral in Eq.~(\ref{T2g0}) in powers of $\Delta^2/k^2$. Yet, the above
discussion points towards another simplification, which is quite useful in practice: within the
interesting regime of soft multiple scattering, all the relevant contributions come from
relatively small transferred momenta $k\ll 1/r$, for which 
one can expand the $r$-dependence as shown in \eq{rexp}. This yields
\beq\label{T2gsmallr}
N_{2g}(\b,\r)\,\simeq\,
\frac{g^2C_F}{2}\,r^i r^j \int \frac{\rmd^2\Delta }{(2\pi)^2}\frac{\rmd^2\k }{(2\pi)^2}  \frac{
\big(k^ik^j-\Delta^i\Delta^j/4\big)\tilde{\mu}(\Delta)}{
[(\k+\D/2)^2+m^2][(\k-\D/2)^2+m^2]} \,\rme^{\rmi\D\cdot \b}\,.
\eeq
We have introduced here the infrared cutoff $m$ as
a `gluon mass'  in the 2-dimensional Coulomb propagator,
\beq\label{Coulombm}
G(\b)=\int \frac{\rmd^2\q}{(2\pi)^2} \,\frac{\rme^{i\q\cdot\b}}{q^2+m^2}\,=\frac{1}{2\pi}
{\rm K}_0(mb)\,,\eeq
where ${\rm K}_0(x)$ is the modified Bessel function of the second kind. After
this modification, the propagator $G(\b)$ shows an exponential decay at large transverse 
separations $mb\gg 1$, which mimics confinement.
As already stressed, the insertion of this `mass' $m$ is not required by infrared
divergences: the integral over $\k$  in \eq{T2gsmallr} is well-defined in the
`infrared' ($k\to 0$) even when $m=0$; and indeed, we shall later study the
limit $m\to 0$ of our results. Rather, the `gluon mass' $m$ is needed in order
to restrict the phase-space allowed to very soft momenta $k\sim \Lambda_{_{\rm QCD}}$, 
which control the physics of the dipole orientation. (In real QCD, this phase-space would
be of course restricted by confinement.) On the other hand,  the integral over $k$ in \eq{T2gsmallr}
develops a logarithmic `ultraviolet' ($k\to\infty$) divergence; it is understood that this divergence is cut off
at the scale $k\sim 1/r$ (see \eq{t2f1} below for details).

It is also interesting to notice that the expansion 
\eqref{rexp} in powers of  $\k\cdot\r$ `does not commute' with the single scattering approximation
studied in the previous section: in the latter, the exchanged momentum $\k$ is identified (via the
Fourier transform) with the final momentum $\p$, hence $\k\cdot\r=\p\cdot\r\sim \order{1}$ and
a finite-order expansion in powers of  $\k\cdot\r$  would be incorrect.
Accordingly, a calculation using $S=\exp\{-N_{2g}\}$ 
together with \eq{T2gsmallr} for $N_{2g}$ cannot reproduce the value
of $v_2$ at very large momenta $p\gg Q_s(b)$ previously obtained in \eq{v2-s}.
More precisely, such a calculation would correctly reproduce the leading-order contribution
$\propto 1/p^4$ to $\tilde{N}_{2g}(\b,\p)$ in \eq{asy}, which is independent of $\phi$, 
but not also its subleading piece $\propto 1/p^6$, which carries the interesting $\phi$-dependence.

The double integral in the r.h.s. of \eq{T2gsmallr} has a relatively simple tensorial
structure, which immediately implies that its result must be written as a linear combination
of the following two rank-2 tensors: $\delta^{ij}$ and $b^ib^j/b^2$. Equivalently, the ensuing
approximation for $N_{2g}(\b,\r)\equiv N_{2g}(b,r,\theta)$ has the following generic structure 
\beqn\label{t2-f3}
N_{2g}(b,r,\theta)=\mathcal{N}_0(b,r)+\mathcal{N}_\theta(b,r) \cos(2\theta)\,,
\eeqn
without higher Fourier components. This is easily verified via direct calculations
of the angular integrals in \eq{T2gsmallr}, which can be analytically completed. This is detailed
in Appendix \ref{app:MS}, from which we here quote the final results:
\beqn\label{t2f1}
\mathcal{N}_0(b,r)&=& \frac{Q_s^2(b)r^2}{4} \ln\left(\frac{1}{r^2m^2}+\rme\right)  \nonumber\\
 &+& \frac{g^2C_F}{4(2\pi)^2} r^2 \int_{0}^\infty \rmd\Delta\,\Delta\,  \tilde{\mu}(\Delta)
\rmJ_0(\Delta b)\nonumber\\
&&\qquad\times \int_{0}^{\infty} \rmd
k\,k\Bigg[\frac{ k^2 -\Delta^2/4 }{\left(k^2+\Delta^2/4+m^2\right) \left( \left(k^2+\Delta^2/4+m^2\right)^2-k^2\Delta^2  \right)^{1/2}}- \frac{k^2}{(k^2+m^2)^2} \Bigg]\,,\nonumber\\
\eeqn
and respectively 
\beqn\label{t2f2}
\mathcal{N}_\theta(b,r)&=&  \frac{g^2C_F}{4(2\pi)^2} r^2 \int_{0}^\infty \rmd\Delta \,\Delta\, 
\tilde{\mu}(\Delta) \rmJ_2(\Delta b) \int_{0}^{\infty} \rmd 
k\,  k\Bigg[\frac{k^2 +\Delta^2/4}
{\left(k^2+\Delta^2/4+m^2\right) \left( \left(k^2+\Delta^2/4+m^2\right)^2-k^2\Delta^2  \right)^{1/2}}
 \nonumber\\*[0.2cm]
 && \hspace{6.cm}  \ \ +\frac{2}{\Delta^2} - 
\frac{2\left(k^2+\Delta^2/4+m^2\right)}{ \Delta^2 \left(\left(k^2+\Delta^2/4+m^2\right)^2-k^2\Delta^2  \right)^{1/2}} \Bigg].
\eeqn
The above expression for $\mathcal{N}_0(b,r)$ has been obtained from \eq{t2f10} in 
Appendix \ref{app:MS} via the following manipulations:
we have first subtracted the dominant behavior of the integrand at high
$k$ and then replaced the subtracted piece via its following, regularized, form:
\beqn \label{qst-mu}
\frac{g^2C_F}{4}\,r^2 \int \frac{\rmd^2\Delta }{(2\pi)^2}\frac{\rmd^2\k }{(2\pi)^2}  
\frac{k^2 \tilde{\mu}(\Delta)}{(k^2+m^2)^2} \,\rme^{\rmi\D\cdot \b}
\,\equiv\, \frac{Q_s^2(b)r^2}{4} \ln\left(\frac{1}{r^2m^2}+\rme\right)\,,
\eeqn
with the impact-parameter dependent  `saturation momentum' $Q_{s}(b)$ defined as
\beq
Q^2_s(b)\equiv \alpha_s C_F \mu(b)=Q_{0s}^2\, \rme^{-b^2/4R^2}\,. \label{qst}
\eeq
$Q^2_{0s}=\alpha_s C_F\mu_0$ is the central value of the saturation momentum at $b=0$.
The coefficient of the logarithm $\ln(1/r^2m^2)$ in the r.h.s. of \eq{qst-mu} unambiguously
follows from the logarithmic  integration over the range $m\ll k\ll 1/r$, 
whereas the constant term under the log specifies our `renormalization' scheme.  Notice
that all the results throughout this paper depend upon the QCD coupling $\alpha_s$, the fundamental
Casimir $C_F$ and the 2-dimensional density $\mu_0$ of color charge squared only via
this quantity $Q^2_{0s}$, to be treated as a free parameter of our model. In spite of our notations,
$Q^2_s(b)$ is not {\em exactly} the saturation scale in the present model, but it is comparable to
it, as we shall shortly argue.

The first piece in the r.h.s. of \eq{t2f1}, proportional to $Q_s^2(b)$, would be the only one to
survive in the case of a {\em homogeneous} target, i.e. when $\tilde{\mu}(\D)\propto\delta^{(2)}(\D)$.
This piece has an apparent logarithmic divergence in the limit $m\to 0$.  However, in the present context, 
where the target is inhomogeneous, this divergence is compensated
by a corresponding divergence generated by the second, integral, term in \eq{t2f1}. This is demonstrated
in Appendix \ref{app:MS}, where we will also show that, when $m\to 0$, the mass parameter $m$ gets 
replaced by $1/R$ within the argument of the logarithm. This being said,
the insertion of a non-zero  `gluon mass' $m$ is still necessary, on physical grounds.

\begin{figure}[t]
 \includegraphics[width=0.45\textwidth]{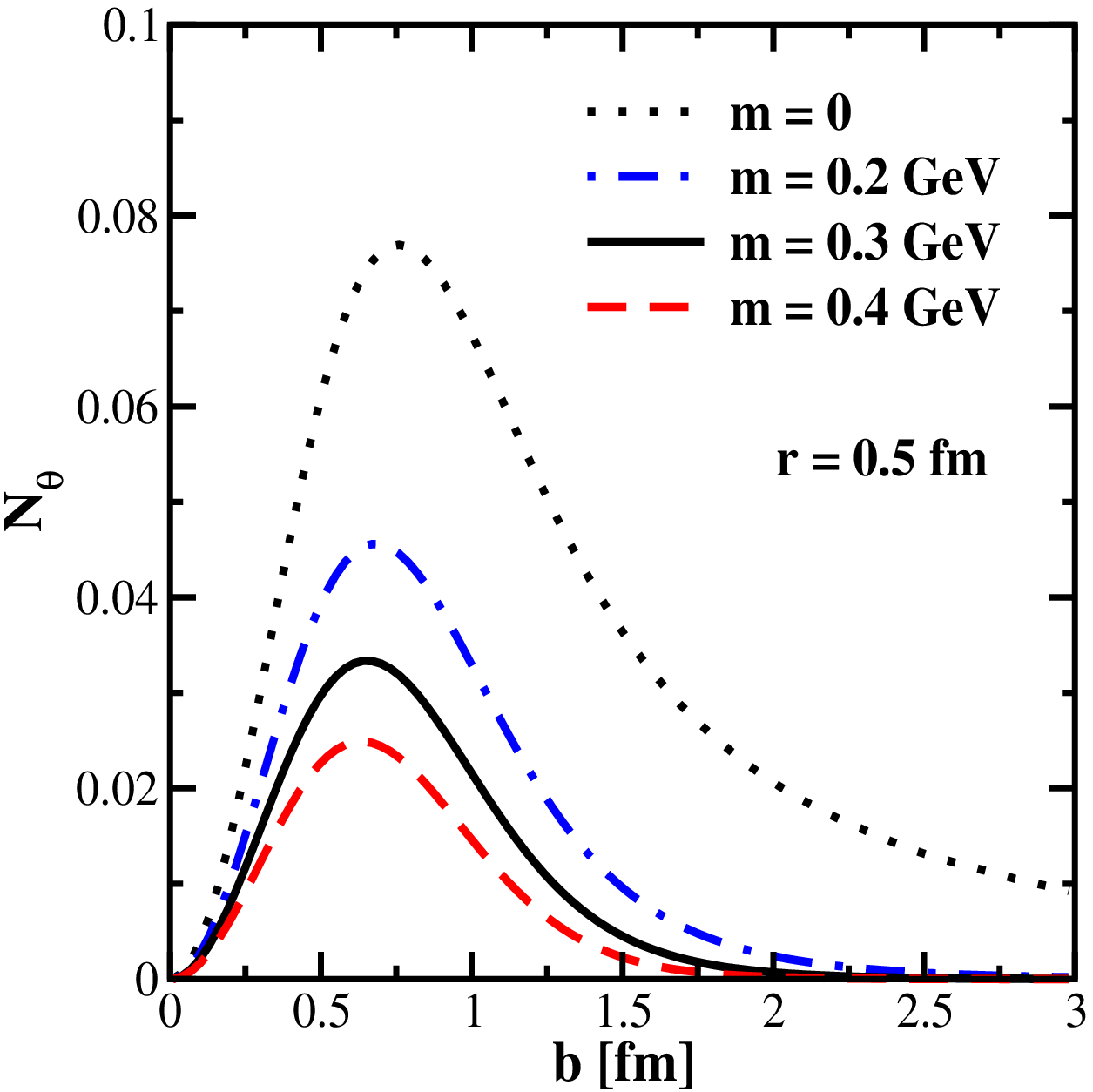} \qquad
 \quad \includegraphics[width=0.45\textwidth]{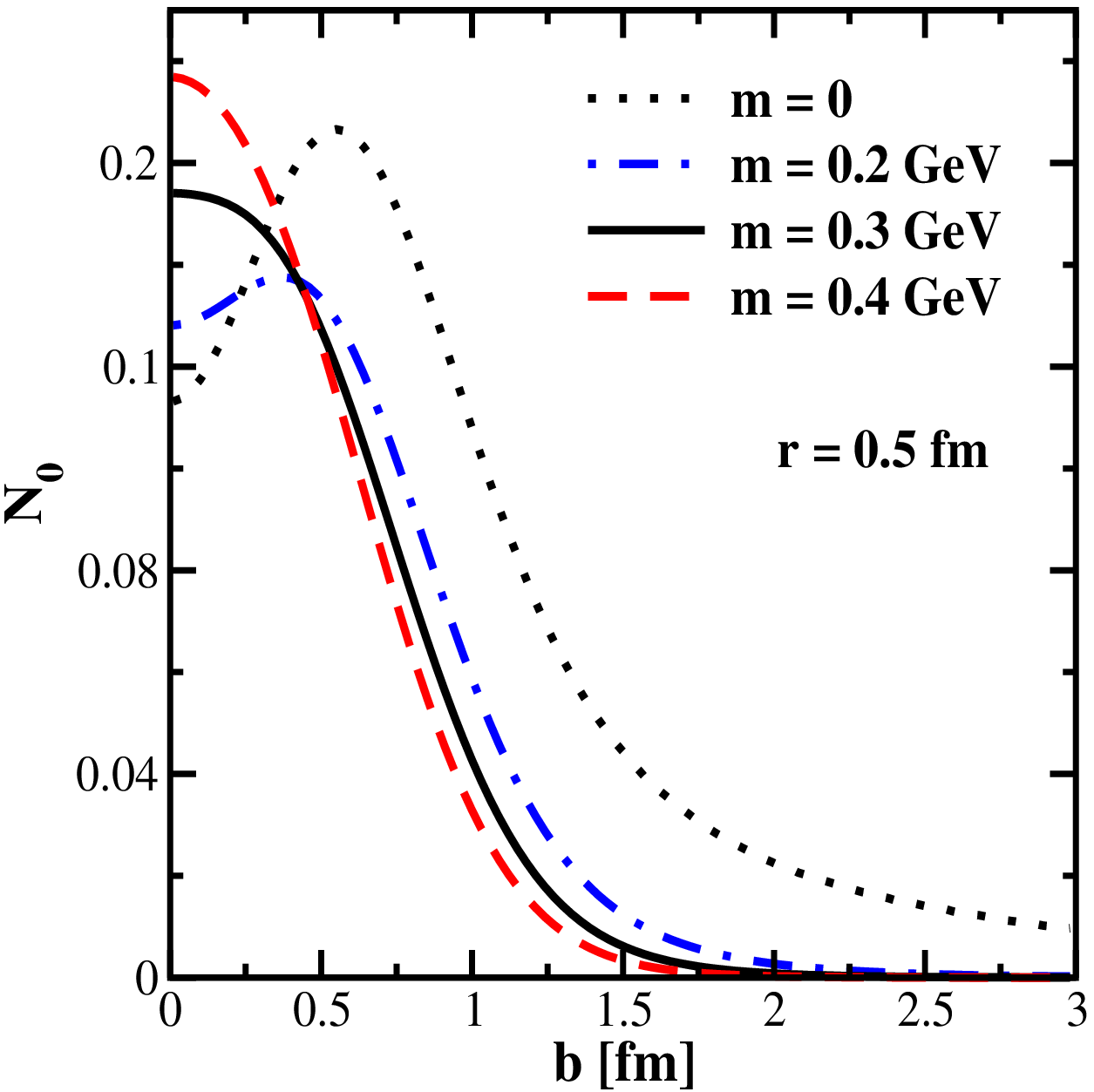}
  \caption{Left: the function $\mathcal{N}_\theta(b,r)$ which encodes the angular dependence
  of the dipole amplitude in the present approximations is numerically computed according to
  \eq{t2f2} and plotted as a function of $b$ for a fixed dipole size $r=0.5$~fm and various values
  of the infrared cutoff $m$ (including $m=0$, cf. \eq{t2f20}).  Right: the corresponding plots for
  the amplitude averaged over the angles $\mathcal{N}_0(b,r)$, cf. \eq{t2f1}.
  All the curves are obtained using $R^2=2\,\text{GeV}^{-2}$ and $Q_{0s}^2=0.165$~GeV$^2$.}
\label{figntheta}
\end{figure}

The saturation momentum is more precisely defined by the condition that the scattering becomes strong:
$N_{2g}(b,r,\theta) \sim 1$. This condition is controlled by the orientation-averaged piece $\mathcal{N}_0(b,r)$,
which is numerically (much) larger than the piece $\mathcal{N}_\theta(b,r)$ encoding the angular dependence.
This is manifest for sufficiently small dipoles $r\ll 1/m$, when the first piece in $\mathcal{N}_0(b,r)$
is enhanced by the large logarithm $\ln(1/r^2m^2)$, but it is generally true for all the
values of $r$ and $m$ of relevance for this work (see e.g. \fig{figntheta}). The actual
saturation momentum in the present set-up, 
to be denoted as $\mathcal{Q}_s(b)$, is conveniently defined by the condition 
$\mathcal{N}_0(b,r)=1$ when $r=2/\mathcal{Q}_s(b)$. This could be numerically extracted
(as a function of $R$ and $m$), if needed, but for the present purposes it will be sufficient to use the
following,  qualitative, estimate, which strictly holds to leading-logarithmic accuracy:
\beq\label{Qsb}
\mathcal{Q}_s^2(b)\,\simeq\,
Q_s^2(b)\,\ln\frac{Q_s^2(b)}{m^2}\,.\eeq

We have previously argued that the angular dependence of the dipole amplitude comes from
relatively soft transferred momenta $k\sim\Delta$. It is interesting to check that at the level of 
\eq{t2f2}. To this aim, let us take the limit $m\to 0$ in that equation. (The corresponding limit for 
$\mathcal{N}_0$ will be discussed in  Appendix \ref{app:MS}.)  Using
 \beq\label{root}
 \left[\left(k^2+\Delta^2/4+m^2\right)^2-k^2\Delta^2  \right]^{1/2}
\longrightarrow
 \big |k^2-\Delta^2/4\big| \quad\mbox{when} \quad m\to 0\,,\eeq
it is easy to see that the expression within the square brackets inside the integrand becomes
\beqn
\frac{1}{\big |k^2-\Delta^2/4\big|} 
+\frac{2}{\Delta^2} - 
\frac{2\left(k^2+\Delta^2/4\right)}{ \Delta^2\, \big |k^2-\Delta^2/4\big|}
=\Theta\big(\Delta/2-k\big)\frac{4}{\Delta^2}\,,
\eeqn
so that the whole contribution to $\mathcal{N}_\theta(b,r)$ indeed comes from soft
momenta $k\le \Delta /2$. As a matter of facts, the ensuing integral over $k$ 
is dominated by its upper limit $\Delta/2$ and the final result for $m=0$ takes
a rather simple form:
\beqn\label{t2f20}
\mathcal{N}_\theta(b,r)\Big |_{m=0}&=& \frac{1}{2}\times \frac{g^2C_F}{4(2\pi)^2} r^2
 \int_{0}^\infty \rmd\Delta \,\Delta\, 
\tilde{\mu}(\Delta) \rmJ_2(\Delta b)\,=\, Q_{0s}^2r^2\,\frac{R^2}{b^2}
\left[1-\left(1+\frac{b^2}{4R^2}\right)\rme^{-\frac{b^2}{4R^2}}\right].
\eeqn
But albeit formally well defined, the limit $m\to 0$ of $\mathcal{N}_\theta(b,r)$ is physically meaningless,
since very soft momenta $k\lesssim m$ are not allowed by QCD confinement. In that sense, the 
`massive' case in \eq{t2f2} is more useful in practice, albeit our current treatment of confinement is 
merely heuristic. 
To illustrate the uncertainty introduced by this treatment, we represent in the left panel of Fig.~\ref{figntheta} the
result of the double integral in \eq{t2f2} as a function of $b$ for several values of $m$, including $m=0$.
As one can see there, the $m$-dependence become stronger with increasing $b$, a feature which is easy
to understand: the integral over $k$ is effectively restricted to values $m\lesssim k \lesssim \Delta$
and larger values for the impact parameter $b$ correspond to smaller values for $\Delta$.

In Fig.~\ref{figntheta}, one also sees that the function $\mathcal{N}_\theta(b,r)$ develops a
maximum at a value of $b$ which is proportional to $R$ and roughly independent of $m$.
For $m=0$, \eq{t2f20} shows that $\mathcal{N}_\theta\propto b^2$ at small $b\ll R$ and we expect
this to remain true for any value of $m$.
Another interesting aspect of the dipole amplitude in \eq{t2f20} is the fact
that it exhibits a power tail $\propto 1/b^2$ at sufficiently large distances $b\gtrsim R$. This 
is in agreement with the discussion after \eq{N2g}: it reflects the fact that the 
angular dependence of the dipole amplitude is controlled by soft gluon exchanges, 
for which there is no confinement in the limit $m\to 0$. For a non-zero `gluon mass' $m$, this power law tail 
will of course be replaced by the decaying exponential $\rme^{-mb}$, which mimics confinement.
This behavior too is visible in Fig.~\ref{figntheta}.

For more clarity, we also plot the angular-averaged amplitude
$\mathcal{N}_0(b,r)$, under the same assumptions as for $\mathcal{N}_\theta(b,r)$ (see 
the right panel of Fig.~\ref{figntheta}). The fact that for small values of the `gluon mass'
$m\le 2$~GeV, the maximum of $\mathcal{N}_0(b,r)$ as a function of $b$ 
appears to be displaced at non-zero values for $b$ is probably an
artifact of the model. But this also shows that the second, integral, term in the r.h.s. of \eq{t2f1}
is indeed important (by itself, this contribution is negative for sufficiently small values of $b$, but it
becomes positive at $b\gtrsim R$). This feature will have no incidence on our subsequent numerical
studies of $pp$ collisions, where we shall restrict ourselves to larger values $m\ge 0.3$~GeV. For 
such values, the maximum of  $\mathcal{N}_0(b,r)$ is located at $b=0$, as expected
on physical grounds.

Using the above results for $N_{2g}$ together with $S=\exp\{-N_{2g}\}$ and the representation \eqref{sv2}  
for the elliptic flow coefficient $v_2$, we finally deduce the following estimate for the latter:
\beqn
v_2(p,b)&=&-\frac{\int r\rmd r\,  \rme^{-\mathcal{N}_0(b,r)}\rmJ_2(pr)\int \rmd\theta\, \rme^{-\mathcal{N}_\theta(b,r) \cos(2\theta)}  \cos(2\theta)}{\int r \rmd r \, \rme^{-\mathcal{N}_0(b,r)}\,\rmJ_0(pr)\int \rmd\theta\, \rme^{-\mathcal{N}_\theta(b,r) \cos(2\theta)} }, \nonumber\\*[0.2cm]
&=&\frac{\int r\rmd r\,  \rme^{-\mathcal{N}_0(b,r)}\,\rmJ_2(pr)\,\rmI_1\left(\mathcal{N}_\theta(b,r)\right)}{\int r\rmd r\,  \rme^{-\mathcal{N}_0(b,r)}\,\rmJ_0(pr)\,\rmI_0\left(\mathcal{N}_\theta(b,r)\right)},\label{v2m-f}\
\eeqn
where $\rmI_0$ and $\rmI_1$ (the modified Bessel functions of the first kind) have been generated via
\beq
\int_0^{2\pi}\rmd\theta  \,
 \rme^{-z \cos(2\theta)}  =2\pi
 \rmI_0(z),\qquad
 \int_0^{2\pi}\rmd\theta  \,
 \rme^{-z \cos(2\theta )}  \cos(2\theta )=-2\pi
 \rmI_1\left(z\right)\,.
 \eeq
$\rmI_1(z)$ is an odd function which has the same sign as its argument. In fact, the quantity
$\mathcal{N}_\theta$ is numerically small in the physical regime of interest (see \fig{figntheta}),
hence one can use the approximation $\rmI_1\left(\mathcal{N}_\theta(b,r)\right)\simeq \mathcal{N}_\theta(b,r)/2$.
This shows that $v_2(p,b)$ is significantly large only for peripheral collisions, i.e. for 
impact parameters $b\gtrsim R$, where lies the peak of the function $\mathcal{N}_\theta(b,r)$.
It furthermore shows that the elliptic flow generated via multiple scattering is  {\em positive}
\footnote{So long as $p$ is not too large, $p\lesssim
Q_s(b)$, the above integrals over $r$ are effectively cut off by the 
Gaussian $\rme^{-\mathcal{N}_0(b,r)}$ at a value $r\sim 1/Q_s(b)$. Hence, within the
range for $r$ which is relevant for the integration, 
the Bessel functions $\rmJ_2(pr)$ and $\rmJ_0(pr)$ remain positive, meaning that 
the sign of $v_2$ coincides with that of the function $\mathcal{N}_\theta(b,r)$ defined in \eq{t2f2}.}
 --- that is, it has the {\em opposite} sign as compared to the case
of a single hard scattering discussed in Sect.~\ref{sec:SSA}.

\begin{figure}[t]
 \includegraphics[width=8 cm]{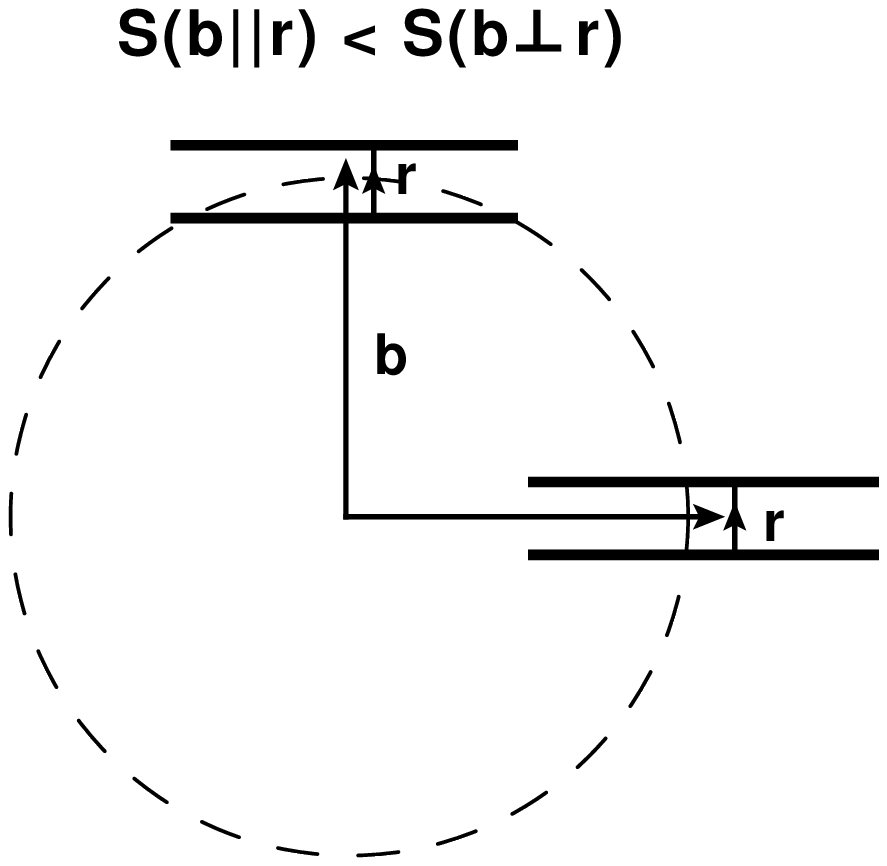}
  \caption{A $q\bar{q}$-dipole with a transverse separation vector $\r$ and impact parameter $\b$ from the center of collisions at two different configurations: $\theta=0$ ($\r\parallel \b$)
and $\theta=\pi/2$ ($\r\perp\b$).}
\label{fS}
\end{figure}

\begin{figure}[t]
 \includegraphics[width=0.45\textwidth]{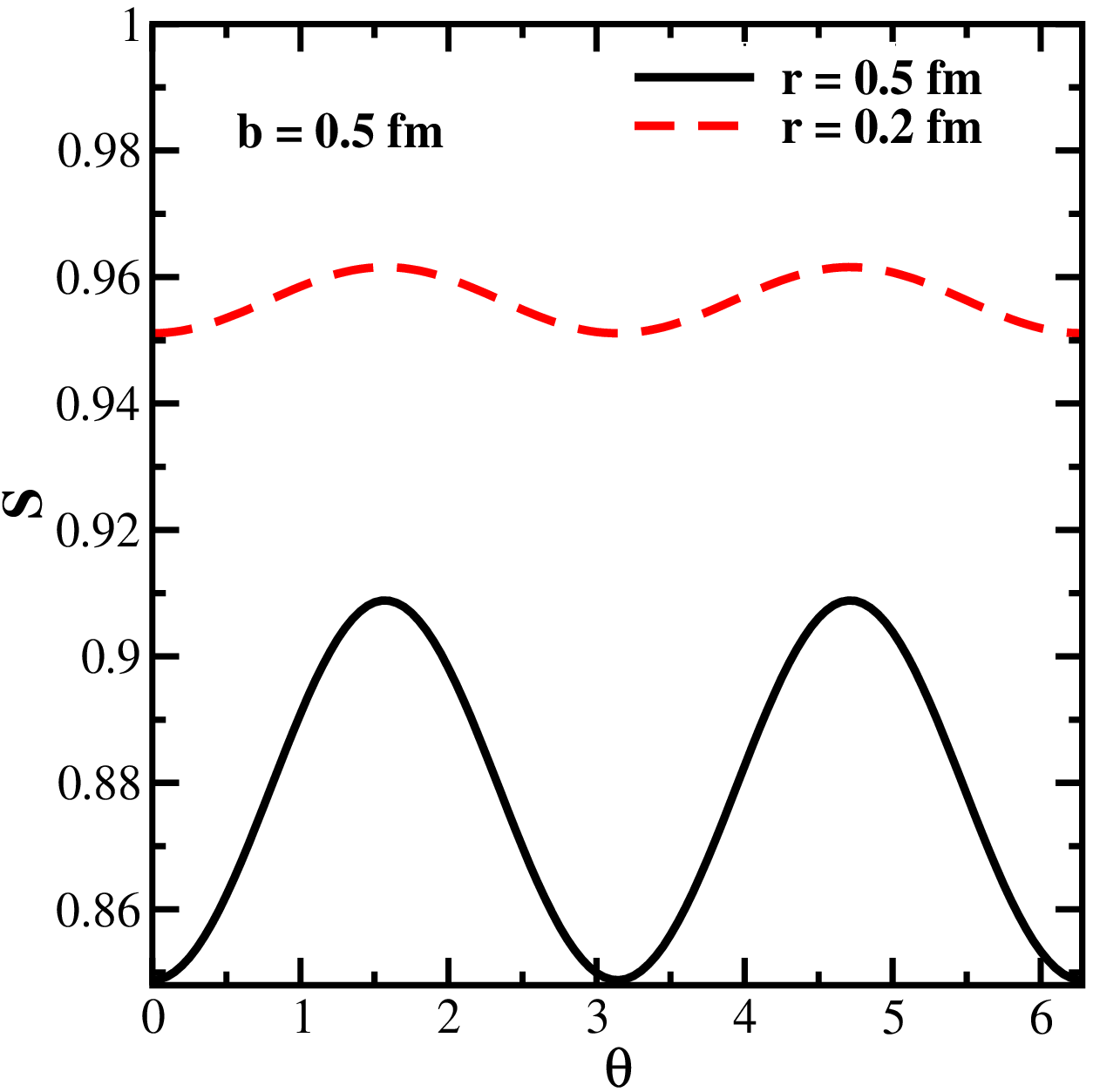}  \qquad
 \includegraphics[width=0.45\textwidth]{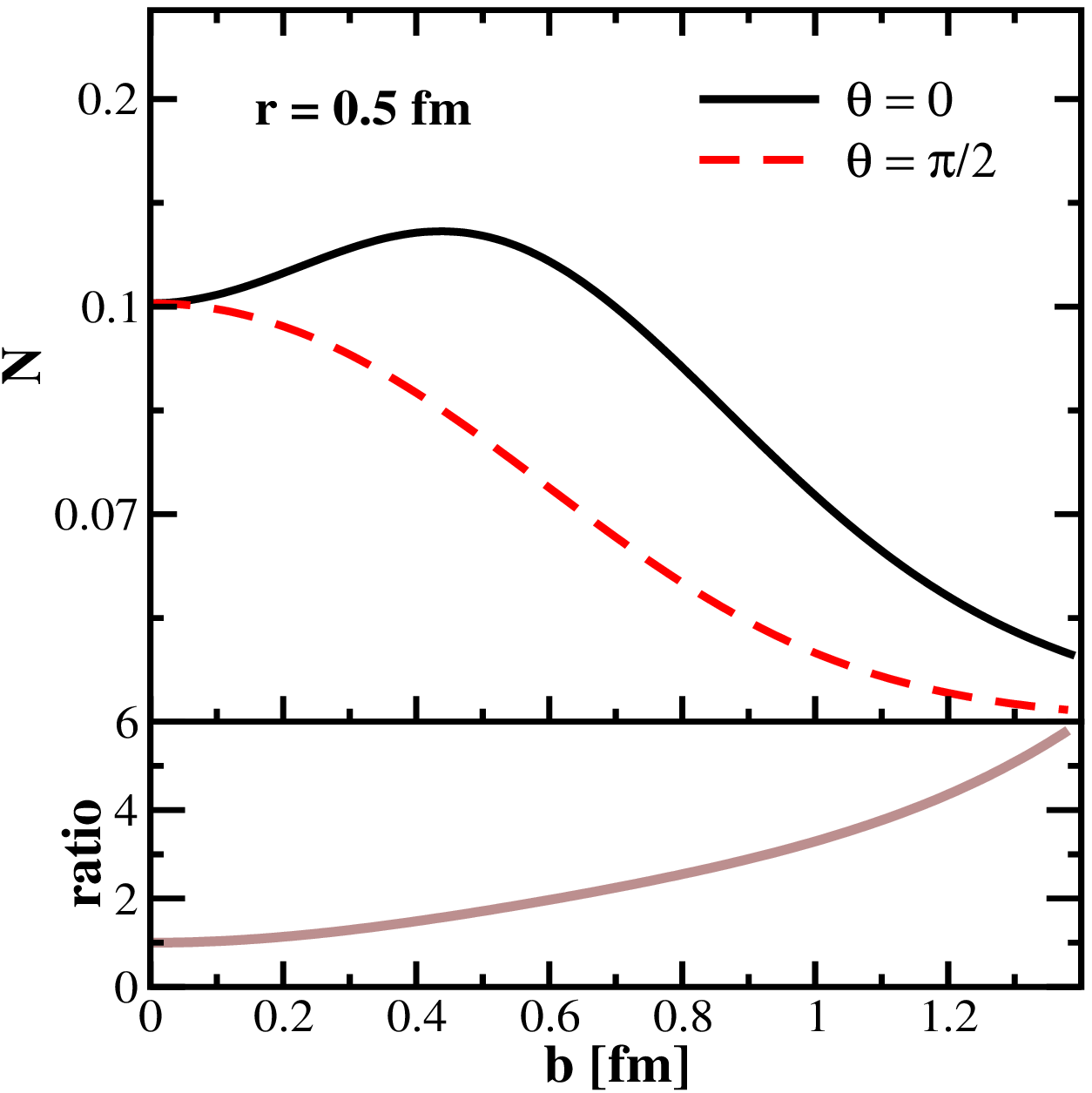}       
 \caption{Left: The dipole `survival probability' (more properly, the $S$-matrix $S=\exp\{-N_{2g}\}$) 
 as a function of
 $\theta$ for a fixed $b$ and 2 values of $r$ (corresponding to rather different scattering strengths).
 Right: the dipole amplitude $N_{2g}(b,r,\theta)$ as a function of $b$ for a fixed value of $r$
and the two extreme possibilities for the orientation:  $\theta=0$ and $\theta=\pi/2$.
The lower insert shows the ratio $N_{2g}(\theta=0)/N_{2g}(\theta=\pi/2)$.
 All curves are obtained by numerically evaluating $N_{2g}$
 according to Eqs.~\eqref{t2-f3}, \eqref{t2f2}, and \eqref{t2f1}, together 
with $m=0.25$ GeV, $R^2=2\,\text{GeV}^{-2}$ (i.e. $R\simeq 0.3$~fm), and $Q_{0s}^2=0.165\,\text{GeV}^2$. 
}
\label{f3}
\end{figure}

Via \eq{t2-f3}, the sign of $v_2$ can be related to properties of dipole scattering. Namely, the fact
that $\mathcal{N}_\theta(b,r)$ is positive implies that the scattering is stronger when the dipole
orientation is (anti)parallel to its impact parameter ($\theta=0$ or $\theta=\pi$) than for a dipole
perpendicular on $\b$ ($\theta=\pi/2$). Equivalently, the $S$-matrix $S(\b,\r)$, which measures
the dipole survival probability, is larger when $\r\perp\b$ than for $\r\parallel \b$ (see Fig.~\ref{fS}).
This property is studied in more detail in Fig.~\ref{f3}: in the left panel, we show the dipole $S$-matrix 
as a function of $\theta$ (for two different dipole sizes and a fixed impact parameter); on the right,
we show the scattering amplitude $N_{2g}(b,r,\theta)$ as a function of $b$ for a fixed value of $r$
and two extreme orientations:  $\theta=0$ and $\theta=\pi/2$.  As one can see, the difference between
`parallel' and `perpendicular' scattering increases with the dipole size $r$ and also with the impact parameter $b$.
These features are intuitively understandable since a point-like dipole should not be sensitive to its orientation.
Besides for very small impact factors $b\lesssim R= 0.3$~fm, the target looks quasi-homogeneous and then
the dipole orientation is irrelevant. We therefore expect the associated $v_2$ to follow a similar trend. This 
will be confirmed by the numerical results to be presented in Sect.~\ref{sec:num}.

Returning to the case of the single scattering approximation, as applying at high $p\gg Q_s(b)$, 
it might be tempting to interpret the negativity of $v_2$ in that case as an opposite trend for the 
dipole scattering, namely $S(\r\parallel \b) > S (\r\perp\b)$. However, we believe that such an
interpretation is truly misleading: in that case, the sign of $v_2$ follows from an analysis that
was performed fully in momentum space. Such an analysis gives one information about the 
unintegrated gluon distribution in
the target (proportional to $ -\tilde{N}_{2g}(\b,\p)$, cf. \eq{asy}), but not about the dipole scattering
as a function of $\r$. To compute the latter, i.e. the function $N_{2g}(\b,\r)$, one needs its Fourier
transform $\tilde{N}_{2g}(\b,\p)$ for {\em all} values of $\p$, and not just for the relatively hard
values for which \eq{asy} applies. In fact, even for small values of $r$, 
the angular dependence of $N_{2g}(\b,\r)$ is controlled by relatively soft
values of $p$ within the inverse Fourier transform $\tilde{N}_{2g}(\b,\p)\to N_{2g}(\b,\r)$
 (cf. the discussion following \eq{T2gexp}).
 
  Finally, let us generalize the previous results to the case where the proton projectile itself
 has a Gaussian distribution in the transverse plane, $x_p q(x_p)\to x_p q(x_p) f(\b)$, with $f(\b)=
 \rme^{-b^2/4R^2}/(4\pi R^2)$. Using this {\em Ansatz} for $f(\b)$ together with the expression
 \eqref{t2-f3} for the dipole amplitude, one can easily perform the angular integrations in
 \eq{v2f} for $v_2$ and thus obtain (the identity $\int_0^{2\pi}\rmd\alpha \,\rme^{-z\cos\alpha}
 \,\cos (2\alpha)=2\pi \rmI_2(z)$ is also useful)  
   \begin{eqnarray}
v_{2}(p,B)&=&\frac{ \int b\rmd b\, \rme^{-(b^2+B^2)/4R^2}\,\rmI_2\big(bB/2R^2)
\int  r\rmd r\, 
\rme^{-\mathcal{N}_0(b,r)}\,\rmJ_{2}(p r)\, \rmI_1\left(\mathcal{N}_\theta(b,r)\right)}
{ \int b\rmd b\, \rme^{-(b^2+B^2)/4R^2}\,\rmI_0\big(bB/2R^2)\int  r\rmd r\, 
\rme^{-\mathcal{N}_0(b,r)}\,\rmJ_0(pr)\,\rmI_0\left(\mathcal{N}_\theta(b,r)\right)}. \label{sv2f}
\end{eqnarray}
We recall that the `dummy' variable $b$ is the impact parameter of a participating quark, whereas
the external variable $B$ refers to the center of the projectile.
The integral over $b$ in the numerator of \eq{sv2f}  is restricted by the support of the function 
$\mathcal{N}_\theta(b,r)$, cf. \fig{figntheta}, hence it receives most of its contribution from relatively
large values $b\gtrsim R$. For the nearly central proton-proton collisions with $B\ll R$,
the overall elliptic flow is negligible, by rotational symmetry: the individual contributions from various
(peripheral) values of $b$ can have any orientation, so they compensate each other. Indeed,
using $\rmI_2(x)\simeq x^2/8$ for $x\ll 1$, it is easy to see that $v_{2}(p,B)$ vanishes as $B^2$ when
$B\to 0$. But for larger impact parameters $B\gtrsim R$, the rotational symmetry
of the interaction region is badly broken (recall Fig.~\ref{fig:pp}) and one expects a non-trivial
net result. Geometrical considerations suggest that $v_{2}(p,B)$ should be proportional
to the eccentricity of the overlapping region, as defined in Eqs.~\eqref{epsilon} or \eqref{epsilonB}, 
which can be more explicitly written as
  \begin{eqnarray}
\varepsilon_{2}(p,B)&=&\frac{ \int\rmd b\, b^3\, \rme^{-(b^2+B^2)/4R^2}\,\rmI_2\big(bB/2R^2)
\int  r\rmd r\, 
\rme^{-\mathcal{N}_0(b,r)}\,\rmJ_{0}(p r)\, \rmI_0\left(\mathcal{N}_\theta(b,r)\right)}
{ \int\rmd b\,  b^3\,\rme^{-(b^2+B^2)/4R^2}\,\rmI_0\big(bB/2R^2)\int  r\rmd r \, 
\rme^{-\mathcal{N}_0(b,r)}\,\rmJ_0(pr)\,\rmI_0\left(\mathcal{N}_\theta(b,r)\right)}\,, \label{epsilonpp}
\end{eqnarray}
and respectively
 \begin{eqnarray}
\varepsilon_{2}(B)=\frac{ \int\rmd b\, b^3\, \rme^{-(b^2+B^2)/4R^2}\,\rmI_2\big(bB/2R^2)}
{ \int\rmd b\,  b^3\,\rme^{-(b^2+B^2)/4R^2}\,\rmI_0\big(bB/2R^2)}\,. \label{epsilon2B}
\end{eqnarray}
Indeed, one can understand this eccentricity as the expectation value 
$\varepsilon_{2}=\langle\cos(2\alpha)\rangle$, where we recall that $\alpha$ is the angle made by 
the impact parameter $\b$ of an individual quark with respect to that, $\B$, of the
center of the projectile (see Fig.~\ref{fig:pp}). Hence larger values for $\varepsilon_{2}$ imply that
all participating quarks have similar impact parameters, hence the produced particles are preferentially
produced along a common direction --- that of $\B$ ---, thus generating a sizable value for the elliptic flow.
And indeed, by inspection of the equations above, it is clear that both $v_2(p,B)$ and $\varepsilon_{2}(p,B)$
[or $\varepsilon_{2}(B)$] are proportional to $B^2$ so long as $B\lesssim R$, hence they are proportional
to each other.  This relation between the elliptic flow and the eccentricity will be further investigated
in Sect.~\ref{sec:num}, where all these quantities will be numerically computed.

 \subsection{Dipole-nucleus scattering: the case of a lumpy target}
 \label{sec:pA}
 
The most straightforward generalization of the previous set-up to the case where the target is
a large nucleus with atomic number $A\gg 1$ would be obtained by assuming that the valence
color charges (and hence the associated gluon distribution) are uniformly distributed throughout the 
nuclear volume --- the so-called ``smooth nucleus''.  Experience with nuclear physics at lower energies
suggests that a reasonable approximation for the 3-dimensional distribution of the nuclear matter 
within a large nucleus is provided by the Woods-Saxon distribution $\rho_A(\vec r)$. By boosting
this distribution and assuming that it also applies to the valence color charges, we conclude that
the case of a  ``smooth nuclear target'' can be obtained by replacing the 2-dimensional
density $\mu(b)$ in \eq{mub} as follows
\beq \label{nuc-0}
\mu(b) \to \mu_A(b)\equiv   \mu_0\,A\,T_A(b),
\eeq
where $T_A(b)$ is the {\em nuclear thickness function},
\beq\label{TA}
T_A(b)\equiv \int\rmd z\,\rho_A(\sqrt{\b^2+z^2}),\qquad\mbox{with}\qquad
\rho_A(\vec r)\,\equiv\,\frac{N_A}{\exp\left(\frac{r-R_A}{\delta}\right)+1}\,,\eeq
where $R_A=(1.12\,{\rm fm})A^{1/3}$ is the nuclear radius and $\delta=0.54$~fm is
the width of the ``nuclear edge'' (the radial distance across which the nuclear density
is rapidly dropping). 
The quantity $\mu_0$ has the same meaning as before --- the color charged squared
for the valence quarks of the {\em nucleon} per unit transverse area ---, hence it
is independent of $A$.
The overall factor of $A$ visible in \eq{nuc-0} reflects the
fact that the density $\rho_A$ is normalized to unity: $\int\rmd^3\vec r \rho_A(\vec r)=1$.
This in turn implies that the  normalization factor $N_A$ scales like $1/A$, hence $T_A
\propto 1/A^{2/3}$ and the color charge density therefore has the canonical
scaling with the number $A$ of nucleons: $\mu_A(b) \propto A^{1/3}$.

Under the above assumption, the formal calculation of the dipole $S$-matrix would proceed in
the same way as before, leading to expressions similar to those already presented in
Eqs.~\eqref{st}, \eqref{t2-f3}, (\ref{t2f1}) and (\ref{t2f2}). The ensuing numerical evaluation however
would likely lead to considerably smaller values for $v_2$, due to combined effect of the
larger value for the nuclear radius $R_A$ and to the fact that the Woods-Saxon
profile is less rapidly varying with $b$ than the Gaussian.

This being said, it is  quite clear that 
a real nucleus is not homogeneous; rather, it is a lumpy superposition of $A$ distinct nucleons
and this lumpiness is known to have important consequences for the phenomenology. 
In particular, it can generate a privileged direction
of motion for the produced particles (for a given impact parameter), via the following mechanism:
the effective dipole, with a given orientation and size $\r$, will scatter off the nucleon which happens
to be located at the dipole impact parameter $\B$. 
(From now on, we shall use $\B$ to denote the impact parameter of the 
dipole w.r.t. the center of the nucleus, and keep $\b$ for its  impact parameter  
w.r.t. the struck nucleon.) As a result, the produced quark will preferentially move along the
direction of the {\em local} impact factor $\b=\B-\b_i$, where $\b_i$ is the position of the struck
nucleon w.r.t. the center of the nucleus. If nucleons are randomly distributed around the given $\B$,
then the information about the orientation of the produced particle will be washed out after averaging
over the nucleon distribution. For large values of $A$, this will likely be the case at impact parameters
$\B$ deeply inside the nucleus, where the nuclear distribution is quasi-homogeneous.
But even in that case,  this cannot happen for impact parameters close to the periphery
($B\sim R_A$), which will therefore generate nonzero contributions to $v_2$. 
These qualitative considerations will be confirmed via an explicit calculation to which we now turn.
  
\begin{figure}[t]
 \includegraphics[width=8 cm]{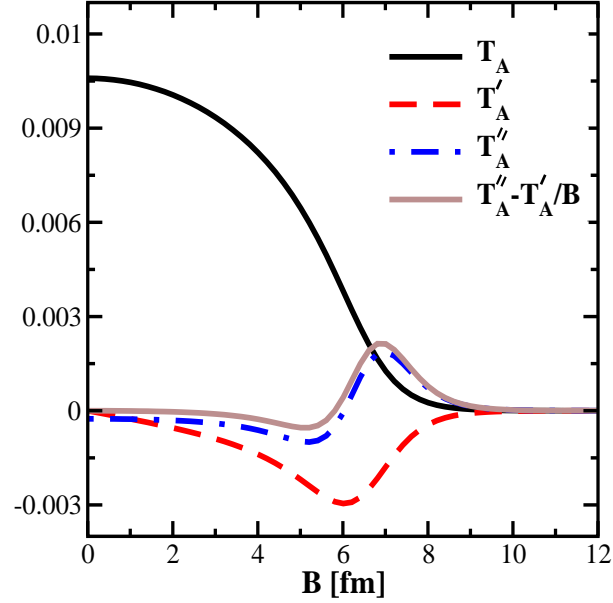}   
 \caption{The nuclear thickness $T_A(B) (\text{in\ fm}^{-2})$ obtained from \eq{TA} with $A=208$, together with
 its first two derivatives 
 $T_A^\prime$ and $T^{\prime\prime}_A$, and the special combination $T^{\prime\prime}_A(B)-T^{\prime}_A(B)/B$
which enters the dipole-nucleus scattering amplitude in \eq{t-exp}.   }
\label{f4-4}
\end{figure}

For a given configuration of the $A$ nucleons inside the nucleus and assuming the dipole to
independently scatter off any of them, the dipole $S$-matrix  should read  (see \cite{Kowalski:2003hm}
for a more complete discussion)
\beq 
S_A(\B,\r)= 
\,\rme^{-\sum_{i=1}^A N_{2g}(\B-\b_i, \r)}.
\eeq
For simplicity, we shall further assume that the various nucleons are distributed 
independently from each other; for each of them, its central position $\b_i$ is
distributed according to the Woods-Saxon thickness function $T_A(\b_i)$. 
The physical observable is then obtained by averaging over all possible configurations of 
the nucleons, as follows
\beq 
S_A(\B,\r)= \int \prod_{i=1}^A\, \rmd^2\b_i\, T_A(\b_i)\,\rme^{-\sum_{i=1}^A N_{2g}(\B-\b_i, \r)}=
\left(\int \rmd^2\b\,T_A\left(|\B-\b|\right)\,\rme^{-N_{2g}(\b,\r)}  \right)^A.
\eeq
The most interesting regime, including for the phenomenology of $pA$ collisions at the LHC,
is such that the scattering between the dipole and a {\em single} nucleon is weak,
$N_{2g}(\b,\r)\ll 1$, yet the overall scattering can be strong (meaning that the $S$-matrix
can be small compared to unity: $S_A(\B,\r)\ll 1$). Under these assumptions, one can expand
the exponential $\rme^{-N_{2g}}$ to lowest non-trivial order, perform the integral over $\b$
and then re-exponentiate the result, to finally obtain (recall the normalization condition
$\int \rmd^2\b\,T_A(\b)=1$)
\beqn 
S_A(\B,\r) \simeq
 \left(1- \int \rmd^2\b\,T_A\left(|\B-\b|\right)\,N_{2g}(\b,\r) \right)^A\,\simeq\,
  \rme^{-AN_{2g}^{A} (\B,\r)},   \label{m-s2}\
\eeqn
with the following definition for the dipole-nucleus scattering amplitude in the two-gluon exchange
approximation (divided by the number $A$ of nucleons)
\beq 
N_{2g}^{A} (\B,\r) = \int d^2\b\,N_{2g}(\b,\r)\,T_A\left(|\B-\b|\right). \label{ta-1}
\eeq
The above integral over $\b$ is effectively restricted (by the support of the dipole-proton amplitude
$N_{2g}(\b,\r)$) to the area of the proton disk, which is small compared to that of the nucleus. In
other terms, one has $b\ll B$ for the most interesting values $B\sim R_A$. In view of this,
one may be tempted to approximate $T_A(\B-\b)\simeq T_A(\B)$, as often done in the literature
\cite{Kowalski:2003hm}. However, this approximation would wash out the
information about the dipole orientation, which is important for us here.
To keep trace of this information, one needs to go one step further
in the small $b/B$ expansion, namely up to quadratic order (the linear term does not contribute
to the integral over $\b$, by parity). We thus write
\beqn\label{TAsecond}
T_A(|\B-\b|)&=&\left(1 -b^i\frac{\partial}{\partial B^i}+
\frac{1}{2}b^ib^j\frac{\partial^2}{\partial B^i\partial B^j}+\dots
\right)T_A\left(|\B|\right), \nn
&\simeq&T_A(B) -\frac{\b \cdot \B}{B} \,T^{\prime}_A(B) +
\frac{b^ib^j}{2}
\left\{
\frac{B^iB^j}{B^2}\,T^{\prime\prime}_A(B)\,+\,\frac{1}{B}\left(\delta^{ij} - 
\frac{B^iB^j}{B^2}\right)T^{\prime}_A(B)\right\}.
\eeqn
Plugging the above expansion and the generic form of $N_{2g}(b,r,\theta)$ given in \eq{t2-f3}  into \eq{ta-1}, one can easily perform the integral over the angle $\theta$ between $\b$ and $\r$ and thus obtain
(from now on, we use $\theta$ to denote the angle made by the vectors $\r$ and $\B$)
\beq 
N_{2g}^{A} (B,r,\theta) = \mathcal{N}^{A}_0 (B,r) +\mathcal{N}^{A}_{\theta} (B,r) \cos(2\theta),  \label{f-ta}
\eeq
where
\beqn 
 \mathcal{N}^{A}_0 (B,r) &=& 2\pi\,\int \rmd b\,b\, \mathcal{N}_0 (b,r)\left\{T_A(B) + \frac{b^2}{4}
  \bigg(T^{\prime\prime}_A(B)+\frac{1}{B}\,T^{\prime}_A(B)\bigg)\right\}, \nonumber\\
 \mathcal{N}^{A}_{\theta} (B,r) &=& \frac{\pi}{4} \int \rmd b\, b^3\,  \mathcal{N}_\theta (b,r)
 \bigg(T^{\prime\prime}_A(B)-\frac{1}{B}\,T^{\prime}_A(B)\bigg).\label{t-exp} 
\eeqn
The $\theta$-dependent piece $\mathcal{N}^{A}_{\theta} (B,r) $ is proportional to the
(first and second) derivatives of the thickness function $T_A(B)$, hence its support is limited
to values of $B$ near the edge of the nucleus, 
within a distance $\Delta B\sim\delta$ around $B=R_A$ (see Fig.~\ref{f4-4}). This is in agreement
with our previous physical discussion and confirms that  
the mechanism under consideration can generate a sizable $v_2$ only in peripheral $pA$ collisions. 
As also illustrated in Fig.~\ref{f4-4}, the special combination $T^{\prime\prime}_A(B)-T^{\prime}_A(B)/B$
which enters $\mathcal{N}^{A}_{\theta} (B,r) $ is positive for most values of $B$ within its support. 
(It can become slightly negative at intermediate values of $B$, but the corresponding value $v_2(p,B)$
is anyway very small, as we shall see.)
Together with the positivity of the respective proton amplitude $ \mathcal{N}_\theta (b,r)$, as numerically
observed in Sect.~\ref{sec:MS}, this implies $\mathcal{N}^{A}_{\theta} (B,r) > 0 $. That is, 
as in the case of a proton target, the scattering is stronger when the dipole orientation $\r$ is 
(anti)parallel to its nuclear impact factor $\B$, rather than perpendicular on it.

At this stage, one could use the integral representations for 
the functions $\mathcal{N}_0(b,r)$ and $\mathcal{N}_\theta(b,r)$, as given in
Eqs.\,(\ref{t2f1}) and respectively (\ref{t2f2}), to numerically perform the integrals in \eq{t-exp}.
This would amount to computing a sequence of three radial integrations, with integrands 
involving the oscillatory Bessel functions. This is indeed possible in practice, but  rather 
tedious and very time-consuming. It turns out that this whole calculation can be efficiently 
reorganized, in such a way to provide fully analytic results for the
nuclear amplitudes $\mathcal{N}^{A}_0 (B,r) $ and $\mathcal{N}^{A}_{\theta} (B,r)$.
This is explained in Appendix~\ref{app:pA}, from which we show the final results
(whose general structure is indeed consistent with \eq{t-exp}):
\beqn \label{NA0}
 \mathcal{N}^{A}_0 (B,r) &=& \pi R^2 Q_{0s}^2 r^2 \ln\left(\frac{1}{r^2m^2}+\rme\right)
 \bigg[T_A(B) + R^2\Big(T^{\prime\prime}_A(B)+\frac{1}{B}\,
 T^{\prime}_A(B)\Big)\bigg]\nn
 &{}&+\frac{\pi}{3}\frac{R^2} {m^2}\,Q_{0s}^2 r^2
 \Big(T^{\prime\prime}_A(B)+\frac{1}{B}\,T^{\prime}_A(B)\Big)\,,
  \\*[0.2cm] \label{NAtheta}
 \mathcal{N}^{A}_{\theta} (B,r) &=&\frac{\pi}{6}\frac{R^2} {m^2}\,Q_{0s}^2 r^2
\Big(T^{\prime\prime}_A(B)-\frac{1}{B}\,
 T^{\prime}_A(B)\Big)\,.
 \eeqn
The first line in the r.h.s. of \eq{NA0} for $\mathcal{N}^{A}_0$, which is proportional to the large logarithm
$\ln({1}/{r^2m^2})$, represents the dominant contribution to the dipole amplitude. Its present calculation
within the MV model is indeed under control (at least for sufficiently small dipole sizes $r\ll 1/m$), 
since this contribution is dominated by relatively large exchanged momenta, $m\ll k\ll 1/r$. Within that 
contribution, the dominant piece is the one proportional to $T_A(B)$. This argument shows that the nuclear
saturation momentum $Q_{sA}$ (the inverse dipole size for which the dipole amplitude 
$AN_{2g}^{A} (\B,\r)$ becomes of order one) can be estimated as
\beq\label{QAdef}
Q_{sA}^2(B)\,\simeq\,AR^2T_A(B)Q_{0s}^2\,.\eeq
This scales like $A^{1/3}$ and has the same impact-parameter dependence as the nuclear thickness
function $T_A(B)$.

The $\theta$-dependent piece $\mathcal{N}^{A}_{\theta}$ of the amplitude, which is the most interesting one for the present purposes, is proportional to $1/m^2$, which demonstrates its non-perturbative origin: 
it has been generated by integrating over soft momenta $k\sim m$. 
In that sense, the above calculation is merely heuristic and in particular strongly dependent upon our recipe for implementing the infrared cutoff $m$. At this point, one may wonder about the difference between the 
small-$m$ behaviors observed in $pA$ and respectively $pp$ collisions: when $m\to 0$,
the $\theta$-dependent piece of the dipole amplitude remains finite for $pp$ collisions,
cf. \eq{t2f20}, whereas it is quadratically divergent in the case of $pA$ collisions, cf. \eq{NAtheta}.
This difference can be traced back to the integral over $b$ which needs to be performed when passing from $pp$
to $pA$ collisions, cf. \eq{t-exp}. When $m\to 0$ and for large $b\gtrsim R$, the respective amplitude 
for $pp$ collisions has a power tail $\mathcal{N}_\theta (b,r)\propto 1/b^2$, as visible in \eq{t2f20}.
Therefore, the integral $ \int \rmd b\, b^3\,  \mathcal{N}_\theta (b,r)$ which enters \eq{NAtheta}
for $\mathcal{N}^{A}_{\theta}$  would be quadratically divergent in the absence of confinement.
After adding the latter in the form of a `gluon mass' $m$, this integral is cutoff at $b\sim 1/m$,
thus yielding $\mathcal{N}^{A}_{\theta}\propto 1/m^2$. 
 
We are finally in a position to compute the elliptic flow coefficient $v_2$ for $pA$ collisions:
by inserting the dipole $S$-matrix obtained according to 
Eqs.\,(\ref{m-s2}) and (\ref{f-ta}) into our master formula \eqref{sv2}, we obtain,
similarly to \eq{v2m-f},
\beq
v_2 (p,B)=\frac{\int r\rmd r\,  \rme^{-A \mathcal{N}^A_0(B,r)}\,\rmJ_2(pr)\,\rmI_1\left(A\mathcal{N}^A_\theta(B,r)\right)}{\int r\rmd r\,  \rme^{-A\mathcal{N}^A_0(B,r)}\,\rmJ_0(pr)\,\rmI_0\left(A\mathcal{N}^A_\theta(B,r)\right)}\,.\label{v2m-f2} 
\eeq
This can be numerically computed using Eqs.\,(\ref{NA0}) and \eqref{NAtheta}, 
with results to be discussed in the next section. The generalization of \eq{v2m-f2} to an extended projectile
is straightforward and will be considered too in Sect.~\ref{sec:num}.

\section{Numerical results for $v_2$ and physical discussion}
\label{sec:num}

In this section we present the numerical results for $v_2$ in $pp$ and $pA$ collisions (with $A=208$) 
as emerging from our model. For more clarity, in the following (and in all plots) 
we shall denote the transverse momentum with $p_T$.
We shall limit ourselves to the scenario which includes the effects of multiple scattering, as discussed
in Sect.~\ref{sec:MS} for $pp$ collisions and in Sect.~\ref{sec:pA} for the $pA$ collisions. For both cases,
$pp$ and $pA$ collisions, we shall exhibit $v_2$ as a function of the transverse momentum $p_T$ of the produced quark, for various choices of the impact-parameter,
the central saturation momentum in the proton $Q_{0s}$, and the
infrared cutoff $m$. 
The only other parameter of our model, i.e. the width  $R$ of the proton 
color charge distribution in the transverse space, is fixed
to the average value emerging from a fit to $t$-distribution of diffractive vector meson
production at HERA, that is $R^2=2\,\text{GeV}^{-2}$. Strictly speaking, such a fit is based on a different    
`saturation model', namely IP-Sat \cite{Rezaeian:2012ji}
but this difference is not essential for the subsequent discussion, which will be mostly qualitative. Note also that the value of $R^2$ extracted using the bCGC model in a fit to the same data \cite{Rezaeian:2013tka} is slightly larger.


\begin{figure}[t]                                       
                                  \includegraphics[width=8. cm] {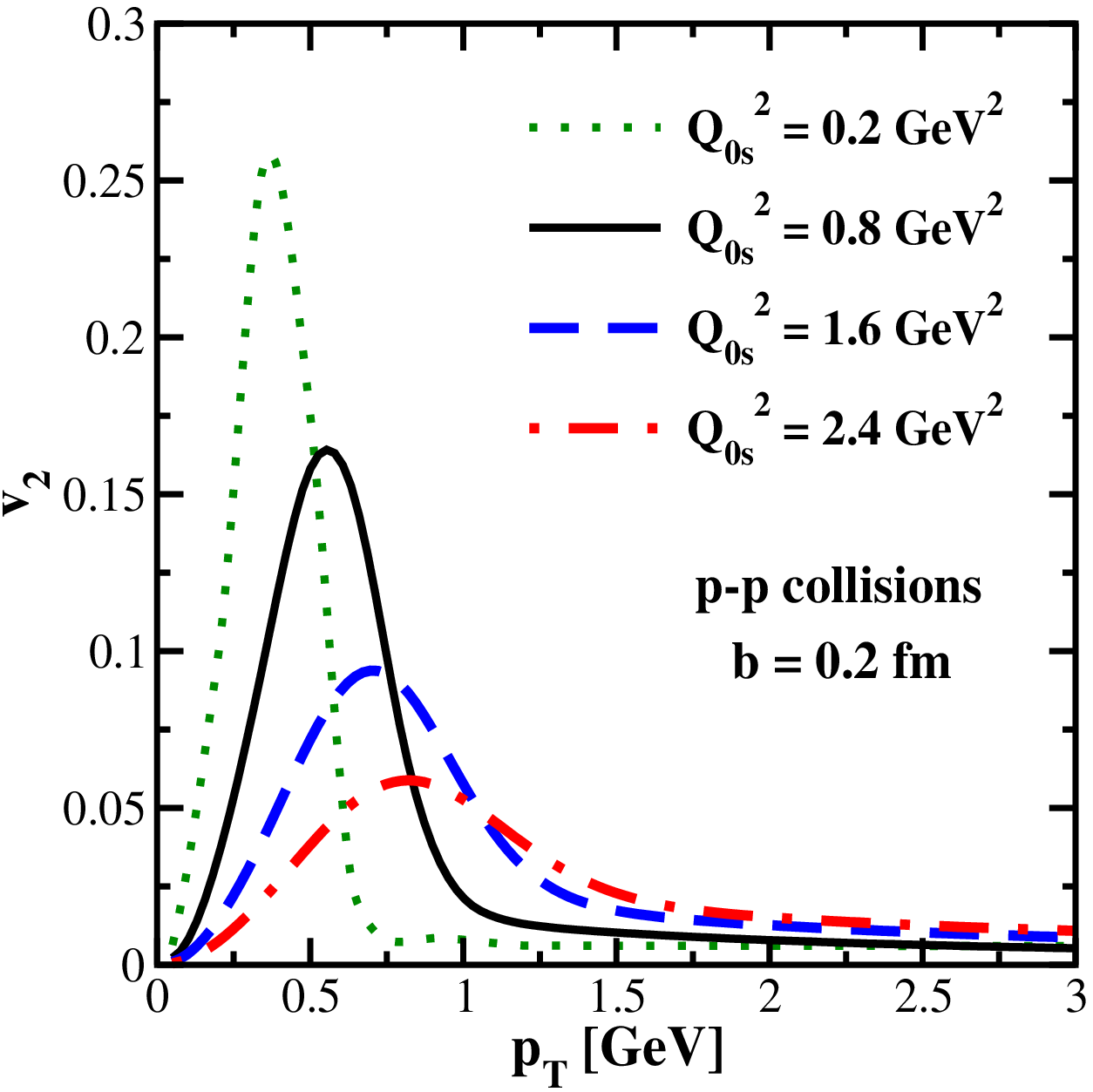}  \qquad                                   
  \includegraphics[width=8. cm] {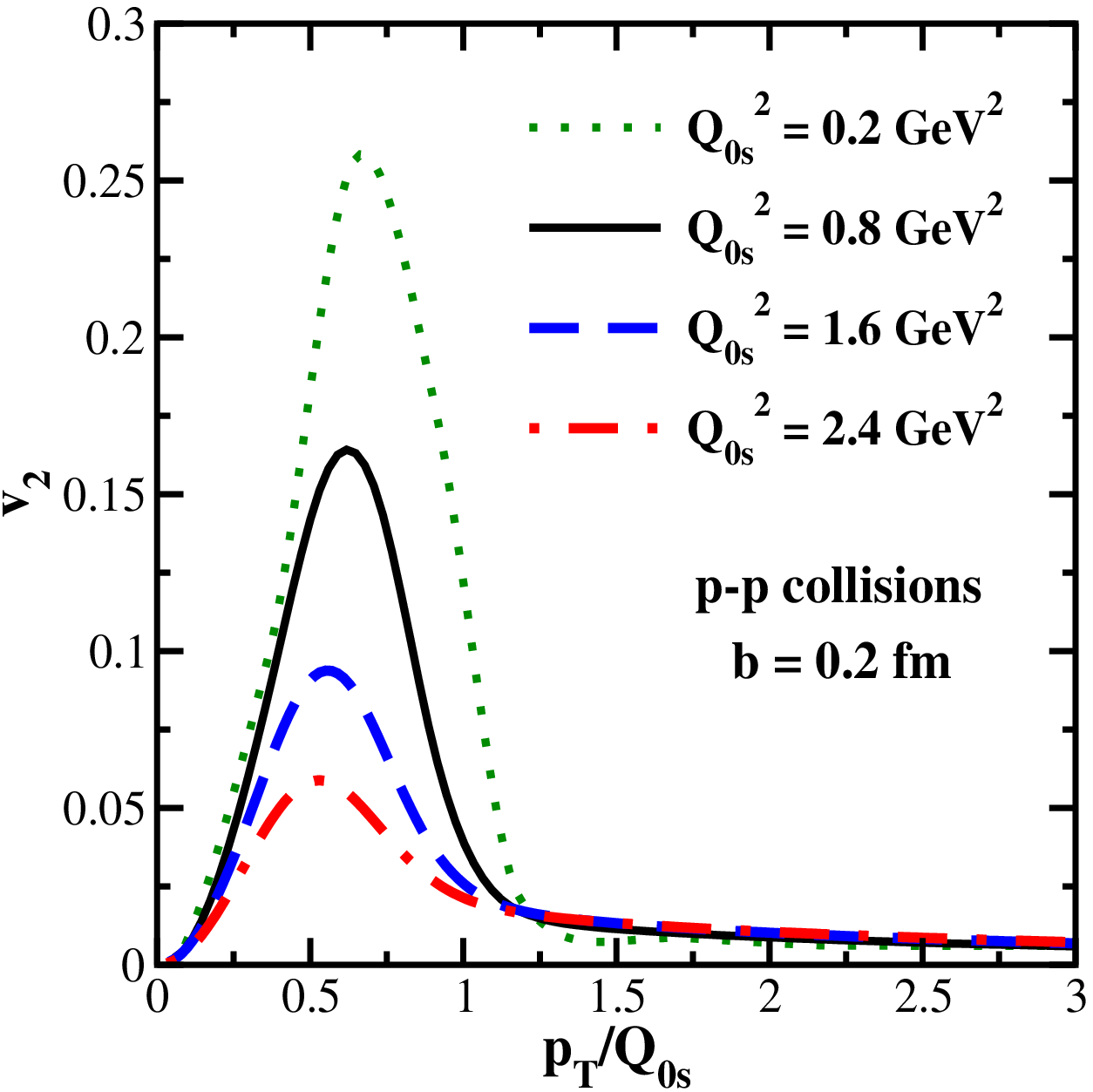}    
\caption{The azimuthal asymmetry $v_2$ in $pp$ collisions with a point-like projectile proton.
Left: 4 values for the central saturation scale $Q_{0s}^2=0.2,\,0.8,\,1.6,\, 2.4 \,\text{GeV}^{2}$ for
a fixed impact parameter $b=0.2$ fm.  Right: the same as 
 the left panel, but plotted as a function of $p/Q_{0s}$.  All these results are obtained by using 
 $m=0.3$ GeV and $R^2=2\,\,\text{GeV}^{-2}$.}
\label{fp-1}
\end{figure}

Concerning $Q_{0s}$ --- the proton saturation momentum at $b=0$ ---, we shall consider a rather wide
range of values, from $Q_{0s}^2=  0.2-0.4\ \text{GeV}^2$ up to $Q_{0s}^2= 2.4\ \text{GeV}^2$. The 
lowest values emerge from phenomenological analyses based on the Balitsky-Kovchegov equation with 
running coupling (rcBK) to either the HERA data \cite{Albacete:2010sy,Iancu:2015joa,Albacete:2015xza},
or to the $pp$ data at RHIC and the LHC \cite{JalilianMarian:2011dt,Albacete:2012xq,Rezaeian:2012ye}.
The highest value could in principle be reached in high-multiplicity events characterized by large fluctuations
\cite{Hatta:2006hs,Avsar:2010rf}. (Notice that the fits to HERA data in \cite{Iancu:2015joa,Albacete:2015xza} use
a more complete version of the BK equation which besides a running coupling, also includes
collinear improvement \cite{Beuf:2014uia,Iancu:2015vea,Iancu:2015joa}.)

\begin{figure}[t]                                                                  
     \includegraphics[width=8. cm] {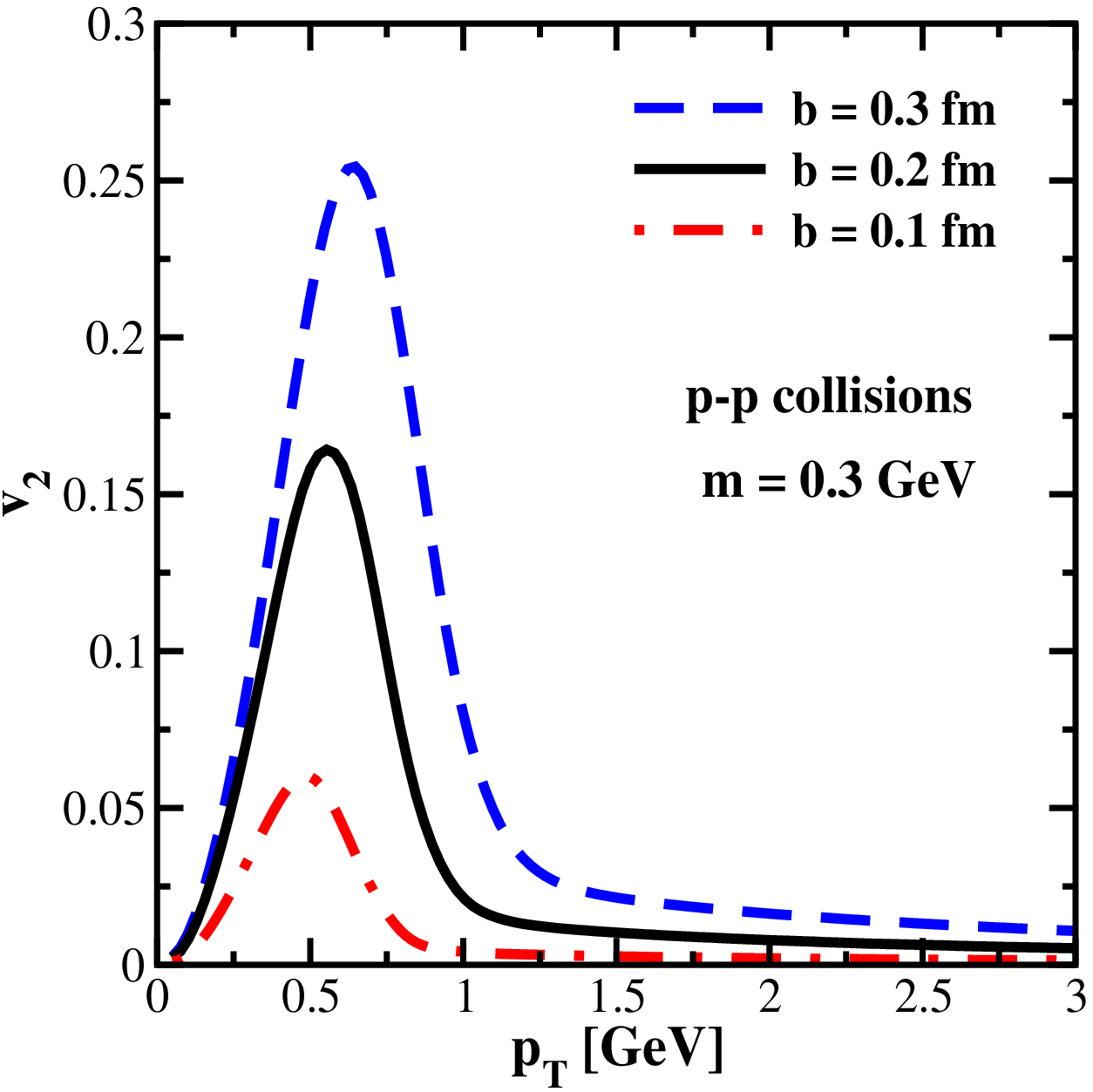}  \qquad
      \includegraphics[width=8. cm] {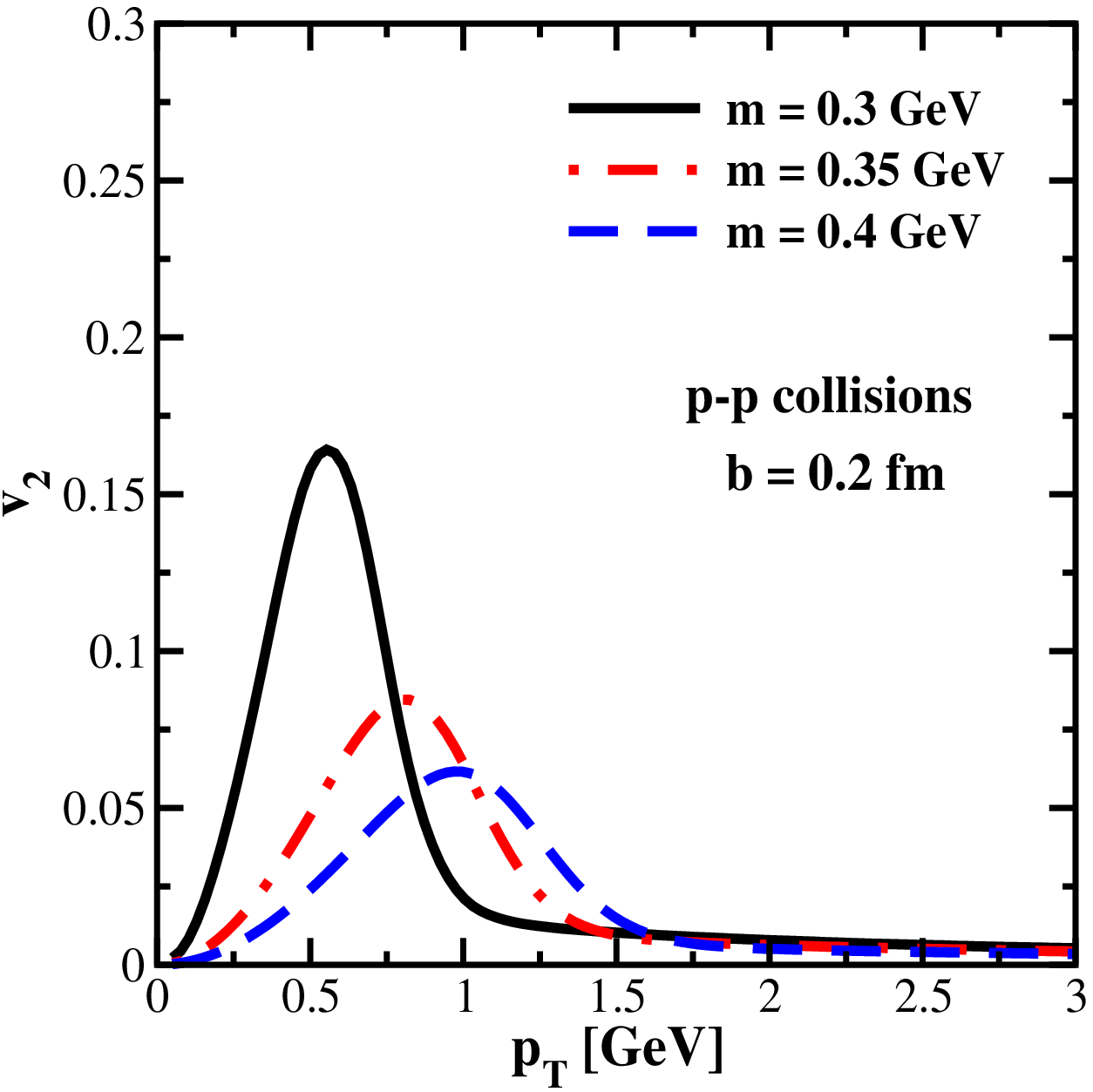}   
\caption{The azimuthal asymmetry $v_2$ in $pp$ collisions with a point-like projectile proton.
Left:  3 different impact parameters:
 $b=0.1,\,0.2,\ \mbox{and}\ 0.3$ fm, for a fixed `mass' $m=0.3$~GeV.
Right: the dependence  upon the infrared cutoff (the `mass' $m$),
for a fixed impact parameter $b=0.2$ fm.  
All these results are obtained by using  $Q_{0s}^2=0.8\,\text{GeV}^{2}$
and $R^2=2\,\,\text{GeV}^{-2}$.}
\label{fp-2}
\end{figure}

We are now prepared to present our numerical results, starting with $pp$ collisions. As stressed
in the Introduction,  we have in mind an asymmetric situation, where one of the protons (`the target') 
looks dense and can be described by the MV model, while the other one (`the projectile') is dilute.
This might be the case for particle production at very forward rapidities and also for rare,
high-multiplicity, events in which the target proton develops `hot spots' via fluctuations in
the high-energy evolution \cite{Hatta:2006hs,Avsar:2010rf}.
We first show our results for the idealized case of a point-like projectile, cf. \eq{v2m-f},
and then for the more realistic case of a projectile which has a non-trivial extent in the transverse plane,
cf. \eq{sv2f}.

In \fig{fp-1} we show the azimuthal asymmetry $v_2$  computed according to Eqs.\,(\ref{t2f1}), (\ref{t2f2}) 
and \eqref{v2m-f} for different choices of the central saturation scale $Q_{0s}$. These plots illustrate the
scaling of the peak position $p_{\rm max}$ with $Q_{0s}$ : when we plot $v_2$ as a function of $p_T/Q_{0s}$,
the peak position $p_{\rm max}/Q_{0s}$ is quasi independent of $Q_{0s}$ and rather close to 1/2.
This scaling property indicates the importance of the saturation physics. 
A larger saturation scale shifts the unintegrated gluon distribution (the integrand of $v_2$) 
to higher transverse momenta.  
In \fig{fp-2} we show the dependence of the azimuthal asymmetry $v_2$ 
upon the impact parameter $b$ (left panel) and upon the infrared cutoff $m$ 
(right panel). As expected, the strength of $v_2$ is increasing with $b$. Remarkably though, 
one see that quite large peak values $v_2(p_{\rm max},b)\gtrsim 0.15$ are obtained already for not
so large impact parameters, $b\lesssim 0.3$~fm, that is, for collisions which are peripheral,
but not {\em ultra}-peripheral. (Recall that the typical transverse size of the color charge distribution in
the target is $2R\sim 0.6$~fm.) It is also interesting to notice that, albeit the height of the peak 
is rapidly increasing with $b$, its position $p_{\rm max}$ changes only slightly when going 
from rather central ($b=0.1$~fm) to more peripheral ($b=0.3$~fm) values.  This observation should
be correlated with the fact that, as manifest on \eq{t2f20}, the piece $\mathcal{N}_\theta(b,r)$ 
of the amplitude which is
responsible for the angular dependence is proportional to the {\em central} value $Q_{0s}$ of the
saturation scale, and not to its {\em local} value at the actual impact parameter.
As anticipated, the $m$-dependence is quite strong:  a slight increase in $m$, from 0.3~GeV
to 0.4~GeV, reduces the peak value of $v_2$ by a factor 3. 

\begin{figure}[t]                                       
                                  \includegraphics[width=8.cm] {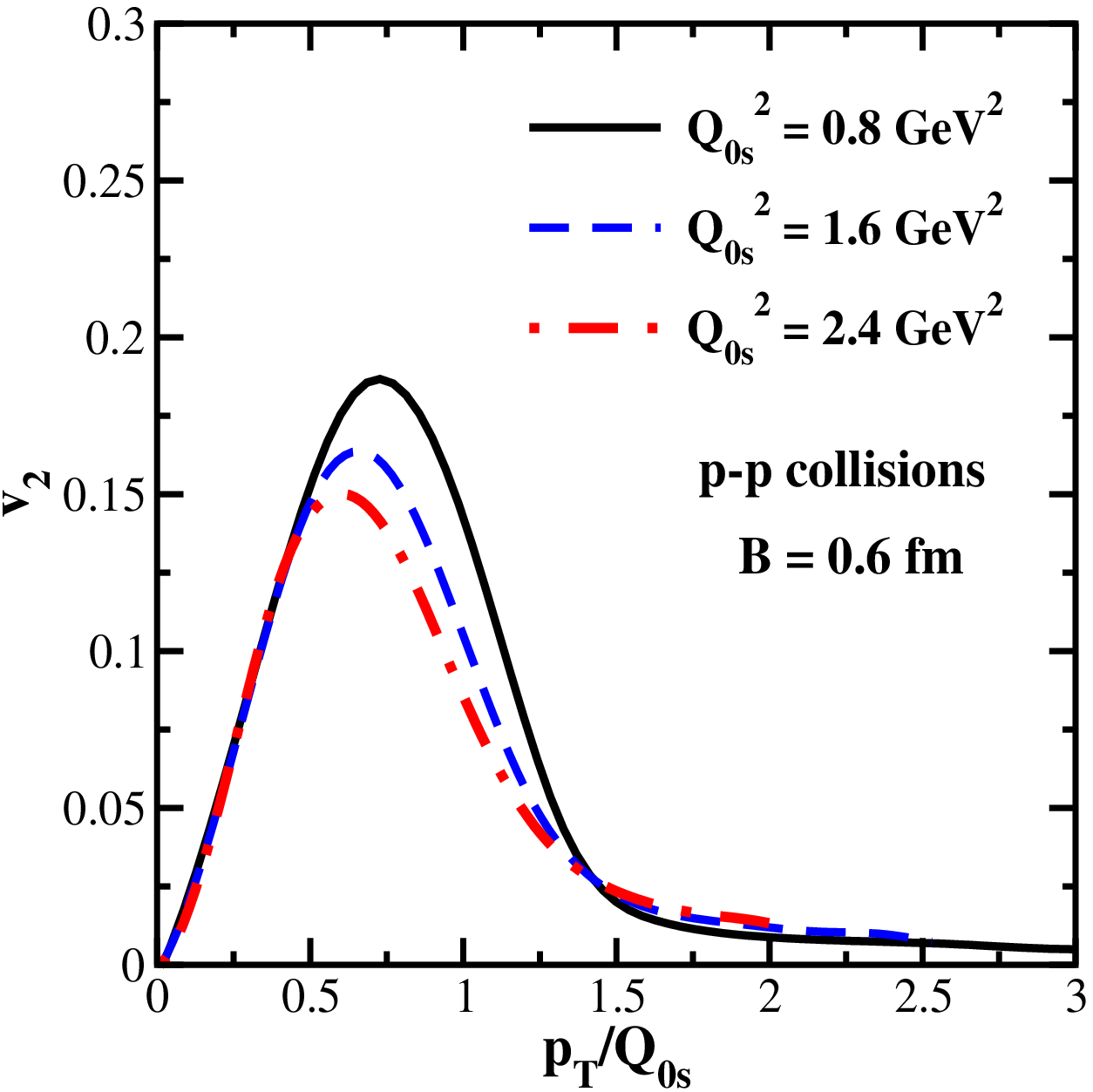}        \qquad                             
  \includegraphics[width=8.cm] {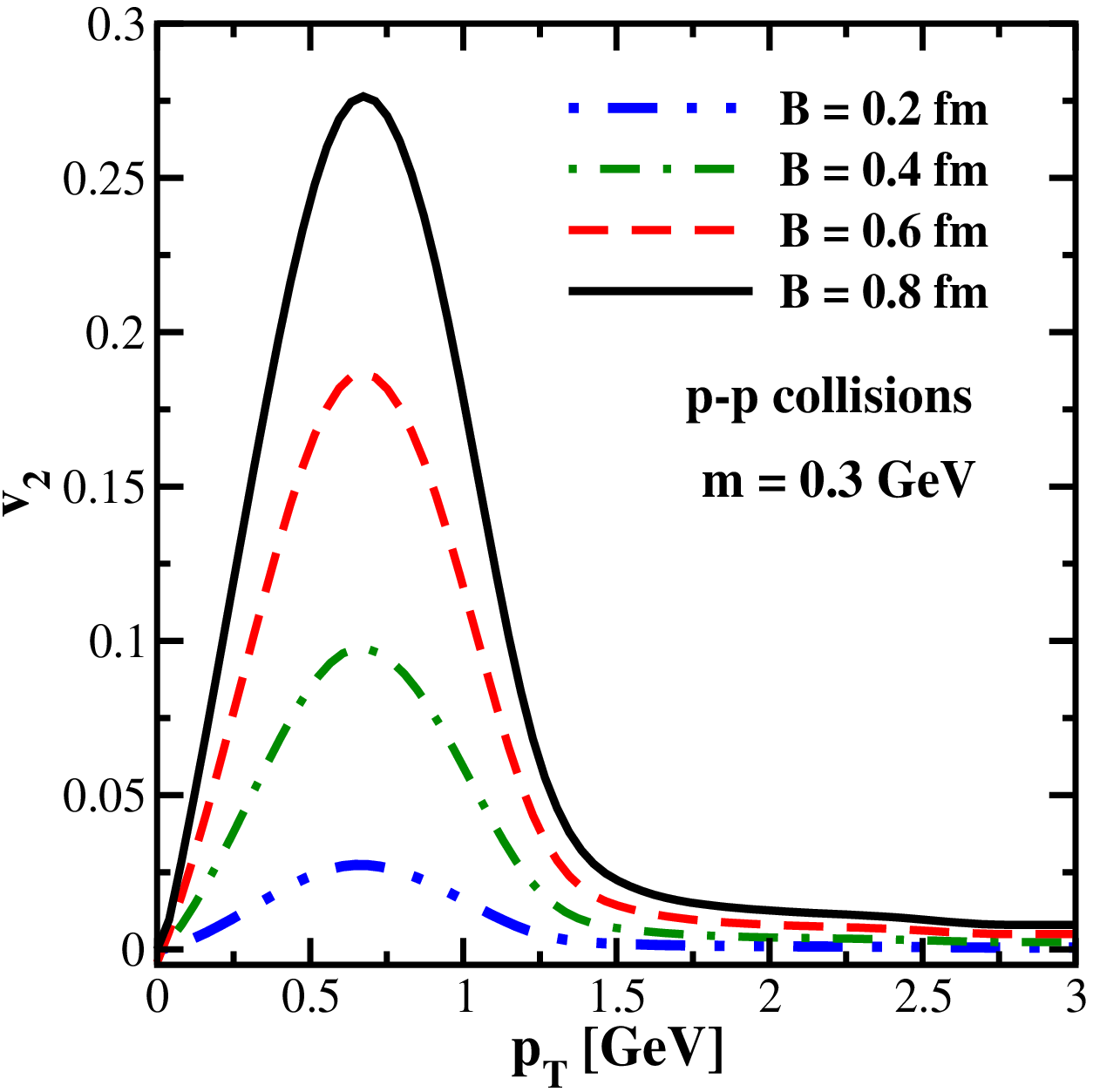}    
\caption{The azimuthal asymmetry  $v_2$ in $pp$ collisions with an extended projectile,
whose center has an impact parameter $B$.
Left: 3 different values for the central saturation scale in the target: 
 $Q_{0s}^2=0.8,\,1.6,\, 2.4 \,\text{GeV}^{2}$, for a fixed 
 $B=0.6$ fm; the results are directly plotted in terms of the `scaling variable'
 $p/Q_{0s}$.  Right: 4 different impact parameters: $B=0.2,\,0.4,\,0.6\ \mbox{and}\ 0.8$ fm, for a fixed
 $Q_{0s}^2=0.8\,\text{GeV}^{2}$.  All these results are obtained by using 
 $m=0.3$ GeV and $R^2=2\,\,\text{GeV}^{-2}$.}
\label{pp-B-Qs}
\end{figure}

\begin{figure}[t]                                       
                                  \includegraphics[width=8.cm] {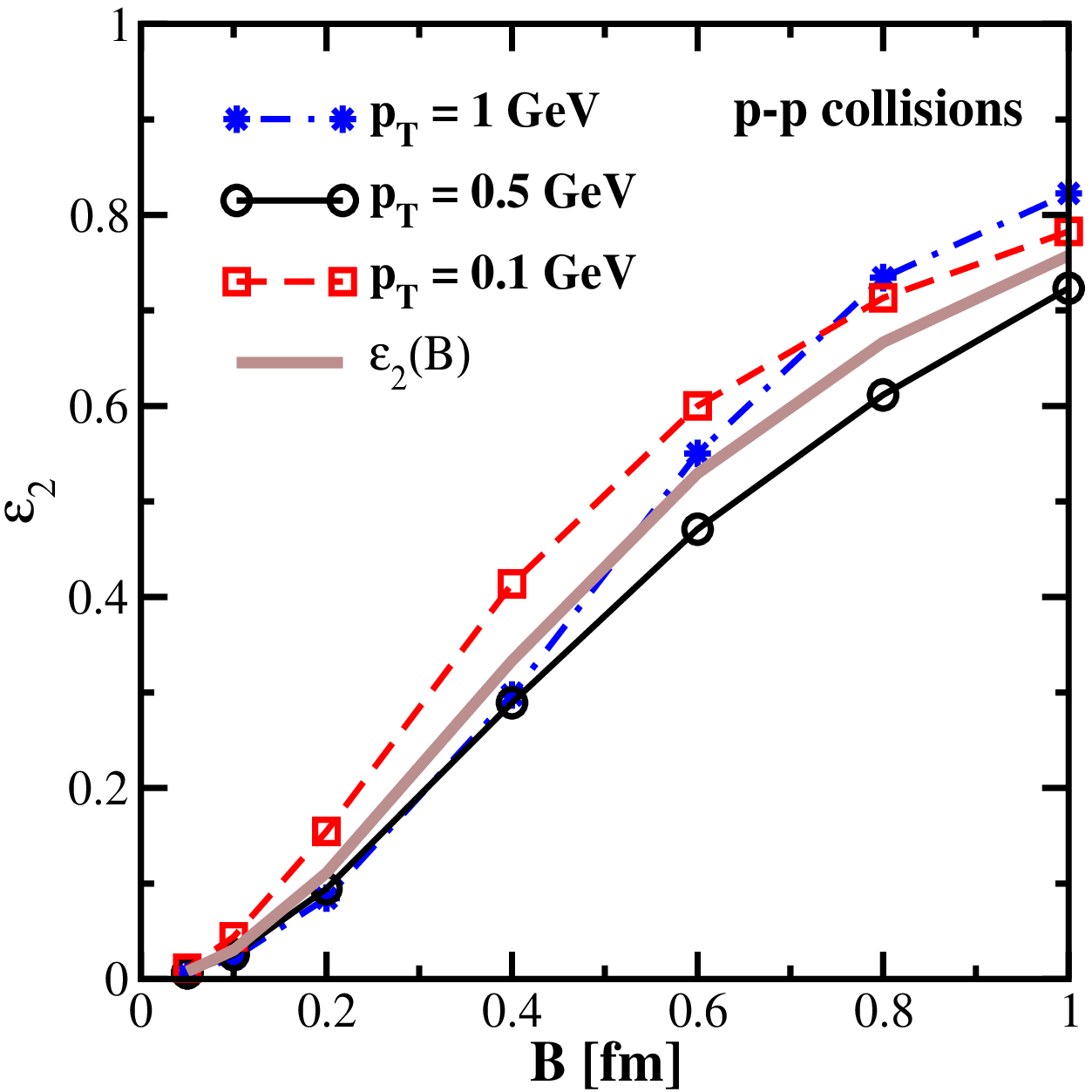}    \qquad                                 
  \includegraphics[width=8. cm] {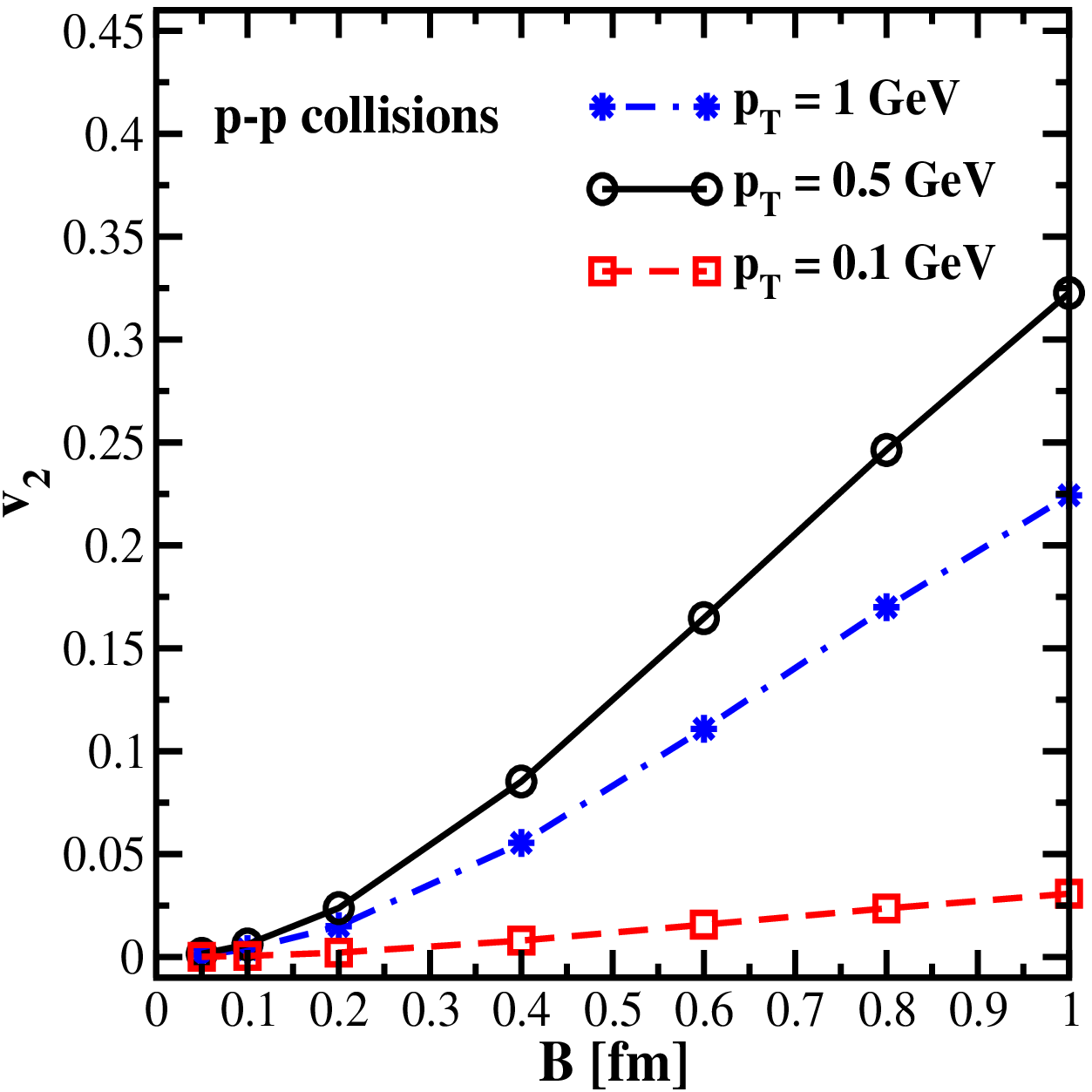}    
\caption{The eccentricity $\varepsilon_2(p_T,B)$ (left panel) and the azimuthal asymmetry $v_2(p_T,B)$ (right panel)
in $pp$ collisions plotted as functions of the impact parameter $B$ of the projectile proton, for 3 values
of the transverse momentum of the produced quark: $p_T=0.1$, $0.5$, and 1~GeV. In the left plot,
we also show the eccentricity integrated over $p_T$, that is, the quantity $\varepsilon_2(B)$
given by \eq{epsilon2B}. All these results are obtained by using  
 $Q_{0s}^2=0.8\,\text{GeV}^{2}$,  $m=0.3$ GeV and $R^2=2\,\,\text{GeV}^{-2}$.}
\label{fepsilon}
\end{figure}

Turning now to an extended projectile with a Gaussian distribution in impact parameter space,
the corresponding $v_2$ is shown  in Fig.~\ref{pp-B-Qs}, for various values 
for $Q_{0s}^2$ and $B$. (We have checked that the $m$-dependence of the results
is similar to that observed for a point-like projectile, cf. \fig{fp-2}.)
One may expect the strength of the azimuthal asymmetry to be reduced, perhaps even significantly,
after averaging over the surface of the projectile, but this is actually not the case: as visible in
Fig.~\ref{pp-B-Qs}, the peak value of $v_2$ remains as large as for a point-like projectile.
To see such a sizable $v_2$, however, one needs to go to larger values for the impact parameter $B$,
which now refers to the center of the projectile  (recall \fig{fig:pp}).
This is in agreement with the discussion at the end of Sect.~\ref{sec:MS}, which also suggests
that the value of $v_2$ should be correlated to the eccentricity $\varepsilon_2$ of the interaction 
region. 

To check this conjecture, 
we have numerically computed $\varepsilon_2(p_T,B)$ and  $\varepsilon_2(B)$
according to Eqs.~\eqref{epsilonpp}--\eqref{epsilon2B}, with the results shown in
\fig{fepsilon} (left panel). These results should be compared to the $B$-dependence of $v_2$,
as exhibited in the right panel of the same figure. These plots confirm
that $v_2$ and $\varepsilon_2$ show a similar trend with $B$: they monotonously increase with $B$ 
--- actually, they are both proportional to $B^2$ so long as $B$ is small enough, $B\lesssim R$. On the other
hand, they show rather different behaviors with $p_T$. The plots for $v_2(p_T,B)$ in the right panel of
\fig{fepsilon} are in agreement with those in the left panel of \fig{pp-B-Qs}: $v_2$ vanishes as
$p_T\to 0$ and has a pronounced peak at $p=p_{\rm max}$ with $0.5 < p_{\rm max} < 1$~GeV.
On the other hand, $\varepsilon_2$ has a rather weak dependence upon $p_T$: the curves
corresponding to different values for the momentum are rather close to each other, and also to the
curve representing the integrated eccentricity $\varepsilon_2(B)$. This 
reflects the fact that the quantity $\varepsilon_2(p_T,B)$ is only weakly sensitive to the dipole scattering,
since mostly controlled by the geometry. 

\comment{
not only is non-zero as $p_T\to 0$, but in fact it shows the opposite
trend: it {\em decreases} with increasing $p_T$ (at least for sufficiently small values of $p_T$).
This can be understood as follows: a momentum $p_T\lesssim Q_{0s}$ gets accumulated via
multiple scattering, hence smaller values for $p_T$ correspond to larger values for
the impact parameter $b$ of the colliding quark. Vice-versa, when increasing $p_T$ (with 
$p_T\lesssim Q_{0s}$ though), the integral over $b$ in \eq{epsilonpp} is controlled by smaller 
and smaller values for $b$, hence the overall result should be decreasing. This is corroborated by
the two plots  in \fig{fepsilon} (left panel) corresponding to $p_T=0.1$ and respectively $0.5$~GeV.
However, larger values $p_T\gg Q_{0s}$ can only be acquired via a single hard scattering, which 
can occur with equal probability at any $b$ with $b\lesssim 2R$. The corresponding range for the 
$b$-integration is therefore quite large and moreover it is increasing with $B$, due to the presence of the modified
Bessel function $\rmI_2\big(bB/2R^2)$ within the numerator in \eq{epsilonpp}. 
This explains the curve for $p_T=1$~GeV in the left panel of \fig{fepsilon}.
}

\begin{figure}[t]                                       
 \includegraphics[width=8.cm] {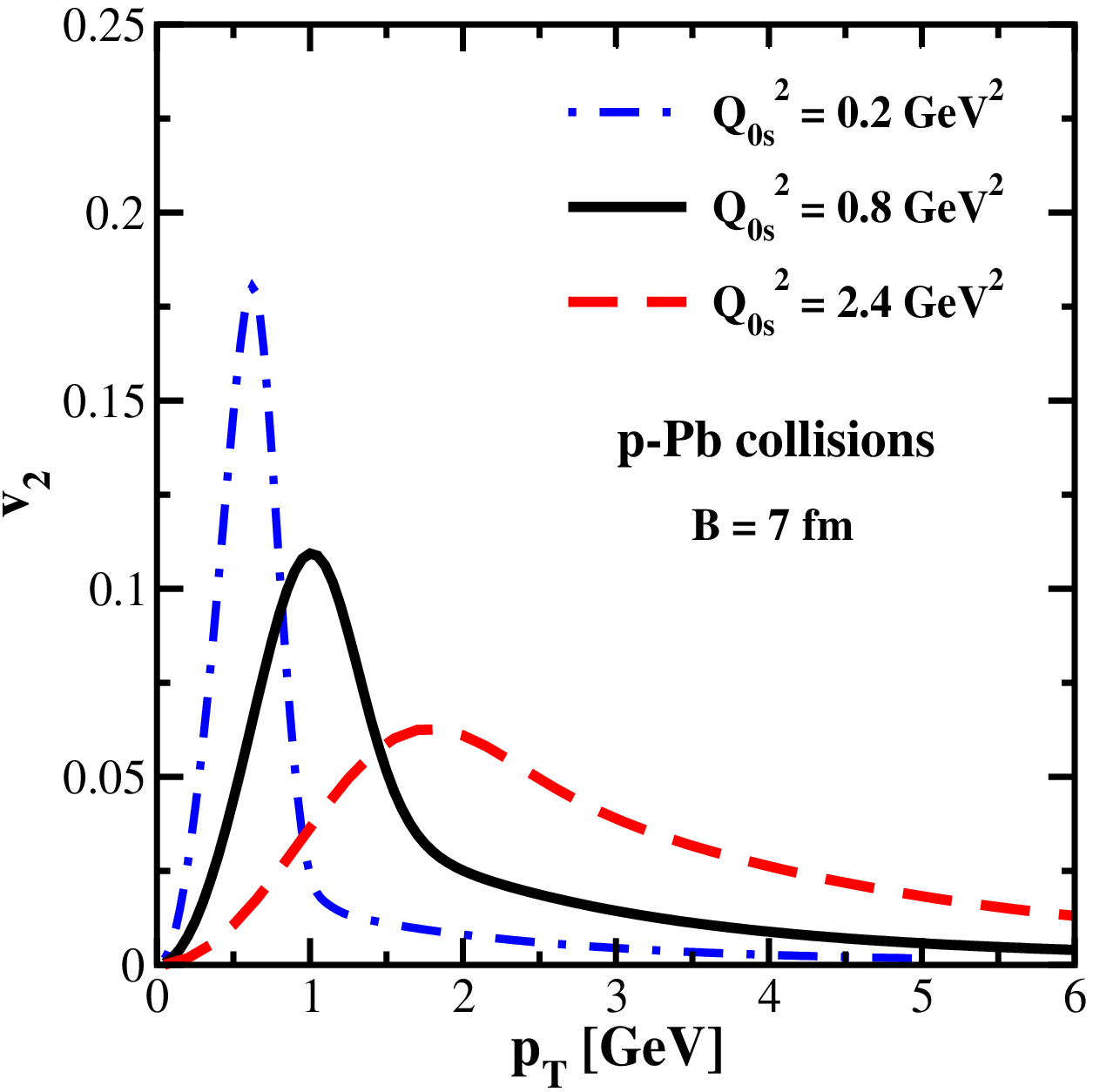}  \qquad
 \includegraphics[width=8.cm] {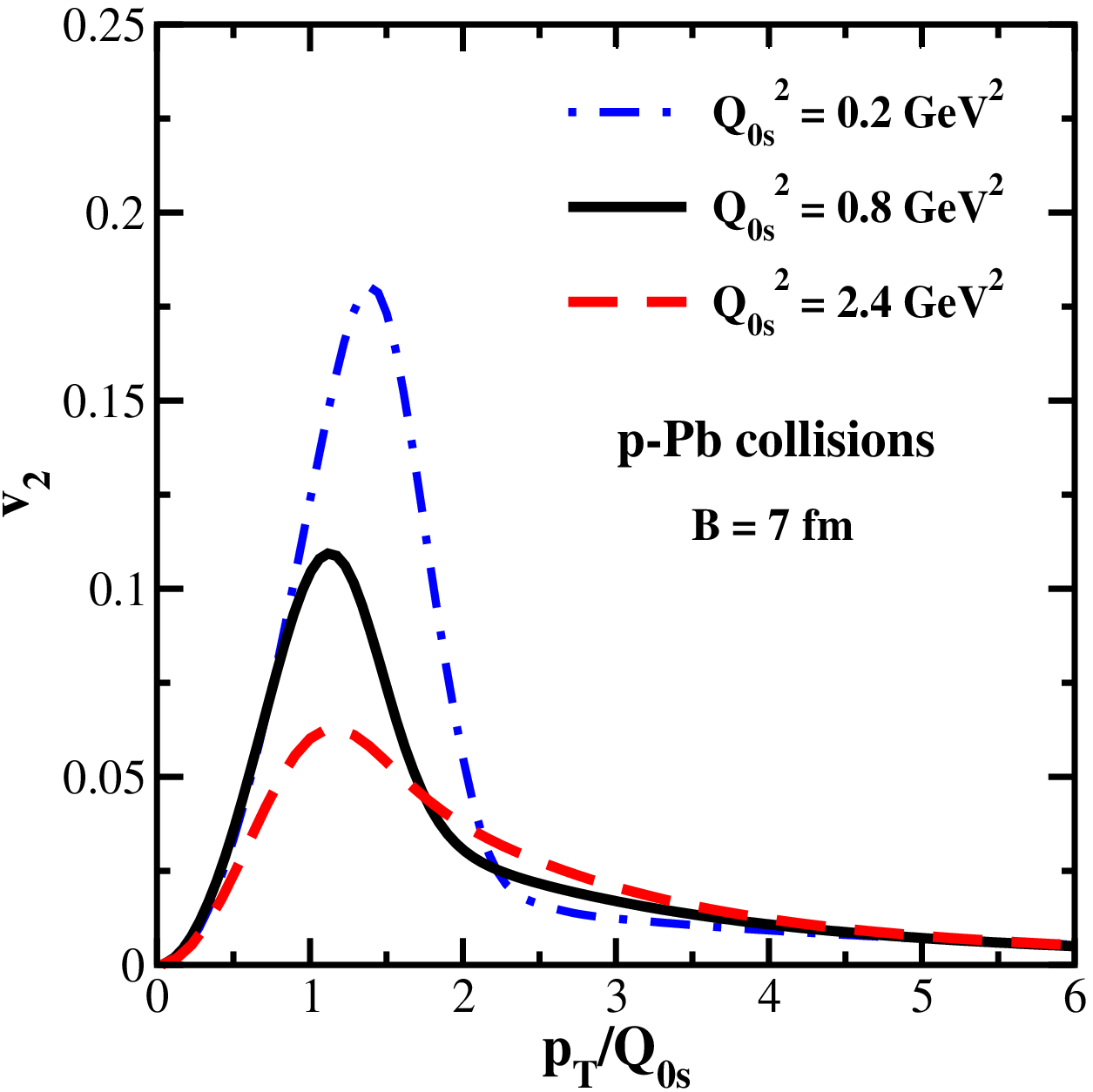}  
\caption{The azimuthal asymmetry $v_2$ in p+Pb collisions for a point-like projectile proton.
 Left: 3 different values for the central saturation momentum in the proton:
 $Q_{0s}^2=0.2,\,0.8,\,\mbox{and}\,2.4\,   \text{GeV}^{2}$ for
a fixed impact parameter $B=7$ fm.
 Right: the same as the left panel plotted as a function of $p_T/Q_{0s}$.  
 All these results are obtained by using  $m=0.2$~GeV and $R^2=2\,\,\text{GeV}^{-2}$.
When interpreting these plots, one should keep in mind
that the nuclear saturation momentum at $B=0$ is $Q_{sA}^2=A^{1/3}Q_{0s}^2$ with $A^{1/3}\simeq 6$.}
\label{fa-1s}
\end{figure}

 \begin{figure}[t]                                       
 \includegraphics[width=8.cm] {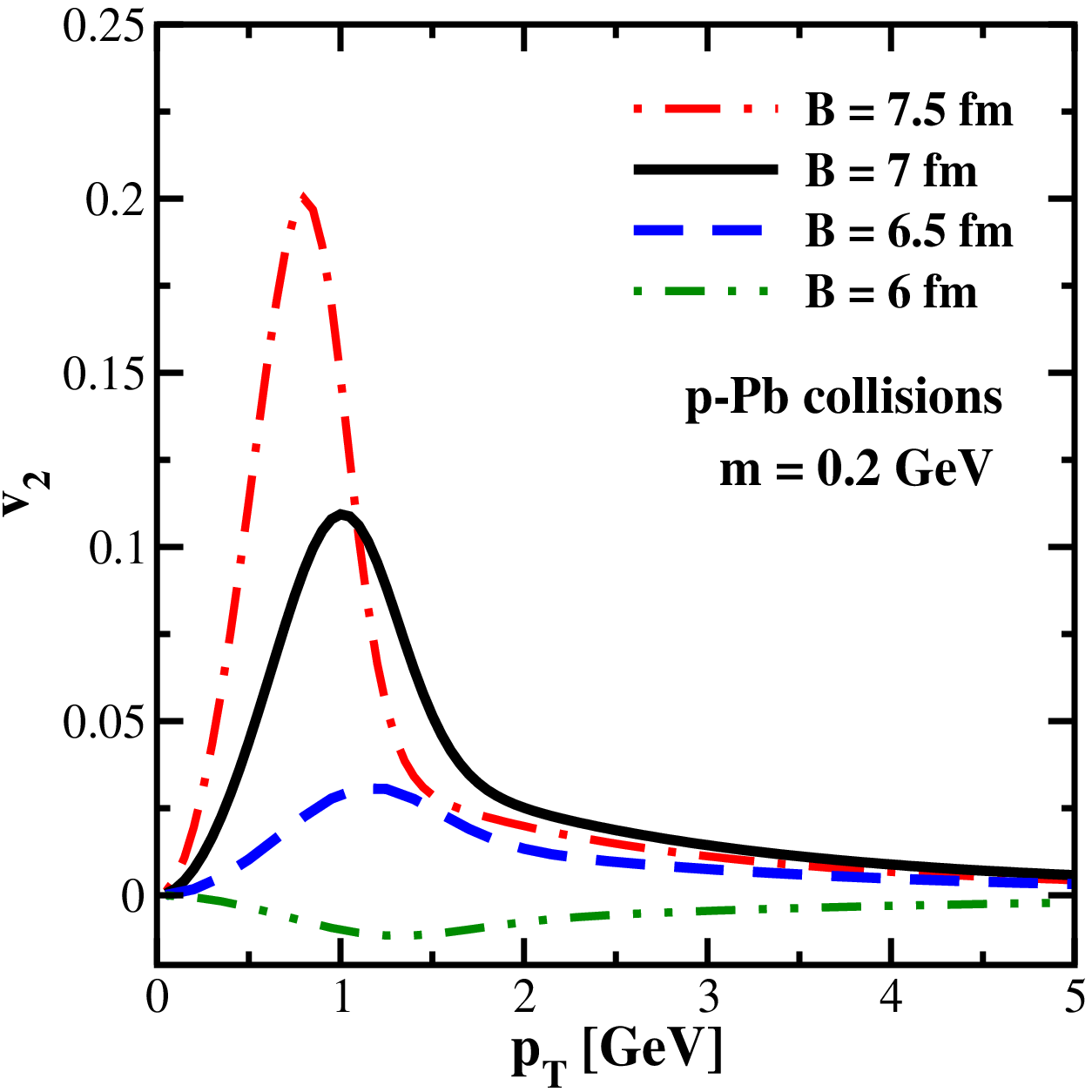}  \qquad
  \includegraphics[width=8.cm] {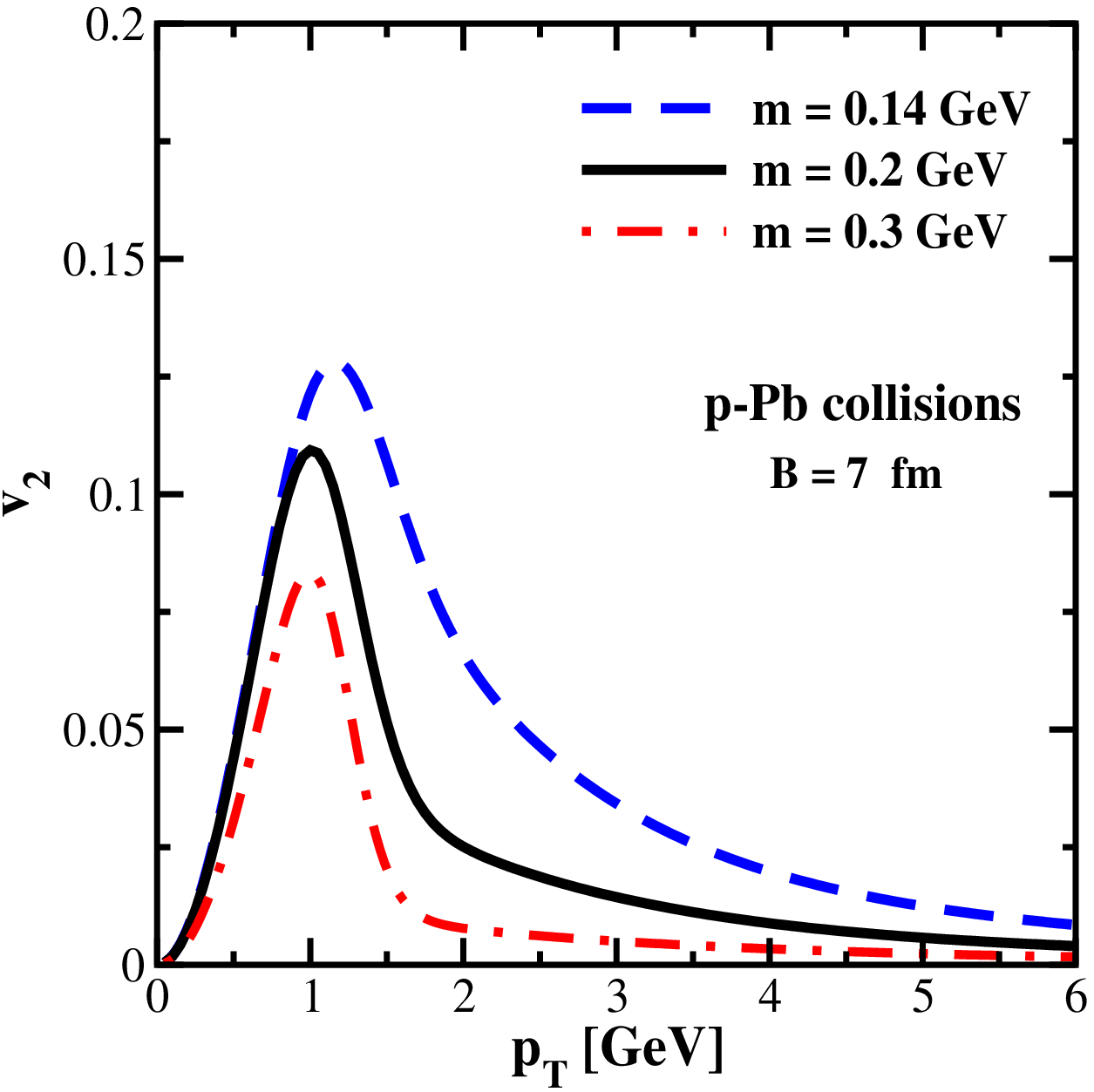}   
\caption{The azimuthal asymmetry $v_2$ in p+Pb collisions for a point-like projectile proton.
Left: 4 different impact parameters $B=6,\,6.5,\,7,\,\mbox{and}
\,7.5$~fm, for a fixed value $m = 0.2$~GeV. Right:  the mass ($m$) dependence of $v_2$ for a fixed impact parameter $B=7$ fm. All these results are obtained by using  $Q_{0s}^2=0.8\,\text{GeV}^{2}$
and $R^2=2\,\,\text{GeV}^{-2}$.}
\label{fa-1}
\end{figure}

We now turn to the case of $pA$ collisions, for which the present formalism is somewhat better justified. 
The respective $v_2$ is computed by numerically integrating \eq{v2m-f2} with the dipole amplitude given
by the analytic results in Eqs.~\eqref{NA0} and \eqref{NAtheta}.
The two plots in \fig{fa-1s}, which exhibit $v_2$ as a function of $p_T$ (left panel) and 
$p_T/Q_{0s}$ (right panel), for different values of the proton saturation scale  $Q_{0s}$, 
are quite similar to the corresponding plots for $pp$ collisions, cf. \fig{fp-1}. In particular,
the peak position appears to respect the expected scaling with the {\em nuclear} saturation momentum
$Q_{sA}=A^{1/6}Q_{0s}$ : indeed,  the maximum occurs at, roughly, $p_{\rm max}/Q_{0s}\simeq 1.2$, 
which is larger by a factor $A^{1/6}\simeq 2.4$
(for $A=208$) than the respective value observed for $pp$ collisions. However, in order
to reach values for $v_2$ which are comparable to those in $pp$ collisions, one now needs to 
go up to much larger values of the impact parameter $B\gtrsim R_A$, there the inhomogeneity
in the nuclear distribution is located (cf. the discussion in Sect.~\ref{sec:pA}). 
The $B$-dependence of the function $v_2(p)$ is illustrated in the left panel of \fig{fa-1}.
This is controlled by the combination $T^{\prime\prime}_A(B)-T^{\prime}_A(B)/B$, cf. \eq{t-exp},
and the results in \fig{fa-1} are indeed in agreement with the previous discussion of
 \fig{f4-4}. Namely, $v_2$ is seen to be sizable and positive for all values $B\ge R_A\approx 6.5$~fm. 

Notice that in the present approximations, the dipole
amplitude (hence our estimate for $v_2$) for the case of a nuclear target depends upon the two scales
 $R^2$ and $Q_{0s}^2$ mostly via their product $R^2Q_{0s}^2$.  (This becomes obvious by
 inspection of Eqs.~\eqref{NA0} and \eqref{NAtheta}.)  Accordingly, the effect of
increasing $Q_{0s}^2$ at fixed $R^2$, as visible in  \fig{fa-1s}, can alternatively be 
associated with increasing $R^2$ for a fixed value $Q_{0s}^2$. 
In the right panel of \fig{fa-1}, we show the dependence of $v_2$ in $pA$ collisions upon the infrared cutoff $m$.
Similarly to the case of $pp$ collisions, one finds that this dependence is rather strong: by decreasing
$m$ from the  `confinement' value $m=0.3$~GeV to the pion mass $m=0.14$~GeV, one increases
the peak value of $v_2$ by a factor of 3.

\begin{figure}[t]                                       
                                  \includegraphics[width=8.cm] {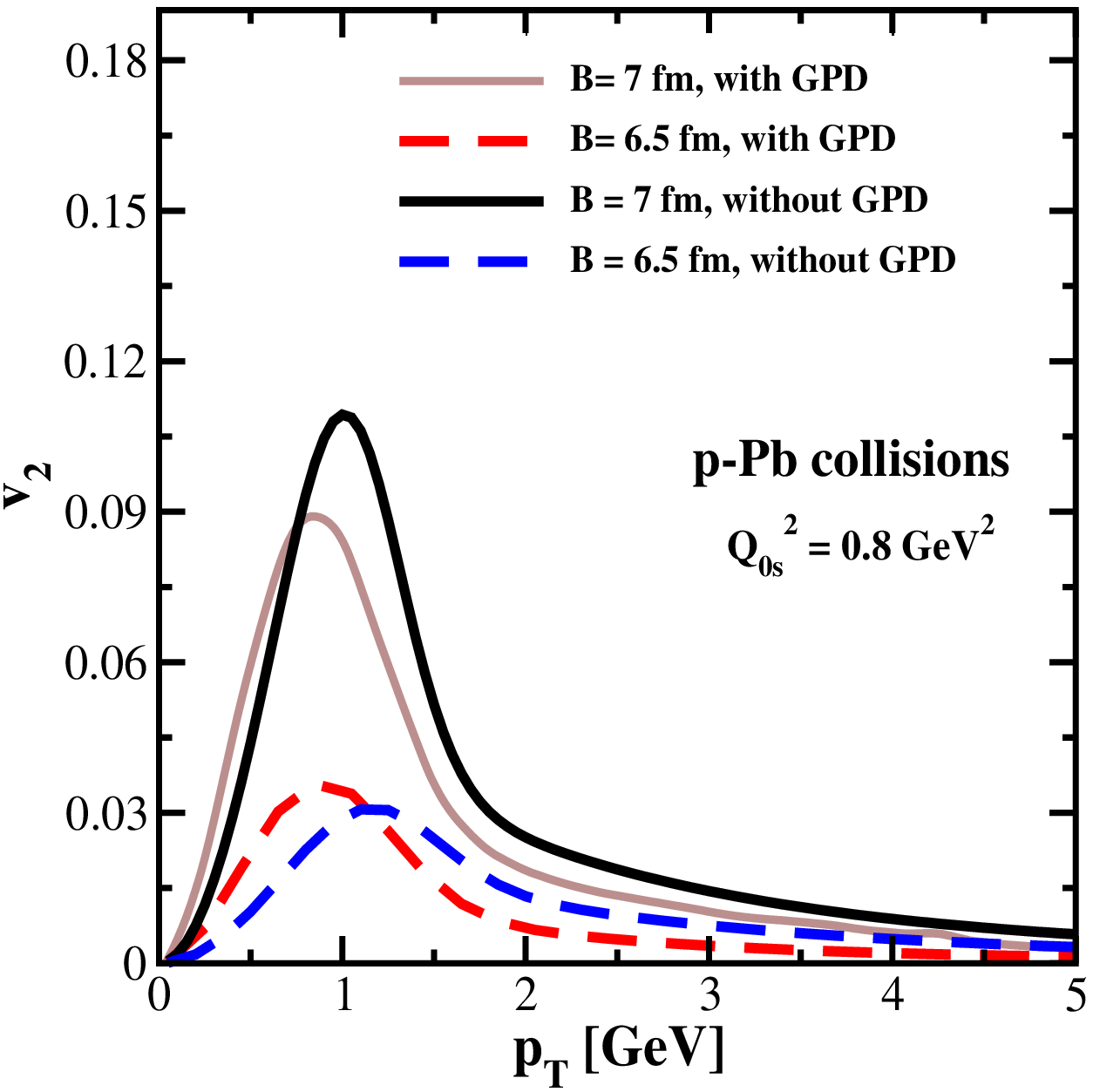}        \qquad                             
  \includegraphics[width=8.cm] {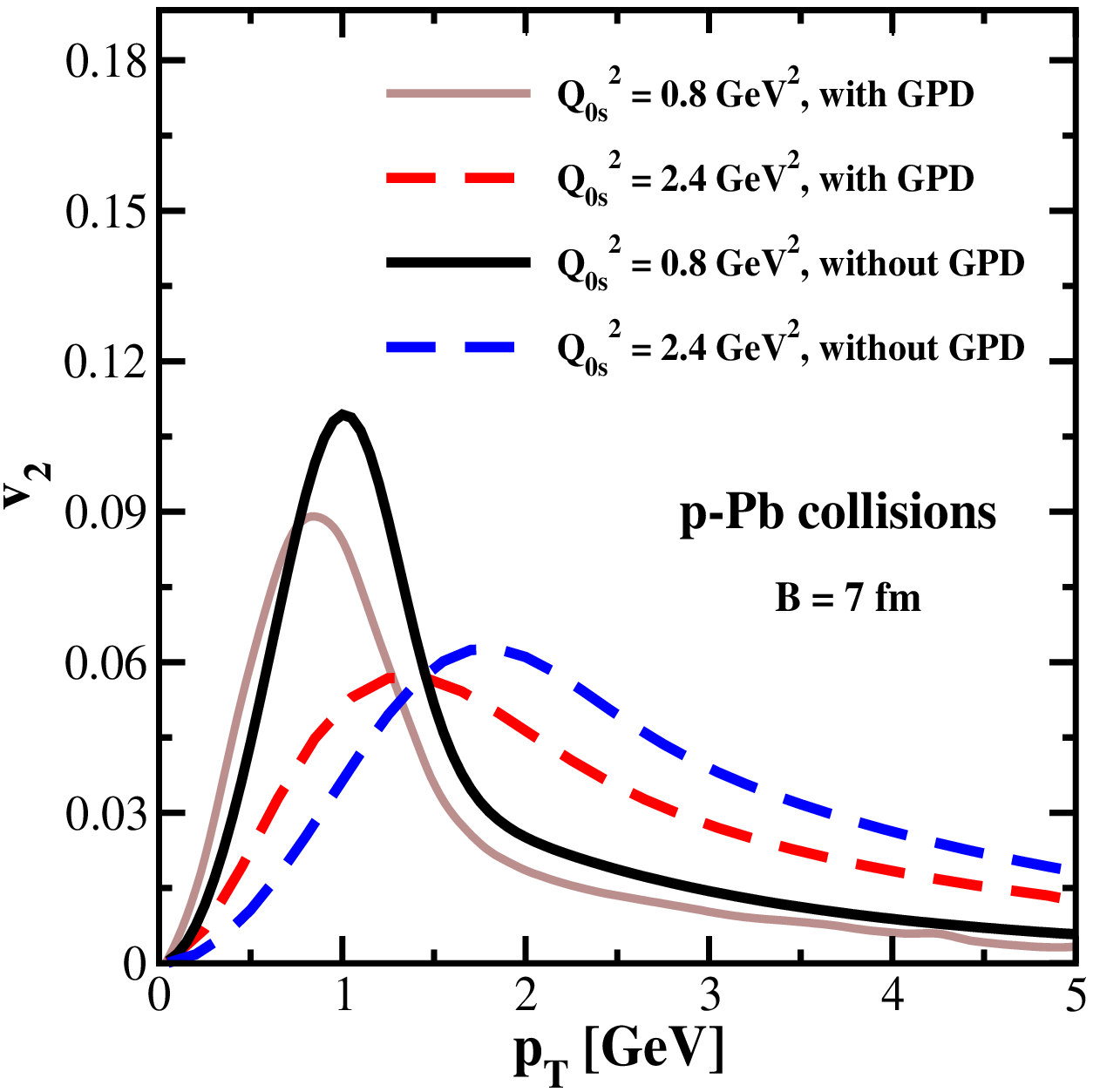}    
\caption{The azimuthal asymmetry  $v_2$ in $pA$ collisions: the results for an extended projectile
(``with GPD'') are compared to those for a point-like projectile (``without GPD'').
Left: 2 different impact parameters: $B=6.5\ \mbox{and}\ 7$ fm, for fixed
 $Q_{0s}^2=0.8\,\text{GeV}^{2}$.  Right: 2 different values for the central saturation scale in the 
proton: $Q_{0s}^2=0.8 \ \mbox{and}\ 2.4 \,\text{GeV}^{2}$, for a fixed 
 $B=7$ fm.  All these results are obtained by using 
 $m=0.2$ GeV and $R^2=2\,\,\text{GeV}^{-2}$.}
\label{pA-B-Qs}
\end{figure}

In \fig{pA-B-Qs} we illustrate the effect of using an extended proton projectile, with Gaussian
distribution in impact parameter. The corresponding formula for $v_2(p_T,B)$ is the straightforward
generalization of \eq{sv2f}, obtained by replacing $\mathcal{N}_0 (b,r) \to A\mathcal{N}^{A}_0 (b,r)$ and
$\mathcal{N}_{\theta} (b,r) \to A\mathcal{N}^{A}_{\theta} (b,r)$. As visible in  \fig{pA-B-Qs}, the effect
is quite small --- at most a change of 20\% in the value of $v_2$ at its peak.

The systematics of the above results for $v_2$ can be physically understood as follows.
First of all, we found that $v_2(p_T)$ is small for both very small and very large values of $p_T$,
but has a maximum at some intermediate value $p_{\rm max}$. In particular, $v_2=0$ for $p_T=0$, as already
obvious by inspection of \eq{sv2}. These features are easy to understand: the angular orientation cannot
play any role when either the momentum $p_T$, or the dipole size $r$, are too small. Since typically $r\sim 1/p_T$,
the second argument explains the rapid decrease of $v_2$ that we observe at high $p_T$. But the detailed shape
of the function $v_2(p_T)$ --- in particular the position, the width, and the height of its maximum --- are strongly
dependent upon the impact parameter and also upon the values of the 3 parameters $Q_{0s}$, $m$, and $R$.

Specifically, as visible in the left panels in both \fig{fp-2} and \fig{fa-1}, 
$v_2$ is negligible for relatively small impact parameters (in particular,
it vanishes as $b, B\to 0$), but it becomes large --- in the sense
that it reaches a peak value $v_2(p_{\rm max})\gtrsim 0.1$ --- when the impact parameter is
comparable to the typical size for inhomogeneity
in the target, that is, $b\sim R \gtrsim 0.2$~fm for a proton and respectively $B\sim R_A\gtrsim 6.5$~fm for a
large nucleus. This is understandable, given that the angular orientation would play no role for a target which
is homogeneous in impact-parameter space. We recall that, for the mechanism under consideration, 
the elliptic flow is driven by the sensitivity of the color-dipole orientation to the variation in the gluonic or 
nuclear distribution in the transverse plane. 

We furthermore found that the peak in $v_2(p_T)$ moves towards larger values of $p_T$
and becomes broader when increasing $Q_{0s}^2$, see \fig{fp-1} and \fig{fa-1s}.  
This is as expected: the larger saturation momentum in the target,
the larger is the typical momentum of the produced parton and the wider is its distribution in $p_T$.
Interestingly, for both $pp$ and $pA$ collisions we found that the position $p_{\rm max}$ of the peak in
$v_2(p_T)$ is proportional to $Q_{0s}$. A similar observation was recently made in Ref.~\cite{Hagiwara:2017ofm}.
When $v_2(p_T)$ is plotted as a function of $p_T/Q_{0s}$, the
peak position $p_{\rm max}/Q_{0s}$ is quasi-independent of $Q_{0s}$, albeit its height and shape
are still strongly dependent (see the right panels in \fig{fp-1} and \fig{fa-1s}). 
Specifically, the maximal value at the peak $v_2(p_{\rm max})$ appears to increase when decreasing $Q_{0s}$,
i.e. when the target becomes {\em more dilute}. This may seem counter-intuitive since, as already stressed,
the multiple scattering represents an essential ingredient of the mechanism under consideration (it even 
changes the sign of $v_2$ as compared to the single-scattering approximation). However, the importance
of the dipole orientation depends in a crucial way upon the balance between the dipole size and the
size of its impact parameter. The dipole size is fixed by the transverse momentum of the produced
quark, $r\sim 1/p_T$, which in turn is determined by the target saturation momentum: $p_T\sim
Q_s(b)\propto Q_{0s}$. Hence, if one keeps increasing $Q_{0s}$, 
the dipole size eventually becomes much smaller than $b$ and the dipole
orientation plays no role anymore. A similar effect is seen when the saturation momentum increases
as a consequence of the high-energy evolution \cite{Kovner:2012jm,Lappi:2015vha,Lappi:2015vta}.


Finally, given the importance of soft, non-perturbative, exchanges for the angular-dependence
of the dipole amplitude, is should be no surprise that our results for $v_2$ are rather strongly dependent
to the `confinement' scale $m$:  the anisotropy is enhanced when decreasing $m$, since the
phase-space for soft exchanges is rapidly increasing, see the right panels in \fig{fp-2} and \fig{fa-1}. 
For the angular-dependent piece of the dipole amplitude and for a proton target, this dependence
has been already exhibited in Fig.~\ref{figntheta}, whereas for a nuclear target, it is directly
visible by inspection of \eq{NAtheta} for $\mathcal{N}^{A}_{\theta}$.

 \begin{figure}[t]                                                  
 \includegraphics[width=8.5 cm] {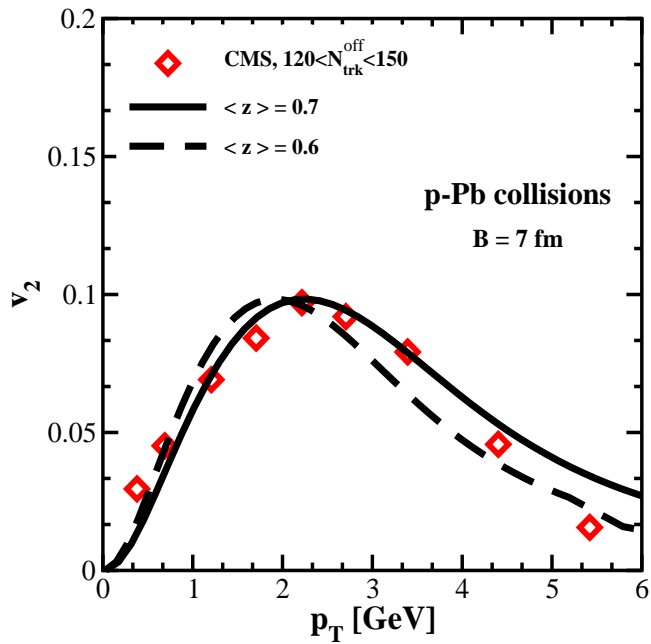}  
\caption{The azimuthal asymmetry $v_2$ in p+Pb collisions at a fixed impact-parameter $B =7$ fm and two different values for the average splitting fraction $\langle z\rangle$. The theoretical results refer to an extended projectile proton, with a Gaussian distribution in impact parameter (GPD). All results are obtained at a fixed $Q_{0s}^2=3\,\text{GeV}^2$ and $R^2=2$~GeV$^{-2}$ and $m=0.1$ GeV.   The experimental data, by the CMS collaboration \cite{Chatrchyan:2013nka}, refer to the second-order elliptic harmonic $v_2\{4\}$ extracted 
from the four-particle cumulant via $v_2\{4\} = [-c_2\{4\}]^{1/4}$. }
\label{fa-3}
\end{figure}

 From the previous considerations in this paper, it should be clear that our current analytic description
 for the mechanism under consideration is too crude to allow for quantitative predictions, or realistic
 applications to the phenomenology. This being said, we would like to show via an example that
 this scenario is not {\em excluded} by the current data. Namely, we will show that, by appropriately
 choosing the values of the impact parameter $B$ and of the free parameters of the models, one can give
 a reasonable description of the $p_T$-dependence of the elliptic flow $v_2(p_T)$ extracted from multi-particle
 azimuthal correlations in p+Pb collisions at the LHC, in a given multiplicity class. This should not be
 confounded with a genuine fit to the data --- it is merely an exploratory comparison.  Given the uncertainties
 inherent in our model, we shall adopt a rather crude strategy for relying the predictions of this
 model to the phenomenology.
 
First, we have not attempted to compute the consequences of our mechanism for multi-particle
azimuthal correlations; rather, we shall make the simplifying assumption that the final particles
are correlated with each other only through the flow correlations with the reaction plane (that is, we 
neglect possible `non-flow correlations'). Under this assumption, we can write $c_2\{2\}\simeq v_2^2$,
and similarly $c_2\{4\}\simeq -v_2^4$, where $c_2\{2\}$ and $c_2\{4\}$ are the second-order
two-particle and respectively four-particle cumulants, as defined e.g. in \cite{Borghini:2000sa,Snellings:2011sz},
and $v_2$ is the usual elliptic flow coefficient, as discussed throughout this paper. More precisely,
our present estimates for $v_2$ refer to a fixed impact parameter $B$, that is, we have a prediction
for the function $v_2(p_T,B)$, whereas the data for p+Pb (and also p+p) collisions are rather classified 
according to the particle multiplicity (the number of reconstructed tracks) in the final state. 
Due to the large multiplicity fluctuations, the correlation between
the multiplicity classes and the cuts in impact parameter is rather loose and not well under control
(see e.g. the discussion in \cite{Adam:2014qja}). 
To cope with that and by lack of any better alternative, we shall simply select
a value of $B$ for which the predictions of our model appear to reasonably agree with the data in a 
given high-multiplicity class. We shall similarly proceed with the free parameters of our model,
which in the case of $pA$ collisions are the dimensionless product $R^2Q_{0s}^2$ and the infrared cutoff $m$.
It is understood that the values for $B$, $m$ and $R^2Q_{0s}^2$ which will emerge from this procedure
must also reflect the influence of other uncertainties or
simplifications inherent in our present approach, like the omission of gluons.
(Gluon production within the present set-up should give rise to a similar $p_T$-dependence in $v_2$,
but only modify its overall magnitude.)

Finally, to have a better comparison with the $p_T$-dependence of the data, which refer to {\em hadrons},
one must take into the account the effect of the quark (or gluon) fragmentation into hadrons. This too will
be implemented in a rather heuristic way,  by assuming that $p_{\text{hadron}}= \langle z\rangle p_{\text{quark}}$ 
where $\langle z\rangle$ is the average value of the splitting fraction and will be treated as a free parameter.


In \fig{fa-3}, we  compare  our results for $v_2$ in peripheral $pA$ collisions (with $A=208$)
with the experimental data from the CMS experiment for 
the second-order elliptic harmonic $v_2\{4\}$  in p-Pb collisions, as extracted from a four-particle cumulant
analysis in events where the number of reconstructed tracks lies in the range $120<N_{\rm trk}^{\rm off}<150$
\cite{Chatrchyan:2013nka}. The two theoretical curves are obtained by using an {\em extended} projectile 
(so the impact parameter $B$ refers to the center of this projectile, as in \fig{pA-B-Qs}),  together with the  
following values for the free parameters:
$B=7$~fm,  $R^2=2\,\,\text{GeV}^{-2}$, $Q_{0s}^2=3$~GeV$^2$, $m=0.1$~GeV, and two values
for $\langle z\rangle$: 0.6 and respectively 0.7. Notice that this value $Q_{0s}^2=3$~GeV$^2$ is in the ballpark
of the theoretical expectations for the proton saturation momentum at the LHC energies. It is quite
remarkable that, in spite of the many simplifications and the crude assumptions, our mechanism appears to be able
to produce the correct $p_T$-dependence of $v_2$ and also its correct size, with reasonable values for
the free parameters. 


 \section{Conclusions and perspectives}
 
In this paper, we have analyzed a less explored, albeit not totally new, mechanism for generating
azimuthal asymmetries in particle production in `dilute-dense' collisions: the dependence of the
cross-section for single-inclusive particle production upon the azimuthal orientation of the momentum
of the produced particle w.r.t. its impact parameter. As compared to previous related studies in the literature
\cite{Kopeliovich:2007fv,Kopeliovich:2007sd,Kopeliovich:2008nx,Levin:2011fb,Zhou:2016rnt,Hagiwara:2017ofm},
we have considered a different model for the gluon distribution in the dense target --- an extension of the 
McLerran-Venugopalan (MV) model --- which combines the proper pQCD tail at high transverse momenta
with a Gaussian profile (inspired by fits to the HERA data) for the distribution of the color charges in the 
impact-parameter space and a `gluon mass' which mimics confinement. A realistic, or at least
physically motivated,  treatment of the non-perturbative aspects related to the transverse inhomogeneity
and to confinement is indeed essential, since, as demonstrated by our analysis, these aspects do actually 
control the azimuthal asymmetries produced by this mechanism.

Within this set-up, we have given a systematic, semi-analytic, study of the angular dependence 
of the dipole scattering amplitude and we have used the result to compute the elliptic flow coefficient
$v_2$, separately for proton-proton and proton-nucleus collisions 
(the main difference being the inclusion of lumpiness effects in the case of a nuclear target).
We thus found that, as a function of the impact parameter, $v_2$ is rather strongly peaked near
the edge of the target, where the transverse inhomogeneity is more pronounced. Hence the present
mechanism will significantly contribute to azimuthal asymmetries only via peripheral collisions.  Furthermore,
as a function of the transverse momentum of the produced particle, $v_2$ shows a maximum at some
intermediate, semi-hard, value which is proportional to the target saturation momentum at $b=0$. 
This maximum broadens and lowers with increasing the 
saturation momentum. The overall  $p_T$-dependence is quite similar to that observed in the phenomenology 
of p+Pb collisions at the LHC. In fact, a semi-quantitative agreement with the data can be obtained with
reasonable choices for the model parameters, but this agreement should be taken with a grain of salt, 
because our model is quite crude and additional  approximations are performed when comparing with the data.

To convincingly demonstrate this mechanism, further studies are necessary. First, one would like to
understand its evolution with increasing energy. To that aim, one should solve the Balitsky-Kovchegov 
(BK) equation  \cite{Balitsky:1995ub,Kovchegov:1999yj} with initial conditions at low energy provided by
our present model. In the respective solutions, one must keep trace of the impact-parameter dependence 
(including the angular dependence) of the dipole amplitude and one
must enforce confinement within the soft gluon emission kernel, preferably by using the same infrared regulator 
(`gluon mass') $m$ as in the initial condition. Similar solutions have been already considered in
\cite{Motyka:2009gi,Berger:2011ew,Berger:2012wx}.

Furthermore, since the orientation of the reaction plane cannot be experimentally measured, one must
compute the imprint of the azimuthal asymmetries generated by the present mechanism on multi-particle correlations.
For instance, one can study the simultaneous production of 2 quarks in $pA$ collisions. The calculation can be 
simplified by assuming that the two quarks (originally collinear with the projectile) scatter 
independently of each other and by taking the multicolor limit $N_c\to\infty$. 
(We recall that the multi-particle correlations generated by the present 
mechanism survive in the large-$N_c$ limit.) As anticipated in the Introduction, we expect the
multi-particle correlations to be important only for 
sufficiently peripheral collisions and to lead to {\em flow} --- a collective motion of particles which are
produced independently from each other, but which are all correlated with the reaction plane
defined by their average impact parameter. Note that even for independent partons in the projectile at large $N_c$, the double-patron-scattering (DPS) Hanbury Brown and Twiss (HBT) correlations may lead to anisotropy \cite{Kovner:2017vro}. This effect is not accounted for in the present mechanism.

Given the prominence of the peripheral physics for the physical problem at hand, we furthermore expect that 
the fluctuations in the shape of the colliding hadrons should play an important role: they should  amplify
the inhomogeneity in impact parameter space and thus enhance the azimuthal asymmetries. The importance
of such fluctuations is supported by a recent analysis of the HERA data for incoherent exclusive diffractive vector meson production in deep inelastic scattering \cite{Mantysaari:2016jaz}. It would be
therefore interesting to redo our present analysis of the angular dependence of the dipole scattering for the case
where the (projectile and/or target) proton has strong shape fluctuations, say as described by the models used
in  \cite{Mantysaari:2016jaz} and which are constrained by the HERA data.

\begin{acknowledgments}
We would like to thank Y. Hagiwara, Y. Hatta, B.-W. Xiao, and F. Yuan for sharing with us a preliminary
version of their recent work \cite{Hagiwara:2017ofm} and for related discussions. We are grateful to T. Lappi, J.-Y. Ollitrault and L. McLerran for comments on the manuscript.
 This research was supported in part by ECOS-Conicyt C14E01.
 The work of E.I. was supported in part by the Agence Nationale de la Recherche project 
 ANR-16-CE31-0019-01. 
  The work of A.R. was supported in part
 by Fondecyt grant 1150135,  Anillo ACT1406 and Conicyt PIA/Basal FB0821.
\end{acknowledgments}

 \appendix
\section{Computing the dipole amplitude in the two-gluon exchange approximation}
\label{app:MS}

In this Appendix, we present details on the calculation leading to  Eqs.\,(\ref{t2f1}) and (\ref{t2f2}) for the
dipole scattering amplitude in the two-gluon exchange approximation. The starting point is \eq{T2gsmallr}
for $N_{2g}(\b,\r)$, which, we recall, has been obtained from the general MV-model formula \eqref{T2g0}
by expanding to second order in $\k\cdot\r$.

Specifically, let $\beta$ and $\theta_k$ denote the angles between $\D$ and $\k$ 
and, respectively, between $\b$ and $\k$. Then we can rewrite \eq{T2gsmallr} as
\beqn\label{t2app}
N_{2g}(b,r,\theta)&=& \frac{g^2C_F}{2} r^2 \int\int  \frac{\Delta\rmd\Delta }{(2\pi)^2}\frac{k\rmd k }{(2\pi)^2}\,  \tilde{\mu}(\Delta) \nonumber\\*[0.2cm]
&\times& \int \int  \rmd\beta \rmd\theta_k\,
 \frac{ \big[k^2\cos^2\left(\theta-\theta_k \right) 
 - ({\Delta^2}/{4})
\cos^2\left(\beta+\theta-\theta_k \right)\big]
 \rme^{\rmi\Delta b \cos(\theta_k -\beta)}}{\left[k^2+\Delta^2/4 +k\,\Delta \cos\left(\beta\right)+m^2\right]   \left[k^2+\Delta^2/4 -k\,\Delta \cos\left(\beta\right)+m^2\right] }\,.  \
\eeqn
We first perform the integral over the angle $\theta_k$ using the identity in \eq{id},  to obtain
 \beqn\label{id2}
\int_0^{2\pi}  \rmd\theta_k\,  \rme^{\rmi\Delta b \cos(\theta_k -\beta)}\, \cos^2\left(\theta-\theta_k \right)
&= &\pi \Big[\rmJ_0(\Delta b)-\cos(2(\theta-\beta))\,\rmJ_2(\Delta b)\Big] , \nn
\int_0^{2\pi}  \rmd\theta_k\,  \rme^{\rmi\Delta b \cos(\theta_k -\beta)}\, \cos^2\left(\beta+\theta-\theta_k \right) &=& \pi \Big[\rmJ_0(\Delta b)-\cos(2\theta)\,\rmJ_2(\Delta b)\Big] . 
\eeqn
The ensuing integrals over  $\beta$ can be now performed by using the following identities 
\beqn\label{id3}
\int_0^{2\pi}  \rmd\beta\, \frac{1}{A^2-B^2\cos^2(\beta)}&=&\frac{2\pi}{A \left(A^2-B^2 \right)^{1/2}}\,, \nonumber\\
\int_0^{2\pi}  \rmd\beta\, \frac{\cos(2\beta)}{A^2-B^2\cos^2(\beta)}&=&-\frac{2\pi}{A \left(A^2-B^2 \right)^{1/2}}-\frac{4\pi}{B^2}+\frac{4\pi A}{B^2 \left(A^2-B^2 \right)^{1/2}}\,.  \
 \eeqn
with $A=k^2+\Delta^2/4+m^2$ and $B=k\,\Delta$.  We thus confirm the general structure
in \eq{t2-f3} and at the same time obtain more explicit expressions for the two functions 
$\mathcal{N}_0(b,r)$ and $\mathcal{N}_\theta(b,r)$, as double
radial integrals to be numerically computed. Specifically, we find
\beqn\label{t2f10}
\mathcal{N}_0(b,r)&=& \frac{g^2C_F}{4(2\pi)^2} r^2 \int_{0}^\infty \rmd\Delta \int_{0}^{\infty} \rmd k\,
\frac{ \tilde{\mu}(\Delta)\,\rmJ_0(\Delta b) \,k\Delta
 \left(k^2 -\Delta^2/4\right)}{\left(k^2+\Delta^2/4+m^2\right) \left( \left(k^2+\Delta^2/4+m^2\right)^2-k^2\Delta^2  \right)^{1/2}}\,,
\eeqn
together with \eq{t2f2} for $\mathcal{N}_\theta(b,r)$.
It is easy to check that the double integral yielding $\mathcal{N}_\theta(b,r)$ is well defined 
as is stands, so its numerical evaluation poses no special problem. 
 On the other hand, the integral over $k$ giving $\mathcal{N}_0(b,r)$ has a
logarithmic ultraviolet divergence and needs to be cut off at $k\sim 1/r$, as already explained.
The precise implementation of this UV cutoff introduces some ambiguity, that we shall fix by replacing
 \eq{t2f10} with  \eq{t2f1}, as discussed in the main text.
 
As also explained in Sect.~\ref{sec:SSA}, the limit $m\to 0$ of $N_{2g}$ is well defined,
because of the explicit transverse inhomogeneity in the target. We have already studied this
limit for the piece $\mathcal{N}_\theta(b,r)$ which controls the angular dependence, cf. \eq{t2f20}.
In what follows we shall perform the corresponding study for the other piece $\mathcal{N}_0$.
To that aim, it is more convenient to use \eq{t2f10}, in which we set $m=0$ and simply cut the
integral over $k$ in the ultraviolet at $k\sim 1/r$. Using \eq{root}, we find
\beqn\label{t2f11}
\mathcal{N}_0(b,r)\Big |_{m=0}&=& \frac{g^2C_F}{4(2\pi)^2} r^2 \int_{0}^\infty \rmd\Delta\,\Delta\,
 \tilde{\mu}(\Delta)\,\rmJ_0(\Delta b)
 \int_{0}^{1/r} \rmd k\,\frac{k}{k^2+\Delta^2/4}\,\big[\Theta(k-\Delta/2)-\Theta(\Delta/2-k)\big]\,,\nn
 &=& \frac{g^2C_F}{8(2\pi)^2} r^2 \int_{0}^\infty \rmd\Delta\,\Delta
 \tilde{\mu}(\Delta)\,\rmJ_0(\Delta b)\,\ln\frac{1}{r^2\Delta^2}\,.
\eeqn
This is strictly true for $1/r \gg \Delta$, meaning $r\ll R$ (recall that the integral over $\Delta$
is restricted to $\Delta \lesssim 1/R$ by the transverse inhomogeneity in the target). At this point,
it is convenient to separate the logarithm as $\ln({1}/{r^2\Delta^2})=\ln(R^2/r^2) + \ln(1/R^2\Delta^2)$,
where the first piece in the r.h.s. is the `large logarithm' (since typically $r\ll R$), whereas the argument
of the second logarithm is of order one. One then easily finds
\beqn\label{t2f12}
\mathcal{N}_0(b,r)\Big |_{m=0}&=&\frac{Q_s^2(b)r^2}{4} \ln \frac{R^2}{r^2}\,+
 \frac{g^2C_F}{8(2\pi)^2} r^2 \int_{0}^\infty \rmd\Delta\,\Delta\,
 \tilde{\mu}(\Delta)\,\rmJ_0(\Delta b)\,\ln\frac{1}{R^2\Delta^2}\,,\nn
 &\simeq & \frac{Q_s^2(b)r^2}{4} \ln \frac{R^2}{r^2}\,+\,Q_{0s}^2r^2\,\frac{R^2}{b^2}
\,,
\eeqn
where in the second line we kept only the power-law tail at large $b$, as developed by
the integral over $\Delta$ from the first line. (This power law tail can be easily obtained by letting 
$ \tilde{\mu}(\Delta)\to  \tilde{\mu}(0)$ inside the integrand, as appropriate for large $b\gg R$.)
Notice that for very large impact parameters $b\gg R$, the first term proportional to the local saturation
momentum $Q_s^2(b)$ is exponentially suppressed and the scattering amplitude (which is small anyway)
is controlled by the power-law tail.  However, this is not the situation that we have considered throughout 
this work; indeed, we have looked at impact parameters $b\lesssim R$, where
the scattering amplitude is dominated by the first term $\propto Q_s^2(b)$.

\section{More details on the case of a lumpy nuclear target}
\label{app:pA}

In this Appendix, we shall present more details on the derivation of Eqs.~\eqref{NA0} and \eqref{NAtheta}
for the dipole scattering off a lumpy nucleus in the two-gluon exchange approximation. Namely,
we would like to compute the integral over $\b$ in \eq{ta-1} with the nuclear thickness function
given by the second-order expansion in \eq{TAsecond}.
To that aim, it is useful to introduce the following Fourier transform
\beq
\tilde N_{2g}(\D,\r)= \int {\rmd^2\b} \,\rme^{-\rmi\b\cdot\D}\,N_{2g}(\b,\r)\,.\eeq
To the order of interest, $N_{2g}(\b,\r)$ is given by \eq{T2gsmallr}, which immediately implies
\beqn\label{T2gnew}
\tilde N_{2g}(\D,\r)\,\simeq\,
\frac{g^2C_F}{2}\,r^l r^m \int \frac{\rmd^2\k }{(2\pi)^2} 
\,\frac{\big(k^lk^m-\Delta^l\Delta^m/4\big)\tilde{\mu}(\Delta)}{[(\k+\D/2)^2+m^2][(\k-\D/2)^2+m^2]}\,.
\eeqn
The interesting integrals over $\b$ can be related to the behavior of $\tilde N_{2g}(\D,\r)$ near $\Delta=0$.
We first have
\beq\label{intN}
 \int d^2\b\,N_{2g}(\b,\r)=\tilde N_{2g}(0,\r)=\frac{g^2C_Fr^2}{4}\int \frac{\rmd^2\k }{(2\pi)^2} 
\frac{k^2\, \tilde\mu(0)}{(k^2+m^2)^2}=
\pi R^2 Q_{0s}^2 r^2 \ln\left(\frac{1}{r^2m^2}+\rme\right)\,.
\eeq
We have also used here $k^lk^m\to (k^2/2)\delta^{lm}$ together with the renormalization
prescription in \eq{qst-mu} and
$\tilde\mu(0)=4\pi R^2\mu_0$.
Not surprisingly, the integral over $\b$ has generated a result proportional to the proton area $\sim R^2$.

For the terms quadratic in $b$, one similarly finds
\beqn\label{intNsecond} \int d^2\b\,N_{2g}(\b,\r)\,b^ib^j&=&-\, \frac{\partial^2}{\partial \Delta^i\partial \Delta^j}
\,\tilde N_{2g}(\D,\r)\Big |_{\D=0}\nonumber\\*[0.2cm]
&=&
\frac{g^2C_F}{2}\int \frac{\rmd^2\k }{(2\pi)^2} 
\frac{\tilde{\mu}(0)}{(k^2+m^2)^2}
\Bigg\{\frac{r^ir^j}{2}+2\delta^{ij}R^2 (\k\cdot \r)^2+ \frac{(\k\cdot \r)^2}{k^2+m^2}
\Bigg[\delta^{ij}-\frac{2k^ik^j}{k^2+m^2}\Bigg]\Bigg\}\,,\nonumber\\*[0.2cm]
&=&
\frac{\pi R^2 Q_{0s}^2}{m^2}\left\{
\frac{1}{3}\Big(2r^ir^j +\delta^{ij}\,r^2\Big)+ 2\delta^{ij}\,r^2(mR)^2
\ln\left(\frac{1}{r^2m^2}+\rme\right)\right\},
\eeqn
where in going from the second to the third line we have used the fact that, under the integral over
$\k$, one can replace 
\beq
k^ik^jk^lk^m \,\longrightarrow\,\frac{k^4}{8}\Big(\delta^{ij}\delta^{lm}+\delta^{il}\delta^{jm}
+\delta^{im}\delta^{jl}\Big).\eeq
By combining the above equations \eqref{intN} and \eqref{intNsecond} 
with the expansion in \eq{TAsecond}, one immediately
finds the results exhibited in  Eqs.~\eqref{NA0} and \eqref{NAtheta}.

\bigskip







\end{document}